\documentclass[journal=pasa, manuscript=research-paper, year=2025, volume=?]{cup-journal}

\usepackage[nopatch]{microtype}
\usepackage{booktabs}

\usepackage{lscape}

\usepackage{float,xcolor}
\usepackage[T1]{fontenc}

\usepackage{graphicx}	% Including figure files
\usepackage{amsmath}	% Advanced maths commands
\usepackage{float}
\usepackage{amssymb}	% Extra maths symbols
\usepackage{geometry}
\usepackage{hyphenat}
\usepackage{mwe}
\usepackage{blindtext}
\usepackage{wrapfig}
\usepackage[figuresright]{rotating}
\usepackage{gensymb}
\usepackage{natbib}
\usepackage{tablefootnote}

\usepackage{caption}
\DeclareCaptionFormat{bold}{\textbf{#1 #2 }#3}
\DeclareCaptionLabelSeparator{dot}{.}
\usepackage{svg}
\usepackage{array}
\usepackage{booktabs}
\usepackage{subcaption}
\usepackage{setspace,hyperref}
\hypersetup{colorlinks=true,
            linkcolor=blue,
            urlcolor=blue,
            linktoc=all,
            citecolor=blue}

\title{SPHERE/ZIMPOL insights into discs around evolved stars: arcs, asymmetries and dust properties}
\author{Kateryna Andrych}
\affiliation{School of Mathematical and Physical Sciences, Macquarie University, Balaclava Road, Sydney, NSW 2109, Australia}
\alsoaffiliation{Astrophysics and Space Technologies Research Centre, Macquarie University, Balaclava Road, Sydney, NSW 2109, Australia}
\email[Kateryna Andrych]{kateryna.andrych@mq.edu.au}

\author{Devika Kamath}
\affiliation{School of Mathematical and Physical Sciences, Macquarie University, Balaclava Road, Sydney, NSW 2109, Australia}
\alsoaffiliation{Astrophysics and Space Technologies Research Centre, Macquarie University, Balaclava Road, Sydney, NSW 2109, Australia}
\alsoaffiliation{INAF, Observatory of Rome, Via Frascati 33, I-00077 Monte Porzio Catone (RM), Italy}

\author{Hans Van Winckel}
\affiliation{Instituut voor Sterrenkunde, K.U.Leuven, Celestijnenlaan 200D bus 2401, B-3001, Leuven, Belgium}

\author{Akke Corporaal}
\affiliation{European Southern Observatory, Alonso de Córdova 3107, Vitacura, Santiago, Chile}

\author{Toon De Prins}
\affiliation{Instituut voor Sterrenkunde, K.U.Leuven, Celestijnenlaan 200D bus 2401, B-3001, Leuven, Belgium}

\author{Daniel Price}
\affiliation{School of Physics and Astronomy, Monash University, Clayton, Vic 3800, Australia}

\author{Steve Ertel}
\affiliation{Department of Astronomy and Steward Observatory, University of Arizona, 933 N. Cherry Avenue, Tucson, AZ 85721-0065, USA}
\alsoaffiliation{Large Binocular Telescope Observatory, University of Arizona, 933 N. Cherry Avenue, Tucson, AZ 85721-0065, USA}

\author{Jacques Kluska}
\affiliation{Instituut voor Sterrenkunde, K.U.Leuven, Celestijnenlaan 200D bus 2401, B-3001, Leuven, Belgium}

 %% First letter not capped
\begin{document}

\begin{abstract}
Second-generation circumbinary discs around evolved binary stars, such as post-Asymptotic Giant Branch (post-AGB) binaries, provide insights into poorly understood mechanisms of dust processing and disc evolution across diverse stellar environments. We present a multi-wavelength polarimetric survey of five evolved binary systems — AR\,Pup, HR\,4049, HR\,4226, U\,Mon, and V709\,Car — using the Very Large Telescope SPHERE/ZIMPOL instrument. Post-AGB discs show significant polarimetric brightness at optical and near-IR wavelengths, often exceeding 1\% of the system's total intensity. We also measured a maximum fractional polarisation of the scattered light for AR\,Pup of $\sim$0.7 in the $V$-band and $\sim$0.55 in the $I$-band. To investigate wavelength-dependent polarisation, we combine the SPHERE/ZIMPOL dataset with results from previous SPHERE/IRDIS studies. This analysis reveals that post-AGB discs exhibit a grey to blue polarimetric colour in the optical and near-IR. Along with high fractional polarisation of the scattered light and polarised intensity distribution, these findings are consistent with a surface dust composition dominated by porous aggregates, reinforcing independent observational evidence for such grains in post-AGB circumbinary discs. We also find evidence of diverse disc geometries within the post-AGB sample, including arcs, asymmetries and significant variations in disc size across optical and near-IR wavelengths for some systems (U\,Mon, V709\,Car). Combining our findings with existing multi-technique studies, we question the classification of two systems in our sample, HR 4226 and V709 Car, which were originally identified as post-AGB binaries based on their near-IR excess. On comparing post-AGB discs to circumstellar environments around AGB stars and YSOs, we found that post-AGB systems exhibit a higher degree of polarisation than single AGB stars and are comparable to the brightest protoplanetary discs around YSOs. Overall, our results reinforce the importance of polarimetric observations in probing dust properties and complex circumbinary structures. We also highlight the importance of combining multi-wavelength and multi-technique observations with advanced radiative-transfer modelling to differentiate between the various evolutionary pathways of circumbinary discs.

\end{abstract}

%\noindent 

\section{Introduction}
\label{sec:paper3_intro}

At the end of their lives, low- to intermediate-mass stars (0.8-8M$_\odot$) experience significant mass loss as they evolve along the asymptotic giant branch (AGB). This mass loss plays a key role in the chemical enrichment of the interstellar medium, although the details of how mass-loss occurs are not yet fully understood \citep{Habing2003agbs.conf.....H, Khouri2020A&A...635A.200K}. In binary systems, mass loss can be amplified through stellar interactions, often resulting in the formation of dusty circumbinary discs during the AGB and early post-AGB phases \citep{VanWinckel2003ARA&A..41..391V, Kamath2015MNRAS.454.1468K}. Studying these second-generation circumbinary discs around post-AGB binary stars provides insights into the still-mysterious processes that govern dust processing and the binary interactions that terminate the giant evolutionary phase of these stars.

Observational studies have revealed key parameters regarding the structure and dynamics of circumbinary discs in post-AGB systems. These binary stars display a distinctive spectral energy distribution (SED) profile, with dust emission peaking in the near-infrared \citep[near-IR, e.g.,][]{DeRuyter2006A&A...448..641D, VanWinckel2009A&A...505.1221V, Kamath2014MNRAS.439.2211K, Kamath2015MNRAS.454.1468K, Gezer2015MNRAS.453..133G, Kluska2022}. Near-IR excess is typically attributed to circumstellar dust and gas within a stable disc with an inner rim near the dust sublimation radius. These discs are generally found to exhibit Keplerian rotation, with angular sizes ranging from roughly 0.5'' to 1'' (100 to 500 AU) as inferred from $^{12}{\rm CO}$ position-velocity mapping \citep[e.g.,][]{Bujarrabal2015A&A...575L...7B, Gallardo_cava2021A&A...648A..93G}. High-resolution near-IR interferometric studies have spatially resolved the hot dust inner rim in post-AGB discs, verifying that it is predominantly located at the dust sublimation radius \citep[$\sim$6\,AU; e.g.,][]{Hillen2016, Kluska2019A&A...631A.108K}. These circumbinary discs are called `full' discs. However, the recent study by \citet{Corporaal2023A&A...674A.151C} confirmed the presence of `transition' post-AGB discs with the dust inner rim up to 7.5 times larger than the theoretical dust sublimation radius. Pilot high-resolution imaging studies of 9 post-AGB circumbinary discs with VLT/SPHERE revealed their complex morphologies and large diversity in terms of disc size and orientation at larger angular scale \citep[$\gtrsim 30$\,mas;][]{Ertel2019AJ....157..110E, Andrych2023MNRAS.524.4168A}. The latter study also identified significant polarised brightness of post-AGB circumbinary discs in near-IR wavelengths, reaching up to $2\%$ of the total system intensity. 

Recently, a detailed multi-wavelength polarimetric imaging study of the post-AGB binary IRAS\,08544-4431 revealed a relatively consistent polarised brightness across optical and near-IR wavelengths, along with substantial forward scattering in optical polarimetry \citep{Andrych2024IRAS08}. These results align with theoretical models that propose large, porous aggregates of submicron-sized monomers as the main dust components in the disc surface layers. Furthermore, \citet{Andrych2024IRAS08} observed wavelength-dependent variations in the IRAS\,08544-4431 disc structure, with the near-IR $H$-band revealing a more extended disc surface compared to the optical $V$ and $I$-bands. The authors suggested that this structural variation across wavelengths could indicate potential disc warping.

Moreover, the presence of a circumbinary disc significantly influences the chemical composition of the post-AGB binary, often leading to a phenomenon known as photospheric chemical depletion of refractory elements 
\citep[\text{usually traced with [Zn/Ti];}][]{Waters1992A&A...262L..37W, VanWinckel1995PhDT........31V, DeRuyter2005A&A...435..161D}. This depletion arises from the re-accretion of chemically fractionated gas and dust in the circumbinary disc \citep{Kama2015A&A...582L..10K}, resulting in unique elemental abundance patterns observed in these systems \citep{Oomen2019A&A...629A..49O, Kluska2022, Mohorian2024MNRAS.530..761M}. The observed orbital parameters of post-AGB binary systems, including their high eccentricities and wide range of orbital periods, do not align with predictions of binary evolution theory \citep[e.g.,][]{Nie2012MNRAS.423.2764N}. Long-term radial velocity monitoring studies have shown that their typical orbital periods range from a hundred to a few thousand days \citep{Oomen2018, VanWinckel2009A&A...505.1221V}. The formation and interaction of circumbinary discs in these systems remain poorly understood, particularly in relation to their influence on the binary evolution of the central stars.

Despite distinct formation histories and lifetimes, post-AGB circumbinary discs \citep[with a lifetime of $\sim10^4-10^5$ years,][]{Bujarrabal2017A&A...597L...5B} and protoplanetary discs (PPDs) around young stellar objects \citep[YSOs; with disc lifetime of up to a few Myr,][]{Benisty2022arXiv220309991B} share many characteristics. Both disc types show comparable IR excesses, dust disc masses \citep[$\sim10^{-3}M_\odot$,][]{Corporaal_IRAS08_2023A&A...671A..15C}, chemical depletion patterns \citep{Kluska2022, Mohorian2024MNRAS.530..761M}, and dust mineralogy \citep{Gielen2011A&A...533A..99G, Scicluna2020MNRAS.494.2925S}. These discs also exhibit similar polarimetric brightnesses and morphologies \citep{Andrych2023MNRAS.524.4168A, Andrych2024IRAS08}. Radiative transfer (RT) modelling efforts for circumbinary discs around post-AGB binaries \citep[e.g.,][]{Corporaal_IRAS08_2023A&A...671A..15C} show that high-angular resolution interferometric data for these systems can be accurately reproduced by using passively irradiated disc models initially developed for PPDs around YSOs.

In this paper, we present multi-wavelength polarimetric imaging results for five evolved binary systems (AR\,Pup, HR\,4049, HR\,4226, U\,Mon, and V709\,Car) observed with VLT/SPHERE adaptive optics (AO) instrument. We aim to investigate the consistency of polarimetric properties, surface morphologies, and dust composition across the post-AGB sample and to compare these with findings from AGB and YSO systems. In Section \ref{sec:sample_and_observations}, we present our target sample and relevant observational details collected from the literature, as well as the observing strategy. In Section \ref{sec:paper3_data_reduction}, we introduce our data reduction methodology. In Section~\ref{sec:paper3_analysis}, we present the analysis of the VLT/SPHERE-ZIMPOL data and corresponding results. In Section~\ref{sec:paper3_discussion}, we discuss the dust properties of post-AGB circumbinary discs and highlight similarities and differences between circumstellar environments of AGB stars, circumbinary discs around post-AGB binaries, and PPDs around YSOs. In Section~\ref{sec:paper3_conclusion}, we present our conclusions.

\section{Target sample and observations}
\label{sec:sample_and_observations}

In this paper, we present multi-wavelength, high-angular-resolution polarimetric observations of the circumbinary discs surrounding a diverse sample of five post-AGB binary systems. This section provides an overview of the sample and the data.

\subsection{Target details}
\label{sec:paper3_target}

\begin{sidewaystable}
    \caption{Stellar and orbital properties of post-AGB binary stars in our target sample relevant to this study.}
    \begin{center}
        \resizebox{1\columnwidth}{!}{%
       \begin{tabular}{ lccccccccccccl }
            \hline
            Name & IRAS & \begin{tabular}[c]{@{}l@{}}Distance \\ {[}pc{]}\end{tabular} & L$_{\rm IR}$/L$_{\ast}$ & E(B-V) & RVb & \begin{tabular}[c]{@{}l@{}}P$_{\rm orbital}$\\ {[}days{]}\end{tabular} &\begin{tabular}[c]{@{}l@{}} T$_{\rm eff}$\\ {[}K{]}\end{tabular} &[Fe/H]&[Zn/Ti] &\begin{tabular}[c]{@{}l@{}}$R_{\rm subl}$\\ {[}mas{]}\end{tabular} & \begin{tabular}[c]{@{}l@{}}$D_{\rm MIR}$\\ {[}mas{]}\end{tabular} & \begin{tabular}[c]{@{}c@{}} $i$ \\  {[}$^\circ${]} \end{tabular} &  \begin{tabular}[c]{@{}c@{}} Disc\\type \end{tabular}\\
            \hline
            \hline
            AR Pup$^{3,11,12,13}$ & 08011-3627 & 661$^{+82}_{-63}$ & 9.15 & 0.4$^{+0.2}_{-0.4}$ & y & 1194 & 5925$^{+1250}_{-725}$&-1&- &1.7$^{+0.3}_{-0.4}$&63$^{+1}_{-1}$ & 75$^{+10}_{-15}$  & full\\
            HR4049$^{1, 6, 11}$ & 10158-2844 & 1500$^{+310}_{-186}$ & 0.26 & 0.21$^{+0.09}_{-0.16}$ & y & 430.6$\pm$0.1 & 7750$^{+525}_{-975}$ &-4.5&-& 6.6$^{+0.9}_{-0.7}$ &42$^{+2}_{-2}$& 49$^{+3.2}_{-3.3}$ &full \\
            HR4226$^{5, 11}$ & 10456-5712 & 1104$^{+35}_{-29}$ & 0.44 & 0.34$^{+0.3}_{-0.34}$ & n & 572$\pm$6 & 4275$^{+600}_{-550}$&0.0&- &7.4$^{+1.2}_{-1.2}$ &58$^{+3}_{-3}$ & - &full$^*$ \\
            U Mon$^{1, 2, 4, 10, 11}$ & 07284-0940 & 800$^{+117}_{-87}$ & 0.23 & 0.18$^{+0.3}_{-0.18}$ & y & 2550$\pm$143 & 5050$^{+450}_{-400}$&-0.8&0.0 &7$^{+0.9}_{-0.9}$ &50$^{+0.5}_{-0.5}$& 58$^{+1.6}_{-1.5}$ &full \\
            V709 Car$^{5, 11}$ & 10174-5704 & 2590$^{+1550}_{-1110}$ & - & 0.88$^{+0.2}_{-0.18}$ & n & 323$\pm$50 & 3500$^{+175}_{-175}$&-&- &3$^{+0.1}_{-0.2}$&140$^{+1}_{-2}$& 33$^{+2.5}_{-2.4}$ & full\\
            \hline
        \end{tabular}%
        }
    \end{center}
    \begin{tablenotes}
    \small
    \item \textbf{Notes:} The distances to the targets were adopted from $Gaia$\,DR3 \citep{Bailer-Jones2021AJ....161..147B}. However, these distances are uncertain because: i) targets are too far away and therefore not flagged as astrometric binaries ii) the orbital motion of the binary results in an angular displacement comparable to the parallax. L$_{\rm IR}$/L$_{\ast}$ represents the infrared luminosity adopted from \citet{Kluska2022}. E(B-V) indicates the total reddening, and T$_{\rm eff}$ represents stellar effective temperature derived from SED fitting \citep{Hillen2017}.  RVb represents the presence of RVb phenomenon with 'y' indicating 'yes' and 'n' indicating 'no'. P$_{\rm orbital}$ represents the orbital period in days. $R_{\rm subl}$ represents theoretical dust sublimation radius and $D_{\rm MIR}$ represents the outer disc diameter as estimated from geometrical modelling of mid-IR interferometric data \citep{Hillen2017}. The disc inclination, $i$, was derived from near-IR interferometric observations \citep{Kluska2019A&A...631A.108K} for all targets except AR\,Pup. For AR\,Pup the disc inclination was estimated using high-resolution imaging \citep{Ertel2019AJ....157..110E}. Disc type represents disc category based on SED and IR-excess features \citep{Kluska2022}. We note that $^{*}$ indicates that for HR\,4226 we infer the disc type from SED shape only, as it had previously remained uncategorized due to the lack of precise infrared photometric observations with Wide-field Infrared Survey Explorer (WISE). More details on the tabulated information can be found in the individual studies mentioned as superscripts in column 'Name': 1 - \citet{Oomen2018}, 2 - \citet{Kiss2007MNRAS.375.1338K}, 3 - \citet{Kiss2017}, 4 - \citet{Giridhar2000ApJ...531..521G}, 5 - \citet{Maas2003A&A...405..271M}, 6 - \citet{VanWinckel1995PhDT........31V}, 10 - \citet{Bodi2019}, 11 - \citet{Hillen2017}, 12 - \citet{Ertel2019AJ....157..110E}, 13 - \citep{Gonzalez1997EPLyrDYOriARPupRSgt}.
    \end{tablenotes}
    \label{tab:paper3_sample}
\end{sidewaystable}

In this study, we focus on five binary systems that are considered to host circumbinary discs: AR\,Pup, HR\,4049, HR\,4226, U\,Mon and V709\,Car. These targets were selected from the mid-infrared interferometric survey for discs around 19 post-AGB stars using VLTI/MIDI \citep{Hillen2017}.  The final selection was based on objects that are observable for Spectro-Polarimetric High-contrast Exoplanet Research/Zurich Imaging Polarimeter \citep[$R_{mag}<11$; SPHERE/ZIMPOL,][]{Schmid2018A&A...619A...9S} and have warm disc diameters greater than 40 mas in near-IR. Initially, all targets were classified as post-AGB binary stars based on their strong near-IR excesses in the SEDs, typically attributed to circumstellar dust near the dust sublimation temperature \citep{DeRuyter2006A&A...448..641D}. In Table\,\ref{tab:paper3_sample}, we present selected observational parameters of our target sample relevant to this study, including their corresponding IRAS names, Gaia DR3 distances, SED characteristics, orbital parameters, and interferometric disc sizes. We also present additional details on individual targets, along with the findings of this study, in Section~\ref{sec:paper3_indiv_cases}.

In summary, our target sample displays diverse properties, particularly regarding atmospheric parameters and elemental abundances of the primary star, as well as binary orbital characteristics and disc orientation. By analysing the polarimetric and interferometric data for these systems, we aim to provide a comprehensive understanding of how binary interactions influence the circumstellar environments of evolved stars.

\subsection{Observations}
\label{sec:paper3_observations}

\begin{table*}
    \centering
    %\caption{SPHERE/ZIMPOL observational details}
    \caption{Observing setup and seeing conditions for SPHERE/ZIMPOL data}
    \label{tab:observations}
    \resizebox{0.9\columnwidth}{!}{%
        \begin{tabular}{lcccllccccc}
            \hline
            Target  & R mag & V mag & Reference star & Observing night & Filters & \begin{tabular}[c]{@{}c@{}}DIT\\ {[}s{]}\end{tabular}& NDIT & ND filter & \begin{tabular}[c]{@{}c@{}}Seeing\\ {[}"{]}\end{tabular}& \begin{tabular}[c]{@{}c@{}}Wind\\ {[}m/s{]} \end{tabular} \\
            \hline
            \hline
            AR Pup   & 9.5  & 9.6  & HD 75885  & 08 Apr 2018 & V, N\_I    & 10  & 2  & ND\_3.5  & 0.30 - 0.98 & 0 - 4.8\\
            HR 4049  & 5.5  & 5.5  & HD 96314  & 07/08 Jan 2019$^*$ & V, Cnt820 & 1.2 & 16 & ND\_3.5  &  0.37 - 0.95 & 1.73 - 8.8\\
            HR 4226  & 6.3  & 6.3  & HD 98025  & 08 Apr 2018 & V, Cnt820 & 1.2 & 16 & No        & 0.30 - 0.98  & 0 - 4.8 \\
            U Mon    & 6.8  & 6.0  & HD 71253  & 29 Dec 2018 & V, Cnt820 & 1.2 & 16 & ND\_3.5  & 0.38 - 1.71 & 1.05 - 7.68 \\
            V709 Car & 9.3  & 11.3 & HD 94680  & 03 Mar 2018 & V, N\_I    & 10  & 2  & ND\_3.5  &  0.33 - 1.2&0.43 - 6.28 \\
             \hline
        \end{tabular}%
    }
    \begin{flushleft}
    \textbf{Notes:} Observations were conducted using SPHERE/ZIMPOL in P1 mode without a coronagraph. We note that $^*$ indicates that HR\,4049 was observed twice, on 7th and 8th January, due to non-ideal conditions. The detector integration time (DIT) and number of detector integrations (NDIT) were chosen based on the target brightness and neutral density filter (ND filter) configuration. DIT$\times$NDIT is the integration for one out of four half-wave plate position angles of a polarimetric cycle, and 20 polarization cycles were taken per target resulting in a total integration time of 40$\times$DIT$\times$NDIT per data set. See Section~\ref{sec:paper3_observations} for more details.
    \end{flushleft}
\end{table*}

The data for this study were obtained using the Zurich Imaging Polarimeter \citep[ZIMPOL,][]{Schmid2018A&A...619A...9S} of the extreme AO instrument SPHERE \citep{Beuzit2019} as part of ESO observational program 0101.D-0752(A) (PI: Kamath). To effectively resolve the scattered light around the post-AGB binary stars and ensure the best accuracy for the multi-wavelength characteristics, we used SPHERE/ZIMPOL in its polarimetric P1 mode without a coronagraph. This mode provides high polarimetric sensitivity and calibration accuracy by averaging out instrumental effects as the sky field rotates with respect to the instrument and telescope pupil \citep{Schmid2018A&A...619A...9S}. In addition to the target systems, reference stars were observed as part of a program. These stars were selected for their similar brightness, colour, and location to the corresponding binary systems. Reference stars were observed immediately after each target, allowing their data to serve as a point-spread function (PSF) during the reduction process. 

We obtained data in both the $V$-band and $I'$-band for all targets in our sample. Specifically, the $V$-band observations were conducted using the ZIMPOL 'V' filter ($\lambda_0 = 554$\,nm, $\Delta\lambda = 80.6$\,nm), while $I'$-band data were obtained using either the 'Cnt820' filter ($\lambda_0 = 817.3$\,nm, $\Delta\lambda = 19.8$\,nm) or the 'N\_I' filter ($\lambda_0 = 816.8$\,nm, $\Delta\lambda = 80.5$\,nm), with broader 'N\_I' filter used for fainter targets. For consistency, we refer to both 'Cnt820' and 'N\_I' as the $I'$-band throughout the text and figures, while their exact names are specified in the tables. 
In total 20 polarimetric cycles were taken for each target and 10 polarimetric cycles for reference stars. Each observational frame contained $\sim10^{5}-10^{6}$ counts, ensuring non-saturated observations \citep{Schmid2018A&A...619A...9S}. In Table~\ref{tab:observations} we present additional observational details for each target. 

In addition to SPHERE/ZIMPOL data, we also include results of near-IR polarimetric observations using the Infra-Red Dual-beam Imaging and Spectroscopy camera \citep[IRDIS,][]{Dohlen2008} of the same instrument for two targets \citep[HR\,4049 and U\,Mon,][]{Andrych2023MNRAS.524.4168A}.

\section{Data reduction}
\label{sec:paper3_data_reduction}

The SPHERE/ZIMPOL polarimetric datasets were reduced following the methodology presented in \citet{Andrych2024IRAS08}.  Here, we provide only a brief overview of data reduction process, highlighting relevant modifications and target-specific details. 

\subsection{Standard Polarimetric differential imaging (PDI) reduction}
\label{sec:pdi}
The initial calibration procedures (bias subtraction, flat fielding, frame centring, polarimetric beam shift correction) and calculation of the Stokes vectors $Q$ and $U$, were performed using the High-Contrast Data Centre (HC-DC, formerly SPHERE Data Centre) pipeline\footnote{This pipeline builds on the core SPHERE ESO software and incorporates additional routines such as improved centring algorithms, automatic frame sorting, and analysis tools for systematic SPHERE data processing.} \citep{Delorme2017sf2a.conf..347D}. 
Instrumental polarisation was accounted for using the \texttt{sz}-pipeline developed at ETH Zurich, which applies corrections based on instrumental calibration data \citet{Schmid2018A&A...619A...9S}. In \ref{sec:paper3_ap_telesc_corr}, we provide details of the measured fractional polarisation ($Q/I_{\rm tot}, U/I_{\rm tot}$, where $I_{\rm tot}$ is the total intensity of the target) before and after correction for each observation cycle, alongside the corresponding telescope polarisation data for each target. Further processing included our custom Python pipeline. To enhance image quality, we retained only the top 75\%\footnote{This value was selected to balance between excluding frames with poorer instrumental performance while still keeping enough frames to benefit from mean combining and improve the data peak signal-to-noise ratio (SNR). This selection resulted in an improvement of peak SNR by $\sim$3 to 15\% for all targets.} of frames in each dataset, selected based on peak-normalised total intensity ($I_{\rm tot}$). This selection helped suppress variability from atmospheric conditions and adaptive optics performance, resulting in an improved signal-to-noise ratio in the final stacked images. Then, we computed azimuthal Stokes parameters $Q_\phi$ and $U_\phi$ \citep[following the formalism of][]{deBoer2020}, total polarised intensity $I_{\rm pol}$, and the Angle of Linear Polarisation $AoLP$. Positive $Q_\phi$ indicates linear polarisation in the azimuthal direction, while negative $Q_\phi$ indicates radial polarisation. $U_\phi$ represents linear polarisation rotated by $\pm45^\circ$ relative to these directions, and $AoLP$ defines the local orientation of the polarisation vector.

\subsection{Additional data reduction steps}

Given the compact angular size of the targets and the unresolved separation between disc features and the central binary, we implemented additional reduction steps specifically aimed at removing the unresolved polarised signal from the final polarimetric images, enhancing the visibility of disc substructures through PSF deconvolution, and recovering the total resolved polarised intensity attenuated by PSF smearing. Unless stated otherwise, all images presented in this paper were processed using the standard PDI reduction and two additional steps: correction for unresolved polarisation (except in the case of AR\,Pup, where the correction could not be reliably applied), and PSF deconvolution. The correction for polarimetric cancellation due to PSF smearing was applied solely for estimating the total polarised flux and was not used in analyses focused on disc substructures, to avoid introducing artificial features or biases.

\subsubsection{Correction of the unresolved central polarised signal}
\label{sec:unres_polarization}

The reduced linearly polarised images for each target display typical signatures of a bright but spatially unresolved section of the circumbinary disc \citep[e.g.,][]{Keppler2018A&A...617A..44K}, including a butterfly pattern in  $Q_\phi$ and a halo in the $I_{\rm pol}$ images (see top row of Figure~\ref{fig:paper3_ap_reduct_HR4049},\,\ref{fig:paper3_ap_reduct_HR4226},\,\ref{fig:paper3_ap_reduct_umon},\,\ref{fig:paper3_ap_reduct_v709},\,\ref{fig:paper3_ap_reduct_AR_pup} in \ref{sec:paper3_ap_reduct}). To remove this unresolved polarisation component, we used the approach presented by \citet{Holstein2020A&A...633A..64V}, measuring the degree and orientation of unresolved component using a circular aperture with a 6-pixel diameter ($\sim22$ mas)\footnote{Similar to \citet{Andrych2023MNRAS.524.4168A}, the 6-pixel aperture was chosen to reliably correct for the unresolved polarised signal while minimizing contributions from the resolved signal.} centred on the stellar position (see Table~\ref{tab:paper3_unresolved}). 

\begin{figure*} 
    \includegraphics[width=0.49\linewidth]{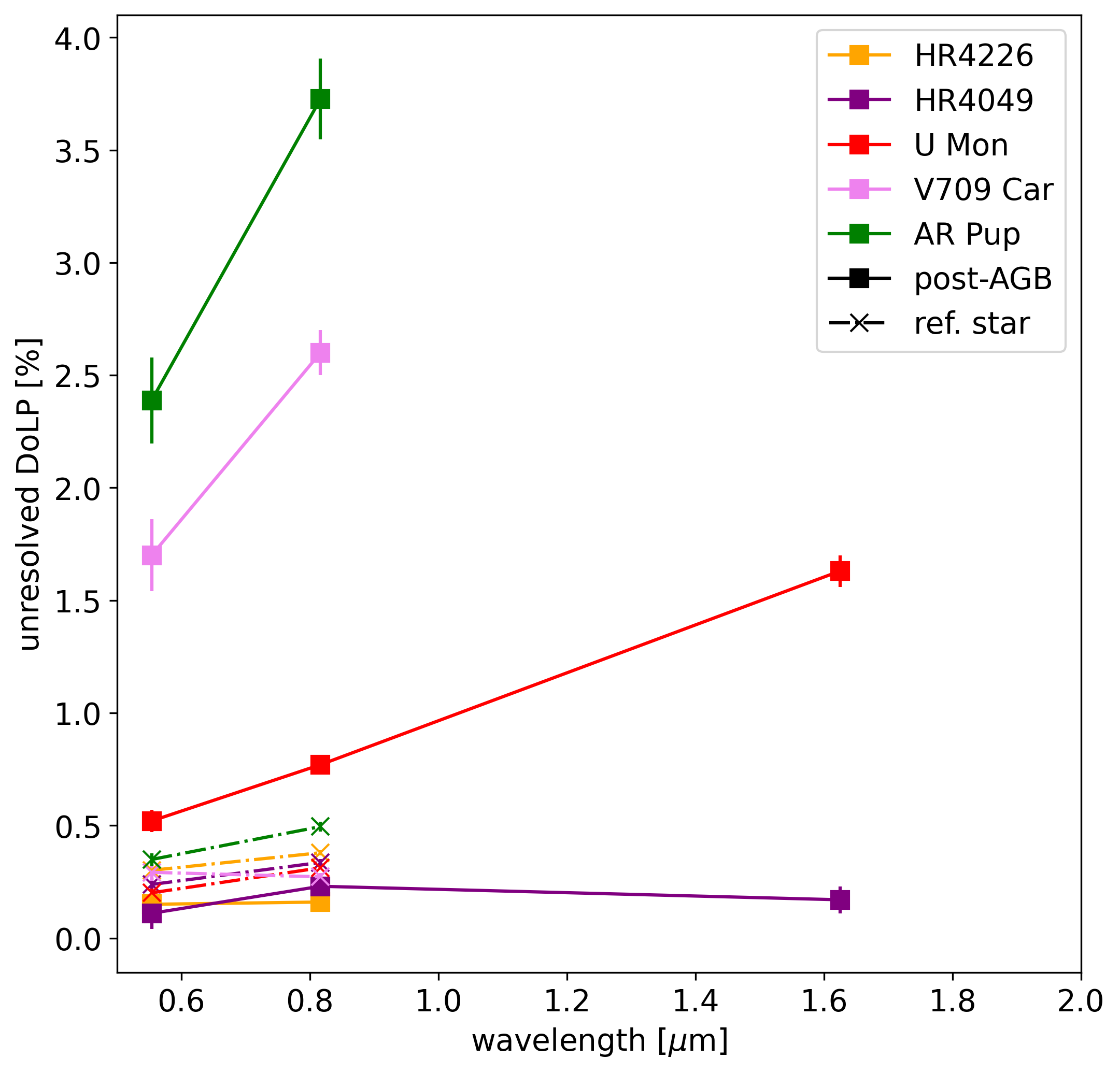}
    \includegraphics[width=0.495\linewidth]{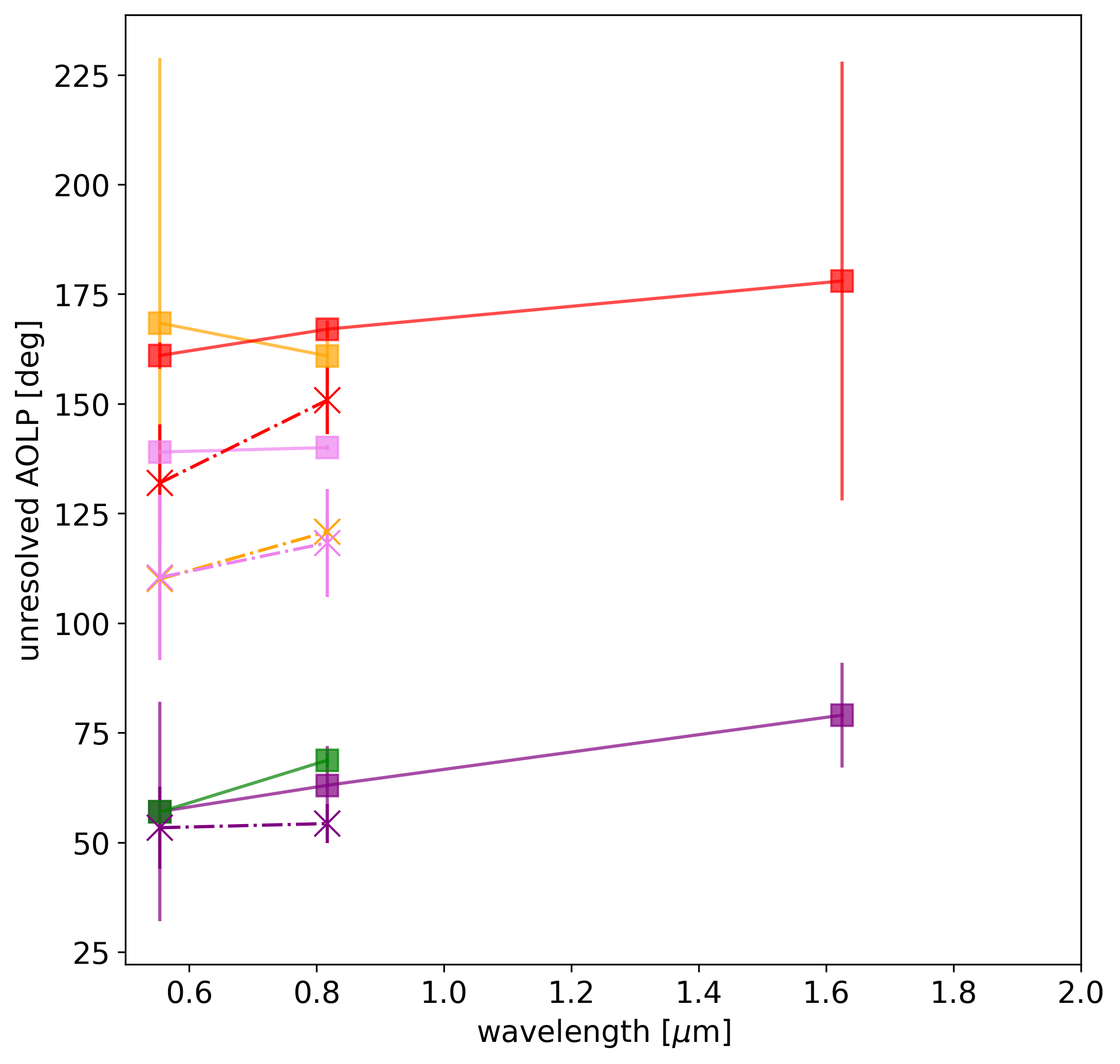}
    \caption[Characteristics of unresolved central polarisation for scientific targets and polarised intensity of reference stars as a function of wavelength.]{Characteristics of unresolved central polarisation for scientific targets and polarised intensity of reference stars as a function of wavelength. The left panel shows the degree of unresolved polarisation (solid lines) relative to the total intensity of each target and the polarised intensity relative to the total intensity for each reference star (dashed lines). The right panel displays the corresponding orientation of unresolved polarisation (AoLP) for the targets (solid lines) and reference stars (dashed lines). Each target binary system and its corresponding reference star are indicated by matching colours. See Section~\ref{sec:paper3_data_reduction} for more details.}
    \label{fig:paper3_wave_dep_unres}
\end{figure*}

To assess the origin of the unresolved signal, we compared measured degree of unresolved central polarisation with the values obtained for reference stars at the same wavelengths (see Figure~\ref{fig:paper3_wave_dep_unres}), as well as with predictions from Galactic interstellar polarisation maps \citep{Heiles2000AJ....119..923H, Versteeg2023AJ....165...87V}. This comparison helps to determine whether the unresolved signal is primarily caused by interstellar polarisation or the disc itself \citep{Andrych2024IRAS08}. We found that for HR\,4049 and HR\,4226 the degree of the unresolved polarisation component closely aligns with the estimated interstellar polarisation. In contrast, for AR\,Pup, V709\,Car, and U\,Mon, it is significantly higher, indicating that the unresolved polarisation is primarily driven by the unresolved portion of the circumbinary disc rather than the polarisation from the diffuse interstellar medium along the line of sight. Where feasible, we subtracted the estimated unresolved signal from the $Q$ and $U$ frames prior to re-deriving $Q_\phi$ and $I_{\rm pol}$. In AR\,Pup, which is observed nearly edge-on, this subtraction could not be reliably performed without affecting the resolved disc morphology. A known side effect of this correction is an artificial decrease in signal within the central 5×5 pixel region of the image. This area was masked in all subsequent analysis steps (see Section~\ref{sec:artefact}). The resulting polarised images are presented in Figure~\ref{fig:paper3_reduced}. 

\begin{table}
    \caption{Characteristics of the unresolved central polarisation.}
    \begin{center}
        \resizebox{0.8\columnwidth}{!}{%
            \begin{tabular}{llcc }
                \hline
                Name & Band & DoLP, \% & AoLP, $^\circ$ \\
                \hline
                \hline
                AR Pup &V & 2.4$\pm$0.2 & 57$\pm$2 \\
                &N\_I & 3.7$\pm$0.2 & 69$\pm$2 \\
                \hline
                HR4049 &V & 0.11$\pm$0.07 & 57$\pm$25 \\
                &Cnt820 & 0.23$\pm$0.04 & 63$\pm$9 \\
                &H$^*$ & 0.17$\pm$0.06 & 79$\pm$12 \\
                \hline
                HR4226&V & 0.15$\pm$0.06 & 168$\pm$62 \\
                &Cnt820& 0.16$\pm$0.04 & 161$\pm$6 \\
                \hline
                U Mon& V & 0.52$\pm$0.05 & 161$\pm$3.4 \\
                &Cnt820 & 0.77$\pm$0.04 & 167$\pm$1.5\\
                &H$^*$ & 1.63$\pm$0.07 & 178$\pm$50 \\
                \hline
                V709 Car& V & 1.7$\pm$0.16 & 139$\pm$0.9 \\
                & N\_I & 2.6$\pm$0.1 & 140$\pm$0.6 \\
                
                \hline
            \end{tabular}%
        }
    \end{center}
    \begin{tablenotes}
    
    \small
    \item \textbf{Notes:}  'DoLP' represents the degree of linear polarisation, 'AoLP' represents the predominant angle of linear polarisation for the unresolved central polarisation. See Section~\ref{sec:paper3_data_reduction} for details.  We note that for AR Pup the orientation of the disc likely leads to a significant overestimation of the unresolved polarized signal (see Section~\ref{sec:paper3_ar_pup}). \\
    \end{tablenotes}
   
    \label{tab:paper3_unresolved}
\end{table}    

\subsubsection{Deconvolution}
\label{sec:deconvolution}

To retrieve the spatially resolved substructures, we deconvolved the resulting linearly polarised images with the reference PSF using the Richardson–Lucy deconvolution algorithm \citep{Richardson1972JOSA...62...55R, Lucy1974AJ.....79..745L}. We note that the deconvolution process converged within 30 iterations for all targets in both the $V$- and $I'$-bands, with pixel-to-pixel changes between successive iterations falling below 1.5\%. However, we note that the same amount of iterations was overcorrecting the AR\,Pup and V709\,Car images (subtracting not only the unresolved polarisation but also part of the resolved disc signal), resulting in degenerate results. Therefore, for AR\,Pup and V709\,Car, we stopped iterations when we achieved pixel-to-pixel variation of 5\% ($\sim10$ iterations). We present the final deconvolved polarised images in Figure~\ref{fig:paper3_reduced}.

\subsubsection{Correction of PSF smearing effect}
\label{sec:psf_smearing}

To accurately measure the total polarised flux from the circumbinary disc, we also corrected resolved $Q_\phi$ and $I_{\rm pol}$ images for polarimetric cancellation due to PSF smearing with the instrument PSF, following the methodology proposed by \citet{Ma2024A&A...683A..18M}. This method uses the observed PSF and the known disc geometry (inclination and position angle) to construct a two-dimensional correction map that estimates the amount of polarised flux lost due to convolution with the instrumental PSF. The resulting map is then used to rescale the observed $Q_\phi$ and $I_{\rm pol}$ distributions, enabling recovery of the intrinsic polarised intensity with relative errors typically below 10\% \citet{Ma2024A&A...683A..18M}.
As this correction significantly increases the strength of the polarised signal, particularly in the inner disc regions where cancellation is strongest, it is essential for accurately recovering the total polarimetric brightness in compact or barely resolved systems \citep{Andrych2024IRAS08}. However, since this correction can also amplify noise and introduce artefacts near the unresolved central binary, it was applied only to estimate the total polarised intensity of targets and was deliberately omitted from analyses of disc orientation and morphological features. 

\subsection{Data quality and artefact analysis}

This section evaluates the quality of the reduced polarimetric data, with particular focus on the limitations imposed by spatial resolution and unresolved emission. We assess the reliability of the $Q_\phi$ quantity, provide an estimation of SNR to define statistically significant regions, and describe image artefacts arising from instrumental and reduction effects rather than physical disc structures.

\subsubsection{Evaluating the reliability of \texorpdfstring{$Q_\phi$}{Qphi} as a measure of polarised intensity}
\label{sec:qphi_test}

For low-inclination ($i \lesssim 40^\circ$) and optically thick discs, single scattering of stellar light on the disc surface produces polarisation vectors oriented azimuthally to the central star, making $Q_\phi$ a reliable measure of polarised flux while minimizing noise biases compared to  $I_{\rm pol}$ \citep{Schmid2006A&A...452..657S, Canovas2015A&A...582L...7C, Simmons1985A&A...142..100S}. To evaluate this approach for our data, we calculated the ratio of integrated polarised signal $Q_\phi$ to $I_{\rm pol}$ within a circular aperture of 3" diameter.
Two out of five targets in our sample (HR\,4049 and HR\,4226) show $Q_\phi/I_{\rm pol}$>85\%, indicating the dominance of single scattering and resolved circumstellar polarisation (see Table~\ref{tab:paper3_disc_orient}). For the remaining targets, lower \( Q_\phi/I_{\rm pol} \) values limit the reliability of \( Q_\phi \) as a measure of polarised intensity and indicate that the contribution of unresolved polarisation is relatively high. For AR\,Pup, the decrease in the $Q_\phi/I_{\rm pol}$ value is likely due to the disc’s nearly edge-on orientation. The significantly lower \( Q_\phi/I_{\rm pol} \) values in U\,Mon and V709\,Car (<70\%) suggest higher disc inclinations and potential multiple scattering, which could alter the orientation of the polarisation vector.
Furthermore, we examined the mean flux distribution in $U_\phi$ in the radial annuli, similar to the analyses of \citet{Avenhaus2018ApJ...863...44A} and \citet{Andrych2023MNRAS.524.4168A}. Although all targets showed signs of an astrophysical signal in $U_\phi$, separating it from instrumental residuals remains challenging. For all targets except HR\,4049 we found that the net $U_\phi$ signal accounts for more than 5\% of $Q_\phi$, supporting the use of $I_{\rm pol}$ for further analysis. For HR\,4049, however, we will use $Q_\phi$.

\subsubsection{Estimation of the SNR}
\label{sec:snr}
To identify the statistically significant area in the final reduced images, we measured the background noise using an annular aperture of the reduced $I_{\rm pol}$ image without the target signal. Based on this, we defined the areas of $I_{\rm pol}$ with a SNR of $\geq 3$ as statistically significant (see second and fourth columns of Figure~\ref{fig:paper3_reduced}). We also estimated the spatial resolution of the data to be $\sim 30$\,mas for all targets except AR\,Pup (with $\sim 40$\,mas, likely due to slightly worse observing conditions or suboptimal selection of the reference star), which is consistent with SPHERE/ZIMPOL specifications \citep{Schmid2018A&A...619A...9S}.

\subsubsection{Data reduction artefacts}
\label{sec:artefact}

\begin{figure*}
    
    \includegraphics[width=0.98\linewidth]{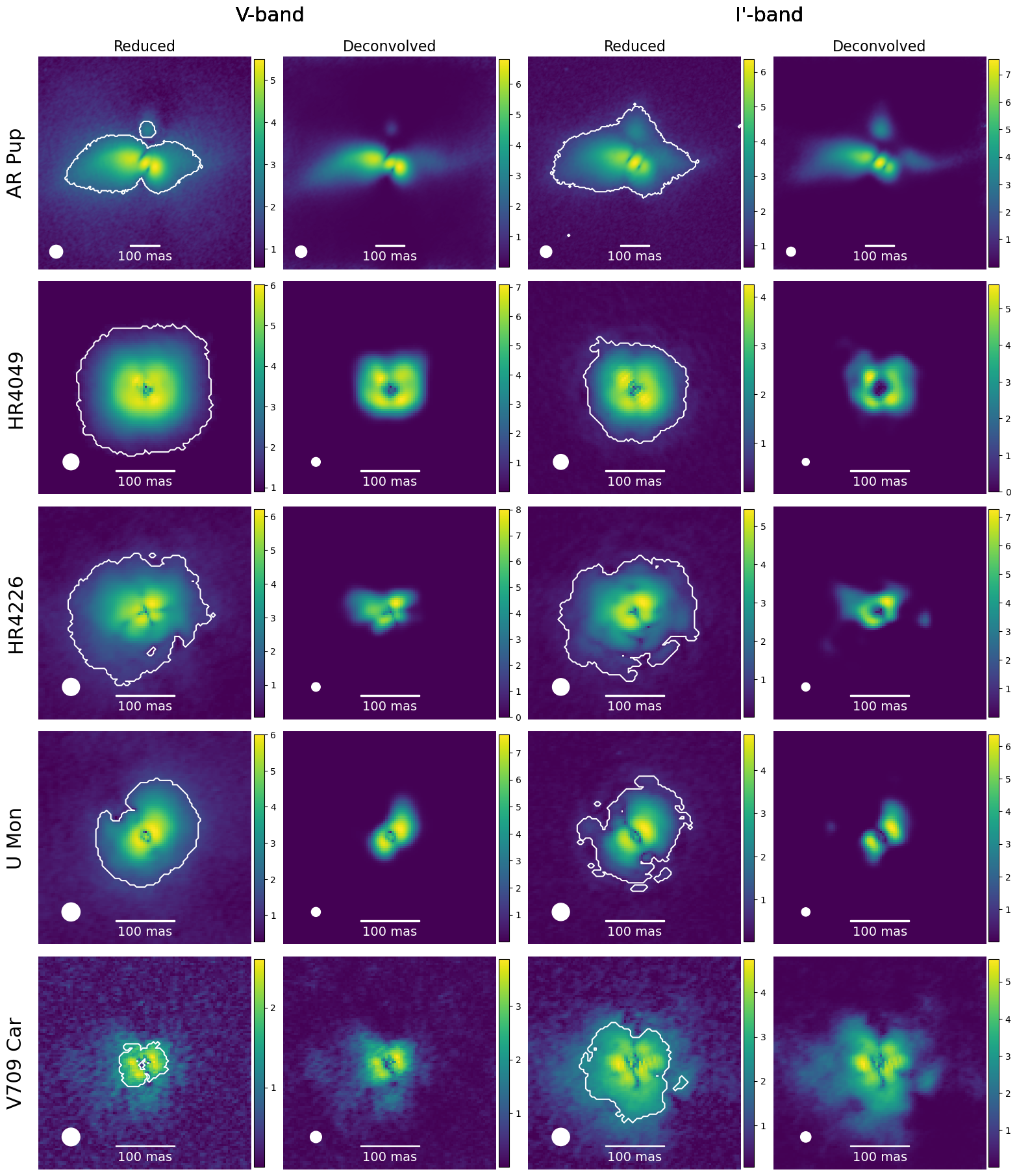}
    \caption[Total polarised intensity of all targets in $V-$ and $I'-$bands.]{Total polarised intensity of all targets in $V-$ and $I'-$bands. The first and third columns display polarimetric images after standard PDI reduction and correction of unresolved polarisation (except for AR\,Pup, where the correction could not be reliably applied), while the second and fourth columns show images after additional deconvolution with PSF. White contours outline regions of statistically significant polarised intensity (SNR=3). White circles in the lower-left corner of each image indicate the size of the resolution element. All images are presented on an inverse hyperbolic scale and oriented North up and East to the left. See Section~\ref{sec:paper3_data_reduction} for more details.}
    \label{fig:paper3_reduced}
\end{figure*}

Some of the features observed in the resulting polarised images (see Figure~\ref{fig:paper3_reduced}) are data reduction artefacts rather than real disc substructures. Below, we describe these artefacts and their possible causes.

During data reduction, we correct for the unresolved central polarisation to retrieve the polarised intensity of the resolved disc (see Section~\ref{sec:unres_polarization}). However, this method produces an unrealistically low intensity in the central 5x5 pixel region of the $I_{\rm pol}$ and $Q_\phi$ images, and any intrinsic $Q_\phi$ component remains undetectable due to limited spatial resolution (see Figure\ref{fig:paper3_reduced}). Therefore, we exclude this region from further analysis, noting that this discrepancy is an artefact of the reduction process.  In addition, the final reduced $V$-band image of U\,Mon shows a bright ring, while the $I'$-band image shows two arcs separated by a dark strip. This dark strip is also an artefact of the unresolved central polarisation subtraction, which inadvertently removes some polarised emission from the disc.

For AR Pup, we do not apply the correction for the unresolved central polarisation due to the inability to clearly separate it from the resolved signal, given the edge-on orientation of the disc (see Section~\ref{sec:unres_polarization}). However, the overlapping of differently oriented polarisation vectors from the unresolved intensity and the resolved disc structure creates two linear ‘shadows’ within 20 mas of the central binary, perpendicular to the disc midplane. We emphasise that these shadows are not physical features of AR Pup but rather artefacts of the observational technique and data reduction.

Additionally, the reduced $I_{\rm pol}$ images of HR\,4226, HR\,4049, and V709\,Car exhibit a thin cross-shaped decrease in polarised intensity, particularly visible in the $V$-band data for HR\,4226. This pattern is not a genuine feature, but rather a reduction artefact, as it is also seen in reference star data for these targets. While the exact cause of this effect is unclear, we emphasise these are not physical characteristics of the circumbinary discs of HR\,4226, HR\,4049, and V709\,Car.

\section{Analysis and Results}
\label{sec:paper3_analysis}

In this section, we present the analysis of reduced linearly polarised images and discuss the results for each target individually, following the approach described in \citet{Andrych2023MNRAS.524.4168A, Andrych2024IRAS08}. The analysis includes measuring the fractional polarisation, determining the orientation of the disc, quantifying the relative disc brightness in polarised light, and examining the wavelength-dependent behaviour of both the polarised intensity and the resolved structure of the circumbinary discs.

\subsection{Measuring fractional polarisation}
\label{sec:paper3_aper_pol}

To investigate geometric variations in the circumbinary discs of our five targets, we analysed how fractional polarisation ($Q/I_{\rm tot}, U/I_{\rm tot}$) varies with distance from the central binary using aperture polarimetry. This method offers insights into the unresolved disc component contributing to the detected central polarisation. Significant tearing or misalignment in the disc would manifest as abrupt shifts in the $Q/I_{\rm tot}-U/I_{\rm tot}$ plane\citep[e.g.,][]{Nixon2013MNRAS.434.1946N, Kraus2020Sci...369.1233K}. 

We measured total fractional polarisation $Q/I_{\rm tot}$ and $U/I_{\rm tot}$ using a series of circular apertures with radii increasing from 0.004" to 0.11" (extended to 0.25" for AR\,Pup), in increments of 0.0036" (corresponding to 1 pixel). The resulting data (see Figure~\ref{fig:paper3_aper_pol} in \ref{sec:ap_aper_pol}) reveal a smooth and gradual variation in fractional polarisation as a function of aperture size for all targets. However, the curves for AR\,Pup, HR 4226, and U\,Mon exhibit a more complex shape, likely due to changes in the orientation of the scattering surface relative to the binary, which affects the resulting polarisation vector. Additionally, interstellar polarisation contributes to the unresolved central polarisation, impacting the fractional polarisation closer to the binary. This effect could explain the changes in polarisation orientation, particularly for HR 4226 and U\,Mon, given their small disc sizes. We discuss these factors in more detail for each target in Section~\ref{sec:paper3_indiv_cases}.

\subsection{Measuring the polarised disc brightness}
\label{sec:paper3_ polarised_bright}

To estimate the polarised disc brightness relative to the total intensity of the system, we calculated the ratio of the resolved polarised emission from the disc to the total unpolarised intensity of each target. The polarised signal was measured within the region with SNR$\geq3$ for each observation, while the total intensity was integrated over a broad 3" aperture. To minimise systematic biases, we used the azimuthal, total polarised, and unpolarised intensity maps ($Q_{\phi}$, $I_{\rm pol}$ and $I_{\rm tot}$) prior to PSF deconvolution, while incorporating corrections for PSF smearing effects during the observations (see Section \ref{sec:paper3_data_reduction}). We note that the resulting values for polarised disc brightness represent a lower limit due to the partial subtraction of the disc polarised signal during data reduction. The final polarised disc brightness ratios ($Q_{\phi}/I$ and $I_{\rm pol}/I$, where $I$ is the total integrated system intensity) for both the $V-$ and $I'-$bands across all targets are presented in Table~\ref{tab:paper3_disc_orient}. For reference, we also provide the polarised disc brightness value without PSF smearing correction ($I^{*}_{\rm pol}/I$).

To ensure an accurate comparison of the available SPHERE/ /ZIMPOL and SPHERE/IRDIS data for our targets, we also applied a similar correction for PSF smearing to the $H$-band SPHERE/IRDIS data of HR\,4049 and U\,Mon, which were previously analysed and presented by \citet{Andrych2023MNRAS.524.4168A}.

    \begin{table*}
        \caption{Summary of derived properties for all targets in V and $I'$-bands.}
        \begin{center}
        \resizebox{1\columnwidth}{!}{%
            \begin{tabular}{ llccccccccc}
            \hline
            Name&Band& \begin{tabular}[c]{@{}c@{}}
            $a$ \\ {[}mas{]}\end{tabular} &  \begin{tabular}[c]{@{}c@{}}$b$\\ {[}mas{]}\end{tabular}&
            \begin{tabular}[c]{@{}c@{}}
            $i$ \\  {[}$^\circ${]}
            \end{tabular} & \begin{tabular}[c]{@{}c@{}}
            $PA$ \\  {[}$^\circ${]}
            \end{tabular}& $e$ &\begin{tabular}[c]{@{}c@{}}
            $Q_{\phi}/I_{\rm pol}$ \\  {[}$\%${]}
            \end{tabular}&\begin{tabular}[c]{@{}c@{}}
            $Q_{\phi}/I$ \\  {[}$\%${]}
            \end{tabular} &\begin{tabular}[c]{@{}c@{}}
            $I_{\rm pol}/I$ \\  {[}$\%${]}
            \end{tabular}&\begin{tabular}[c]{@{}c@{}}
            $I^{*}_{\rm pol}/I$  \\  {[}$\%${]}
            \end{tabular} \\

            \hline
            \hline
            AR Pup&V &80&-&75$\pm$10&50$\pm$5&-&$\sim$75& 4.4$\pm$0.3& 4.4$\pm$0.3& 3.0$\pm$0.2\\
            &N\_I &80&-&75$\pm$10&50$\pm$5&-&$\sim$80&5.1$\pm$0.3& 5.2$\pm$0.2&3.6$\pm$0.2\\ 
            
            \hline
            HR4049&V & 29$^{+2}_{-2}$& 25$^{+2}_{-2}$& 29$^{+8}_{-12}$ & 109$^{+22}_{-24}$ & 0.5&>85& 1.1$\pm$0.3& 1.2$\pm$0.3& 0.5$\pm$0.1\\
            &Cnt820 & 29$^{+1.2}_{-1.2}$& 27$^{+1.4}_{-1.3}$& 23$^{+9}_{-16}$ & 138$^{+28}_{-30}$ &0.4 &>85&0.95$\pm$0.1& 0.95$\pm$0.1& 0.51$\pm$0.05\\ 
            &H$^{**}$ &37$^{+2}_{-2}$ & 35$^{+2}_{-2}$& 17$^{+14}_{-14}$ & 174$^{+28}_{-30}$ & 0.29& -&0.6$\pm$0.15$^{\dag}$& 0.6$\pm$0.15$^{\dag}$&0.3$\pm$0.15\\
            
            \hline
            HR4226&V & 20$^{+5}_{-3}$& 13$^{+2}_{-3}$& 51$^{+13}_{-17}$ & 104$^{+20}_{-20}$ &0.77 &>85& 1.8$\pm$0.1& 1.9$\pm$0.1& 0.51$\pm$0.03\\
            &Cnt820 & 23$^{+1.4}_{-1.4}$& 17$^{+1}_{-1}$& 40$^{+6}_{-9}$ & 106$^{+14}_{-12}$ &0.64&>85&1.6$\pm$0.06& 1.75$\pm$0.06& 0.61$\pm$0.02\\ 

            \hline
            U Mon&V & 19$^{+2}_{-2}$& 13$^{+1}_{-1}$& 48$^{+9}_{-9}$ & 128$^{+12}_{-12}$ &0.75&$\sim$57& 1.2$\pm$0.1&1.8$\pm$0.2& 0.54$\pm$0.05\\
            &Cnt820 &22$^{+9}_{-4}$& 17$^{+4}_{-5}$& 41$^{+25}_{-32}$ & 131$^{+28}_{-30}$ &0.65& $\sim$65&1.05$\pm$0.05& 1.4$\pm$0.07& 0.55$\pm$0.03\\ 
            &H$^{**}$ &32$^{+3}_{-3}$ & 29$^{+2}_{-3}$& 25$^{+14}_{-18}$ & 144$^{+10}_{-15}$ &0.42 &-& 0.57$\pm$0.15$^{\dag}$& 0.66$\pm$0.15$^{\dag}$&0.34$\pm$0.15\\
            
            \hline
            V709 Car&V &19$^{+7}_{-3}$& 16$^{+3}_{-6}$& 27$^{+37}_{-19}$ & 121$^{+28}_{-30}$ &0.47& $\sim$68& 0.5$\pm$0.06& 0.6$\pm$0.09& 0.29$\pm$0.03\\
            &N\_I &26$^{+7}_{-5}$& 23$^{+4}_{-5}$& 28$^{+27}_{-21}$ & 150$^{+28}_{-30}$ &0.47& $\sim$80&0.40$\pm$0.04& 0.49$\pm$0.07& 0.25$\pm$0.03\\

            \hline
        \end{tabular}%
        }
        
    \end{center}
    \begin{tablenotes}
     \small
    \item \textbf{Notes:} $a$ and $b$ represent the major and minor half-axes of the disc in mas, $i$ indicates the inclination, $e$ represents the eccentricity. $Q_{\phi}/I_{\rm pol}$ represent the ratio of azimuthal to total polarized disc brightness of the target. $Q_{\phi}/I$ and $I_{\rm pol}/I$ represent the azimuthal and total polarized disc brightness relative to the total intensity of the target. $I^{*}_{\rm pol}/I$ represents the total polarized disc brightness without correction for the PSF smearing (see Section~\ref{sec:paper3_data_reduction}).
    The position angle ($PA$) is presented in degrees and rises counterclockwise from the vertical axis (North) to the first principal radius (major axis). See Section~\ref{sec:paper3_analysis} for more details.\\

    $^{**}$ indicates the data adopted from \citet{Andrych2023MNRAS.524.4168A}\\
    \end{tablenotes}
       
    \label{tab:paper3_disc_orient}
    \end{table*}

\subsection{Determination of disc orientation}
\label{sec:paper3_disc_orient}

The final reduced polarised images reveal a bright resolved `ring' structure in both $V-$ and $I'-$band observations for HR\,4049, HR\,4226, U\,Mon and V709\,Car (see Figure~\ref{fig:paper3_reduced}). However, for V709\,Car, the `ring' is not fully resolved and the target shows lower polarised intensity compared to the other targets (see Table~\ref{tab:paper3_disc_orient}), so it should be interpreted with caution. Additionally, no `ring' is observed for AR\,Pup due to the nearly edge-on disc orientation (see Sec.~\ref{sec:paper3_target} for details). Although we refer to the brightest resolved structure as a `ring', it is important to clarify that it represents the smallest resolved section of the circumbinary disc close to the central binary rather than the physical dust inner rim of the circumbinary disc, which is too small to be resolved by the SPHERE instrument for post-AGB systems (see Section \ref{sec:artefact}). 

\begin{figure*} 
     \includegraphics[width=0.26\linewidth]{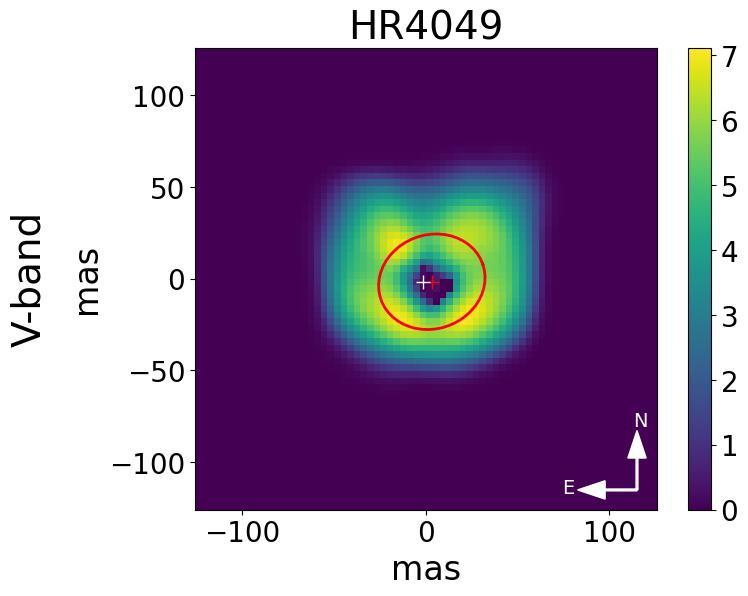}
     \includegraphics[width=0.24\linewidth]{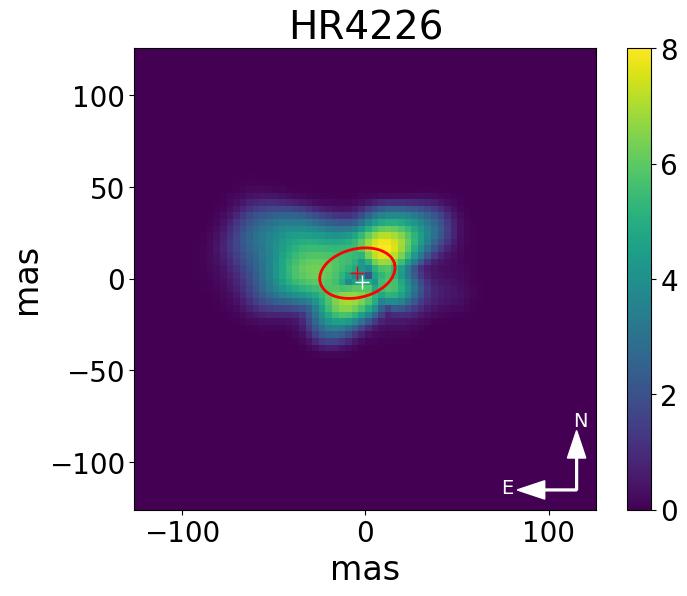}
     \includegraphics[width=0.24\linewidth]{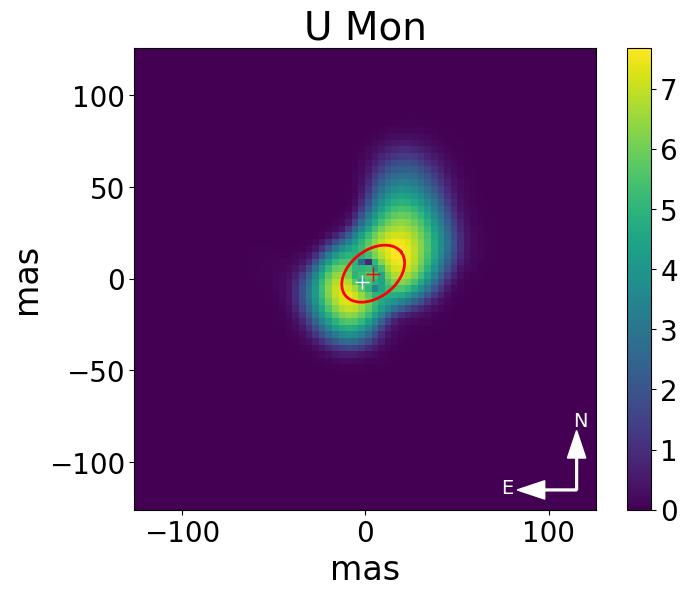}
     \includegraphics[width=0.24\linewidth]{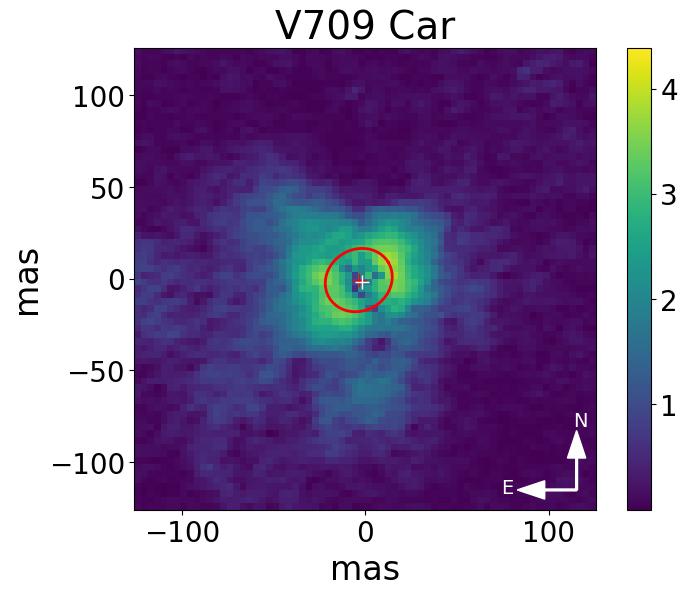}
     \includegraphics[width=0.26\linewidth]{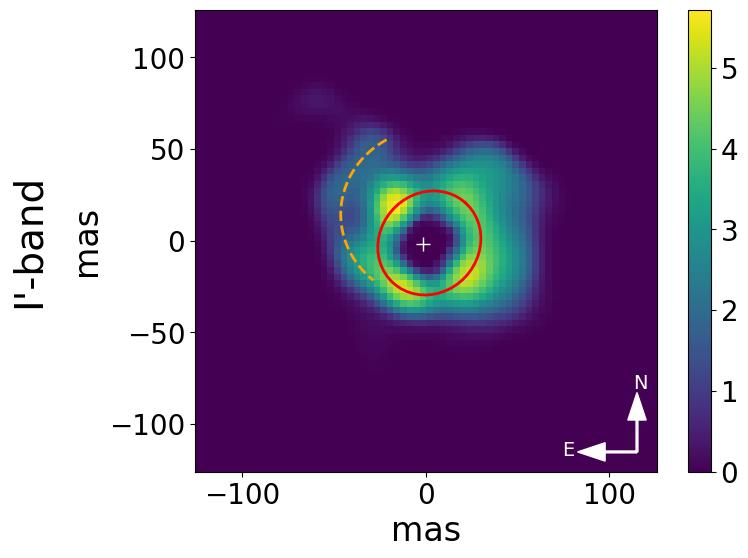}
     \includegraphics[width=0.24\linewidth]{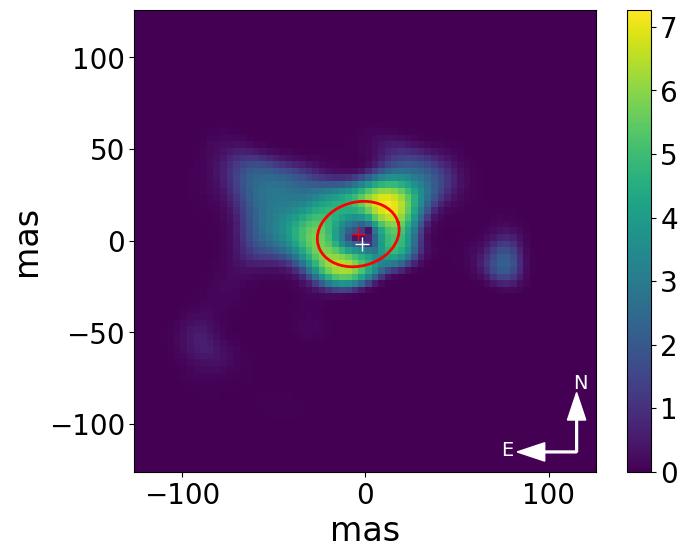}
     \includegraphics[width=0.24\linewidth]{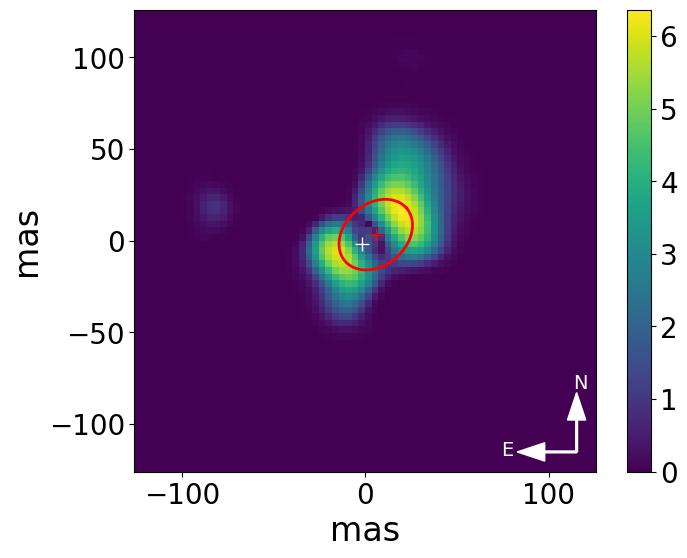}
     \includegraphics[width=0.24\linewidth]{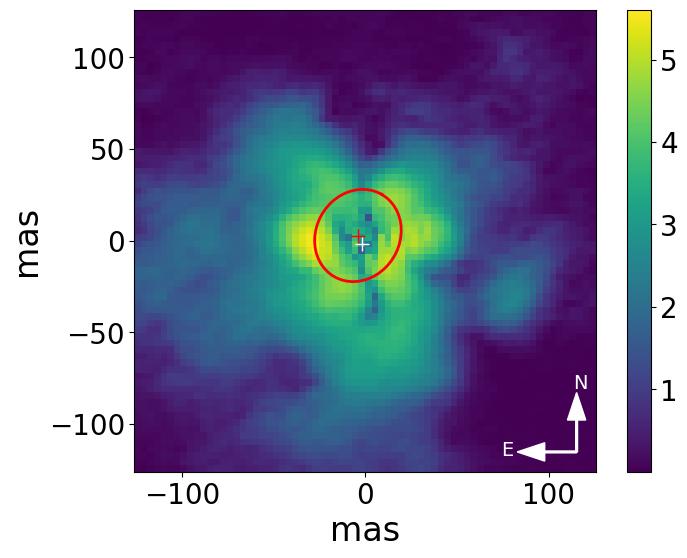}
    
    \caption[Disc orientation results based on the polarimetric images of all targets.]{Disc orientation results based on the polarimetric images of all targets (see Section~\ref{sec:paper3_disc_orient} for more details). The red ellipses illustrate the most plausible PA and inclination of discs, while the dashed orange line highlights a significant substructure for HR\,4049 (see Section~\ref{sec:paper3_substr}). The red cross in the centre of the images represents the centre of the fitted ellipse, while the white cross represents the approximated position of the binary based on the PSF}. The low intensity of the central 5x5 pixel region of each image is a reduction bias caused by over-correction of the unresolved central polarisation (Section~\ref{sec:paper3_data_reduction}). Note: all images are presented on an inverse hyperbolic scale.
    \label{fig:paper3_ellipse_fit}
\end{figure*}

Following the methodology outlined in \citet{Andrych2023MNRAS.524.4168A, Andrych2024IRAS08}, we used the deconvolved linearly polarised images ($I_{\rm pol}$, or $Q_\phi$ in case of HR\,4049, see Section~\ref{sec:paper3_data_reduction}) to estimate the orientation of the resolved disc surface. To do this, we fitted an ellipse to the positions of peak brightness along the `ring' and determined the corresponding semi-major and semi-minor axis lengths, as well as the position angle (PA)\footnote{The PA is measured counterclockwise from the vertical axis (North) to the first principal radius (major axis), with $180^{\circ}$ ambiguity in the disc PA on the sky.} of the disc. Interferometric studies in the near-infrared \citep[e.g.,][]{Kluska2019A&A...631A.108K, Corporaal2023A&A...674A.151C} have shown that in post-AGB circumbinary discs, the dust inner rim typically coincides with or lies beyond the expected dust sublimation radius, rather than being shaped by binary truncation. Given the spatial resolution limits of SPHERE, the inner rim remains unresolved for all targets in our sample, and the observations instead trace the disc’s outer surface layers at larger separations from the central binary. As a result, the disc can be reasonably modelled as intrinsically circular, with its apparent ellipticity (see second and fourth columns of Figure~\ref{fig:paper3_reduced}) originating from projection effects due to disc inclination. Following this assumption, we estimated the corresponding disc inclination for each target using the semi-major and semi-minor axes of the ellipse. The most plausible position and orientation of circumbinary discs, resolved in $V$ and $I'-$bands for each target, are presented in Figure~\ref{fig:paper3_ellipse_fit} and Table~\ref{tab:paper3_disc_orient}. 

\subsection{Exploring the extended disc morphology}
\label{sec:paper3_morphology}

In this section, we investigate the extended disc morphology of our targets by analysing the disc brightness profiles and examining the resolved substructures in both the $V-$ and $I'-$bands.

\subsubsection{Brightness profiles}
\label{sec:paper3_profiles}

To explore the complexity of the resolved disc surface and reveal the spatial distribution of polarised intensity, we calculated linear, azimuthal, and radial brightness profiles following the methodology of \citet{Andrych2023MNRAS.524.4168A, Andrych2024IRAS08}. 

The linear brightness profiles trace the distribution of polarised intensity along the disc major and minor axes, providing insight into its symmetry. The azimuthal brightness profile captures intensity variations along the brightest resolved part of the circumbinary disc (`ring'), starting from the eastern end of the major axis and proceeding counterclockwise. The radial brightness profile describes the variation of disc polarised intensity with distance from the central binary. To account for the impact of projection effects due to disc orientation on the radial intensity distribution, we deproject observed discs to a 'face-on' view using the estimated inclination of the resolved disc surface for both the $V-$ and $I'-$bands (see Section~\ref{sec:paper3_disc_orient} for details on the inclination and \citet{Andrych2023MNRAS.524.4168A} for more information on the methodology). However, we note that we do not perform the deprojection for AR\,Pup (due to the nearly edge-on orientation of the disc) and V709\,Car (due to not fully resolved `ring', see Section~\ref{sec:paper3_disc_orient}).

We also compared the radial brightness profiles to the expected $r^{-2}$ illumination drop-off, typical of scattered light emission, to determine whether the extent of the observed emission is influenced by disc morphology or limited by observational sensitivity (see bottom row of  Figure~\ref{fig:paper3_profiles_hr4049}, \ref{fig:paper3_profiles_hr4226}, \ref{fig:paper3_profiles_umon}, \ref{fig:paper3_profiles_v709car}, \ref{fig:paper3_profiles_ar_pup} in \ref{sec:paper3_ap_profiles}). 
For HR\,4049, HR\,4226, and U\,Mon, we found that beyond $\sim 0.03-0.05"$, the radial brightness profiles drop off more steeply than the $r^{-2}$ trend in both the $V-$ and $I'-$bands, suggesting a significant reduction in surface dust density or a shadowing effect beyond this region at the observed wavelengths \citep[e.g.,][]{Perez2018ApJ...869L..50P}. In contrast, V709\,Car profile follows the $r^{-2}$ illumination, indicating sensitivity limitations in fully resolving the disc's surface. For AR\,Pup, the orientation of the system significantly impacts the radial brightness profile, accounting for the deviation from the $r^{-2}$ trend.

Three types of brightness profiles for each target are presented in \ref{sec:paper3_ap_profiles}, with individual discussions of the results in Section~\ref{sec:paper3_indiv_cases}. 

\subsubsection{Detection of substructures}
\label{sec:paper3_substr}

\begin{figure*}
     \includegraphics[width=0.3\linewidth]{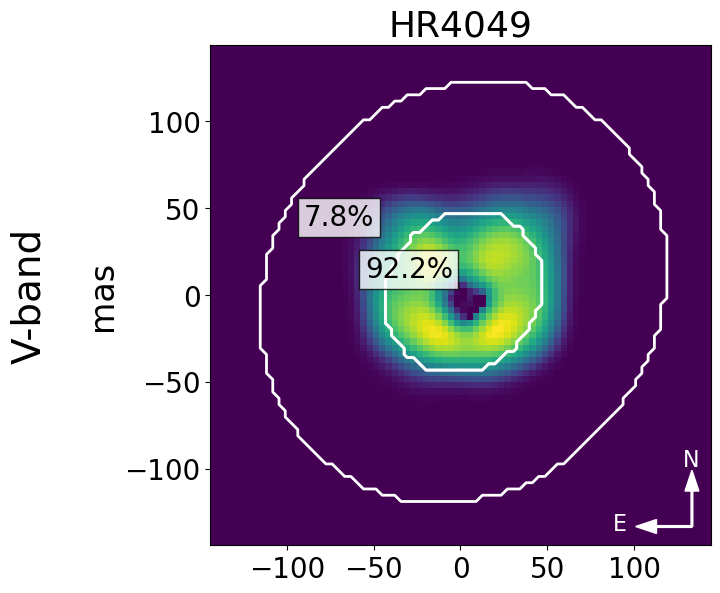}
     \includegraphics[width=0.25\linewidth]{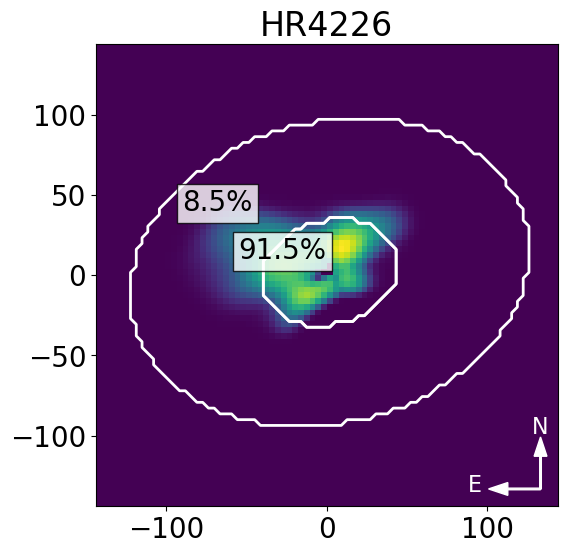}
     \includegraphics[width=0.25\linewidth]{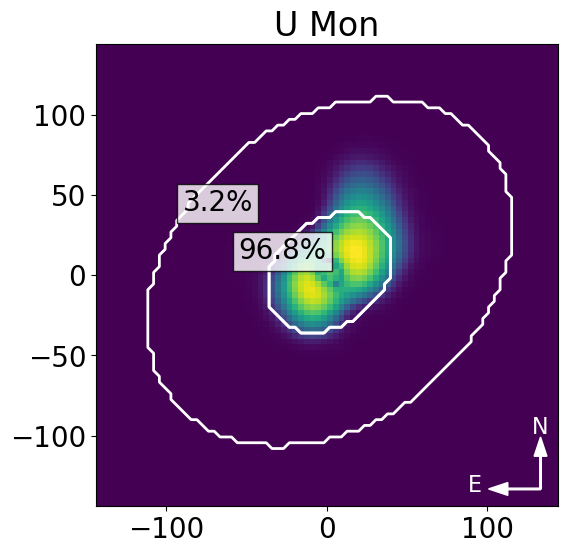}

     \includegraphics[width=0.3\linewidth]{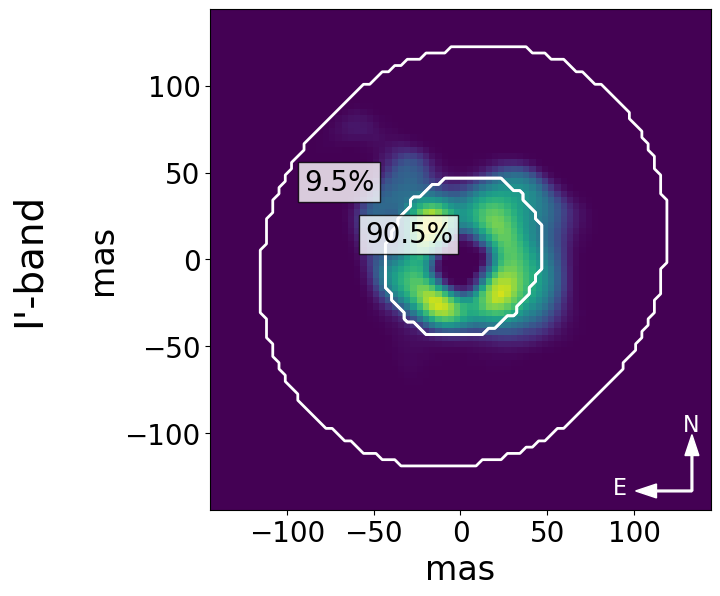}
     \includegraphics[width=0.25\linewidth]{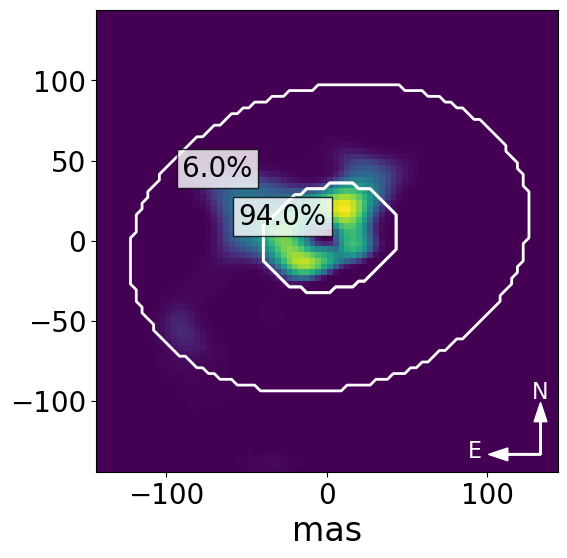}
     \includegraphics[width=0.25\linewidth]{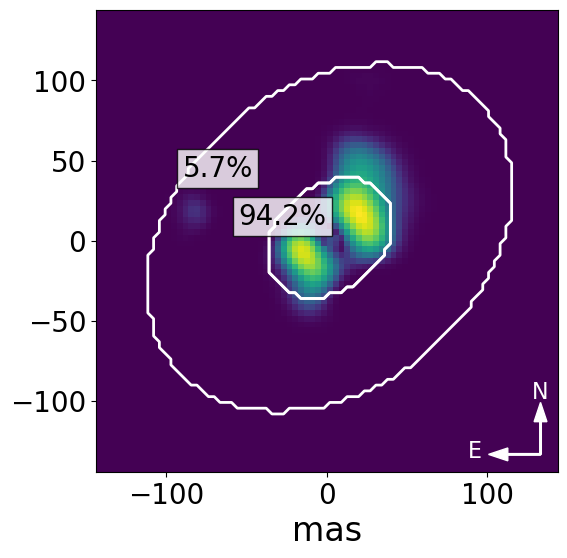}

     \includegraphics[width=0.3\linewidth]{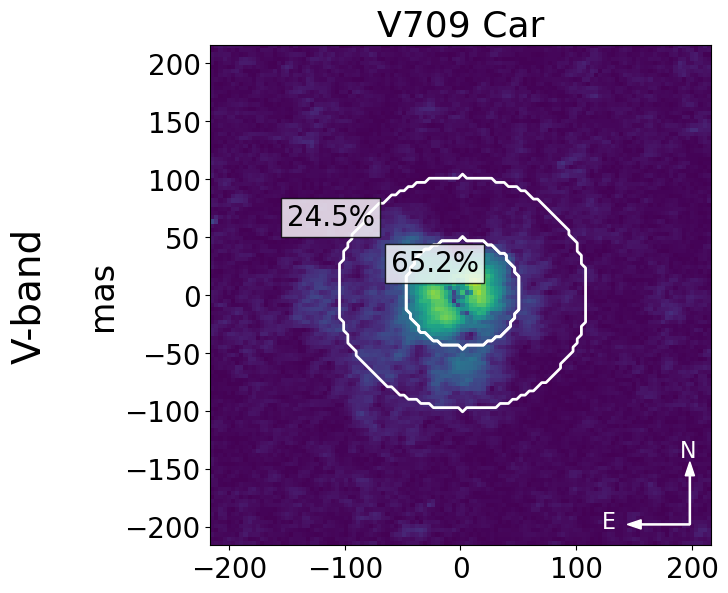}
     \includegraphics[width=0.25\linewidth]{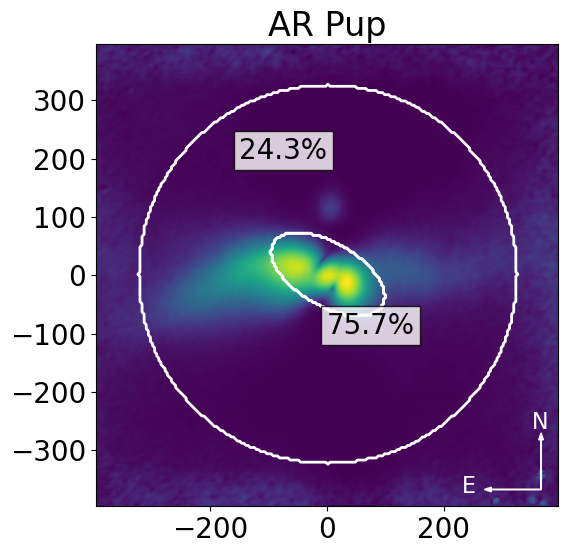}

     \includegraphics[width=0.3\linewidth]{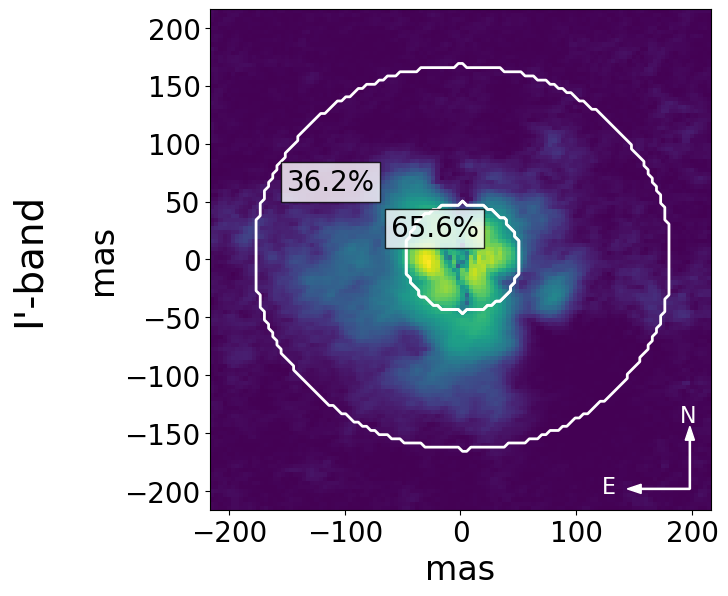}
     \includegraphics[width=0.25\linewidth]{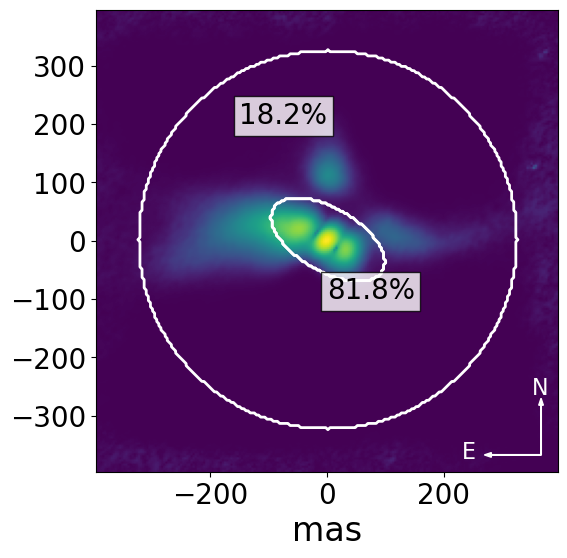}
    
    \caption[Percentage of total polarised disc intensity per resolved structure for all targets in $V-$ and $I'-$bands.]{Percentage of total polarised disc intensity per resolved structure for all targets in $V-$ (top row) and $I'-$ (bottom row) bands (see Section~\ref{sec:paper3_substr}). All images are presented on an inverse hyperbolic scale and oriented North up and East to the left. The low intensity of the central 5x5 pixel region of each image is a reduction bias caused by over-subtracting of unresolved central polarisation (Section~\ref{sec:paper3_data_reduction}).}
    \label{fig:paper3_substr}
\end{figure*}

To identify real features in the linearly polarised images (such as arcs or gaps) we used two criteria: i) the estimated SNR$\geq3$ for the final deconvolved polarimetric images, and ii) the centrosymmetric orientation of the polarisation vector (AoLP, see Figure~\ref{fig:paper3_aolp} of \ref{sec:paper3_ap_aolp}). Using these criteria, we identified reliable substructures, in addition to the bright elliptic `ring', in the polarimetric images of the circumbinary disc for all targets in both the $V$ and $I'-$bands. To quantify the brightness of these substructures, we calculated the percentage of total polarised intensity corresponding to the extended disc substructures and the central ring, excluding the unresolved central polarisation (as shown in Figure~\ref{fig:paper3_substr}). Notably, the substructures appear more pronounced in the $I'-$band compared to the $V-$band for all targets.

\subsection{Disc scattering morphology}
\label{sec:paper3_scatter}

\begin{figure*} 
     \includegraphics[width=0.49\linewidth]{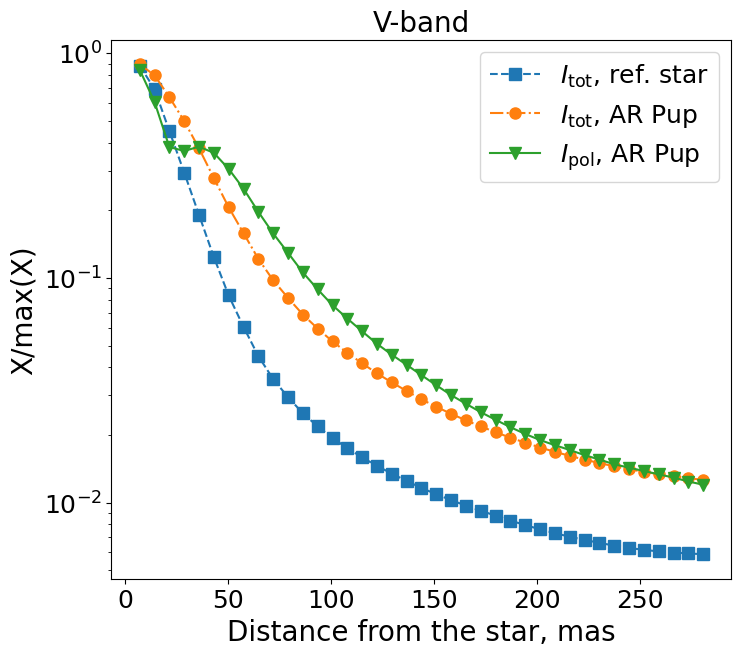}
     \includegraphics[width=0.49\linewidth]{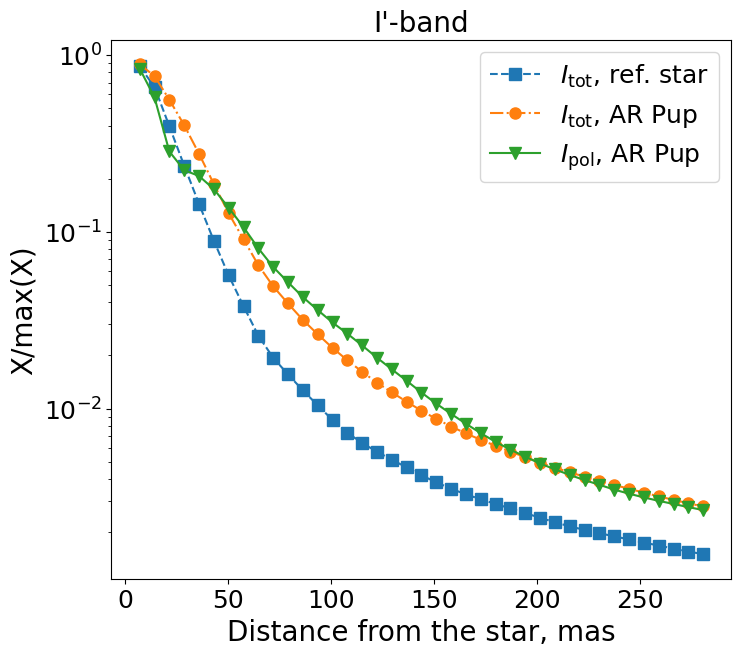}
  
    \caption{Radial profiles of normalized total intensity ($I_{\rm tot}/I^{\rm max}_{\rm tot}$) for AR\,Pup (orange circles) and reference star HD\,75885 (blue squares), and normalized polarised intensity ($I_{\rm pol}/I^{\rm max}_{\rm pol}$) for AR\,Pup (green stars). See Section~\ref{sec:paper3_scatter} for details.}
    \label{fig:paper3_scattered}
\end{figure*}

The total intensity frames ($I_{\rm tot}$) for each target in the sample include both the stellar intensity and scattered light from the disc. \citet{Tschudi2021A&A...655A..37T} proposed a method to disentangle the direct stellar light from the scattered disc component, which requires a clearly resolved separation between the bright inner disc rim and the central star. However, this condition is not met for any of our systems due to the small angular size of the disc's inner rim. To assess whether disentangling of stellar intensity and scattered light is feasible for our binary systems, we computed radial brightness profiles for the total intensity of each post-AGB system and its corresponding reference single star (see Section~\ref{sec:paper3_profiles} for methodology details). The resulting profiles were normalised to the maximum intensity of each target. For four out of five targets (HR\,4049, HR\,4226, U\,Mon, and V709\,Car), the radial intensity profiles closely match those of the reference star, indicating that the image is strongly dominated by the variable PSF of the star. The small angular size of the resolved discs confines their scattered-light signal to regions very close to the PSF core, where it is too faint to be distinguished. As a result, the total intensity images are overwhelmingly dominated by the stellar contribution, and any disc signal remains undetectable under the given PSF halo and seeing conditions. However, this is not the case for AR\,Pup. The radial brightness profile for $I_{\rm tot}$ of AR\,Pup is less steep than that of the reference star (see Figure~\ref{fig:paper3_scattered}). In this system, the disc obscures the central binary (see Section~\ref{sec:sample_and_observations} and Figure~\ref{fig:paper3_reduced}), improving the contrast and allowing the disc to be resolved in total intensity.

\begin{figure}
     \includegraphics[width=0.5\linewidth]{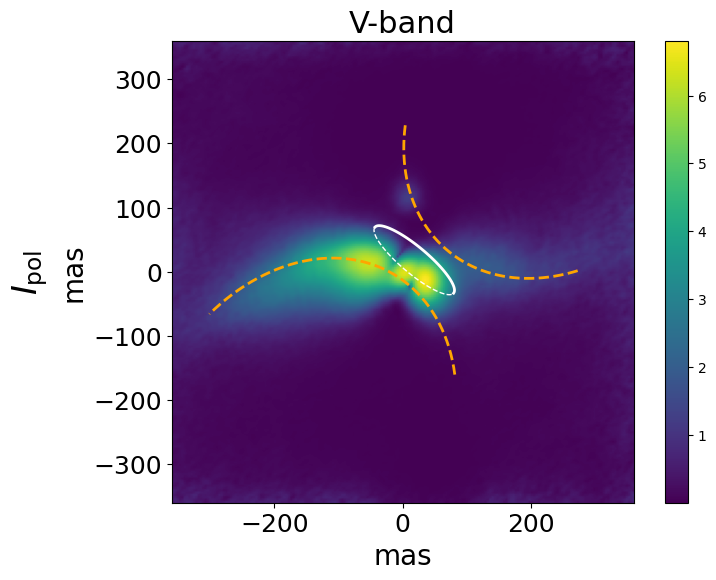}
     \includegraphics[width=0.46\linewidth]{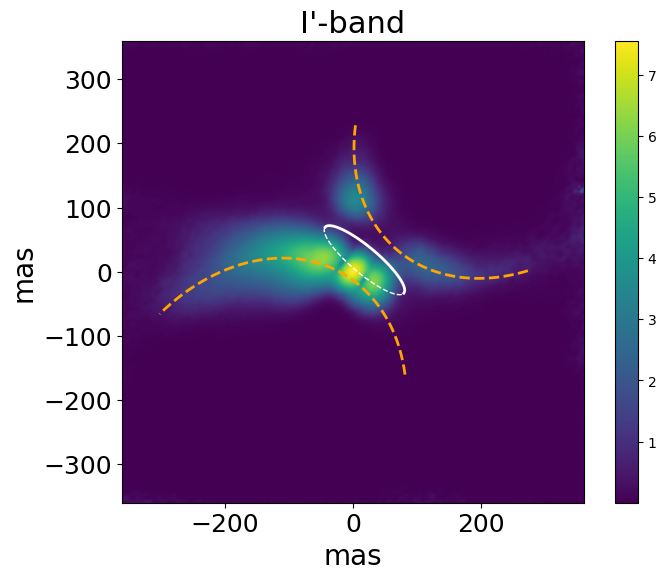}
     \includegraphics[width=0.5\linewidth]{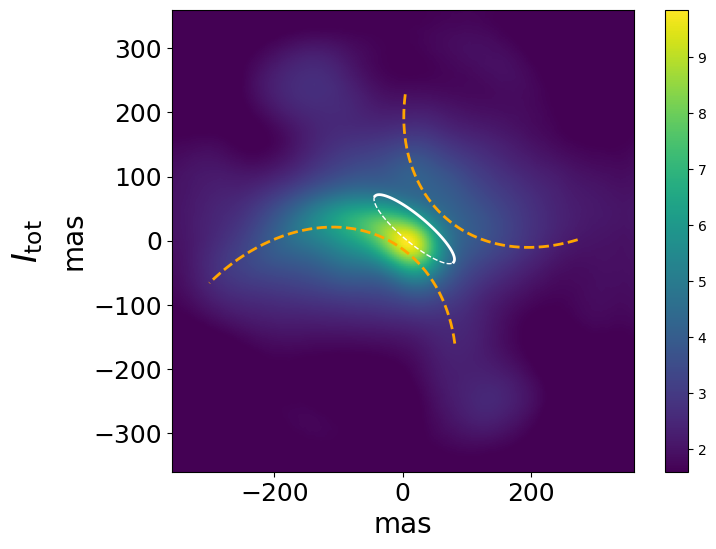}
     \includegraphics[width=0.46\linewidth]{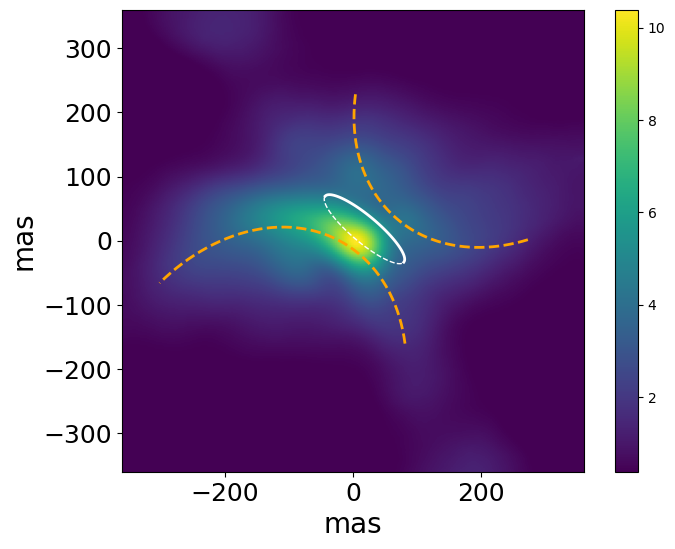}
     \includegraphics[width=0.5\linewidth]{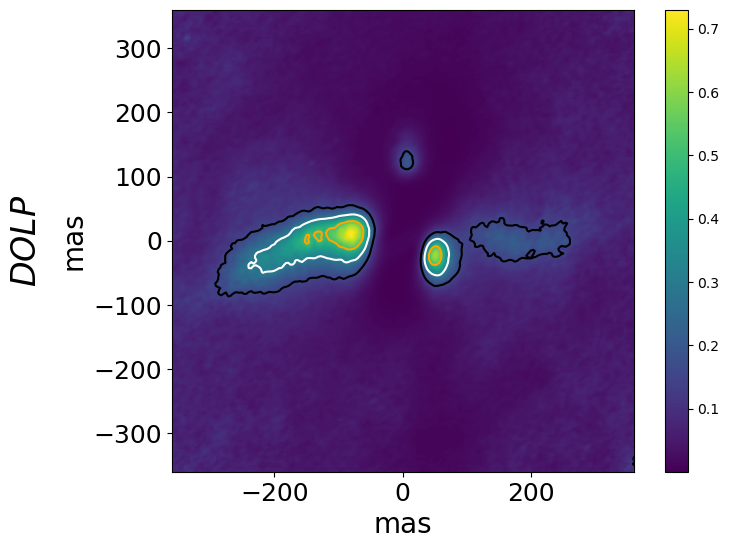}
     %\hspace{0.4cm}
     \includegraphics[width=0.47\linewidth]{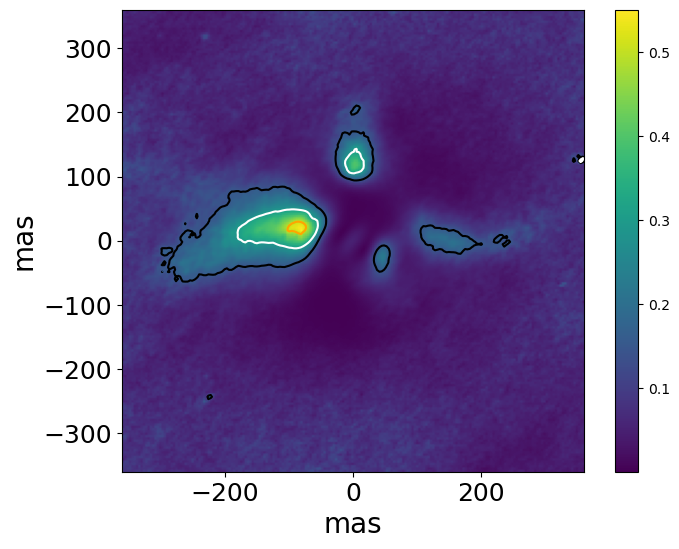}
    \caption{Resulting images of AR\,Pup in $V-$ (left column) and $I'-$ (right column) bands. The top row shows total polarised images with highlighted disc midplane (white ellipse) and flared disc scattering surfaces (dashed orange line). The middle row shows deconvolved total intensity images with highlighted disc midplane and scattering surfaces. The bottom row shows a deconvolved  DoLP map (see Section~\ref{sec:paper3_scatter}) with contours marking regions with polarimetric efficiency of 15\% (black), 30\% (white) and 50\% (orange). Polarised and total intensity images are presented on an inverse hyperbolic scale. All images are oriented North up and East to the left. See Section~\ref{sec:paper3_ar_pup} for details.}
    \label{fig:paper3_AR_Pup}
\end{figure}

Moreover, we computed the degree of linear polarisation (DoLP) map of the resolved scattered emission by calculating the pixel-to-pixel ratio of polarised intensity to the total intensity of the system ($I_{\rm pol}/I_{\rm tot}$). While the total intensity ($I_{\rm tot}$) of AR\,Pup is dominated by direct stellar light close to the central binary, we were able to analyse dust polarisation efficiency at larger separations from the stars. To carefully retrieve the scattered morphology of the disc we also deconvolve the resulting $I_{\rm tot}$ and DoLP map of AR\,Pup with the observational PSF (see Section~\ref{sec:paper3_data_reduction} for details). The final total intensity image and DoLP map of AR\,Pup in $V-$ and $I'-$bands are presented in Figure~\ref{fig:paper3_AR_Pup}. From the DoLP map of scattered light, we directly measured the maximum DoLP of the resolved disc structures (fractional polarisation of the scattered light), finding values of $\sim0.7$ in the $V-$band and $\sim0.55$ in the $I'-$band. These values fall within the broad range of theoretical DoLP values for porous dust aggregates, which can span from $\sim$0.1 to 0.9 at optical wavelengths depending on a complex combination of grain size, porosity, and composition \citep{Tazaki2022A&A...663A..57T}. However, we note that similar polarisation degrees could likely also result from a combination of scattering angles differing from $90^\circ$ and a size distribution of monomers with sizes comparable to the observed wavelengths\citep{Min2012A&A...537A..75M, Tazaki2019MNRAS.485.4951T}.   A more detailed discussion of these results is presented in Sections~\ref{sec:paper3_ar_pup} and~\ref{sec:dust_properties}.

\subsection{Wavelength-dependent polarised intensity and structure of circumbinary disc}
\label{sec:paper3_wavelength}

Polarimetric observations enable estimation of the fraction of stellar light scattered and polarised by the surface layers of circumbinary discs, thereby providing a lower limit on the total reflected light from the disc  \citep[e.g.,][]{Benisty2022arXiv220309991B}. However, accurately constraining dust properties in distant post-AGB discs through polarimetry remains challenging. These difficulties arise from their compact angular size, limitations of current instrumentation, and uncertainties in disc geometry. Despite these constraints, it is still possible to measure the total polarised brightness of the disc at different wavelengths. Assuming that the scattering geometry remains approximately constant across wavelengths, variations in the polarised signal can be attributed primarily to the scattering and absorption characteristics of the dust grains. Consequently, the wavelength dependence of polarised brightness serves as a valuable tool for probing the physical properties of the dust \citep[e.g.,][]{Ma2023A&A...676A...6M, Ma2024A&A...683A..18M}. 

\begin{table}
    \centering
    \caption{Disc polarimetric colours.}
    \label{tab:colours}
    \resizebox{0.6\columnwidth}{!}{%
        \begin{tabular}{lccccc}
            \hline
            Target  & $\eta_{VI}$ & $\eta_{VH}$\\
            \hline
            \hline
            AR Pup   & 0.4 $\pm$ 0.3  & -  \\
            HR 4049  & -0.4 $\pm$ 1 & 0.6 $\pm$ 0.5 \\
            HR 4226  & -0.3 $\pm$ 0.2 & -   \\
            U Mon    & -0.6 $\pm$ 0.4 & -0.9 $\pm$ 0.3\\
            V709 Car &  -0.5 $\pm$ 0.7& - \\
            \hline
        \end{tabular}%
    }
    \begin{flushleft}
    \textbf{Notes:} $\eta_{VI}$ quantifies the polarimetric colour between the $V$- and $I'$-bands, while $\eta_{VH}$ represents the polarimetric colour between the $V$- and $H$-bands. We note that $-0.5 < \eta < 0.5$ is classified as grey colour, $\eta < -0.5$ as blue and $\eta > 0.5$ as red  \citep{Tazaki2019MNRAS.485.4951T, Ma2023A&A...676A...6M}. See Section~\ref{sec:paper3_wavelength} for more details.
    \end{flushleft}
\end{table}

\begin{figure}     
    \includegraphics[width=1\linewidth]{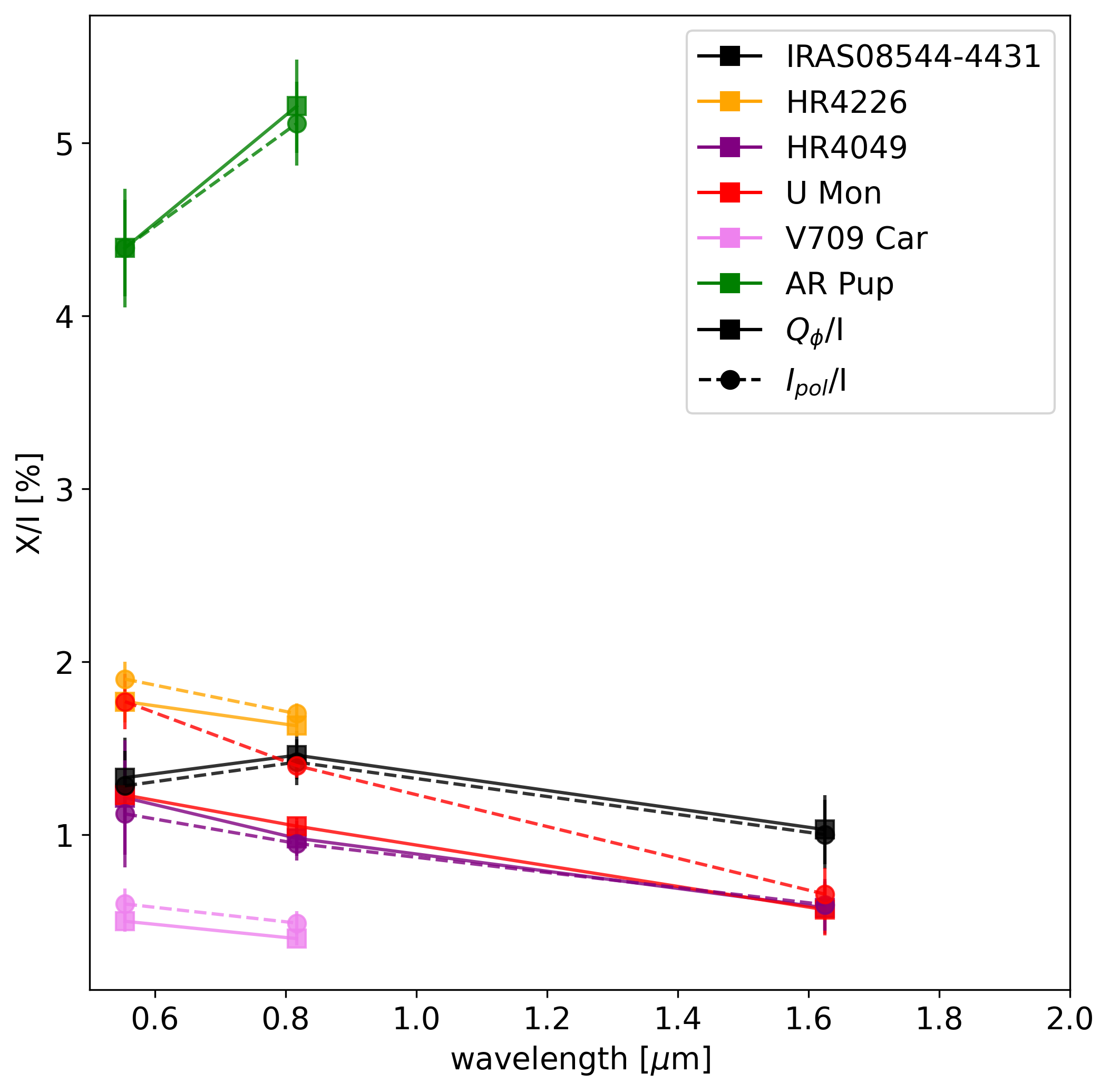}
    
    \caption{Azimuthal ($Q_{\phi}/I$, dashed line) and total ($I_{\rm pol}/I$, solid line) polarised disc brightness relative to the total intensity as a function of wavelength for all targets. See Section~\ref{sec:paper3_wavelength} for more details.}
    \label{fig:paper3_wave_dep}
\end{figure}

To investigate how the disc polarised intensity varies with wavelength, we combined the $V$- and $I'$-band polarimetric results from this study with $H$-band data available for HR\,4049 and U\,Mon \citep{Andrych2023MNRAS.524.4168A}. The polarised brightness values reported in \citet{Andrych2023MNRAS.524.4168A} were not corrected for PSF smearing, which are known to attenuate the polarised flux. Therefore, to ensure consistency across wavelengths, we applied a PSF smearing correction to the $H$-band data, following the procedure described in Section~\ref{sec:paper3_data_reduction}. The resulting wavelength dependence of both the total ($I_{\rm pol}/I$) and azimuthal ($Q_{\phi}/I$) polarised intensity for each system is presented in Figure~\ref{fig:paper3_wave_dep}. In most cases, the two ratios ($Q_{\phi}/I$ and $I_{\rm pol}/I$) are consistent within uncertainties. An exception is U\,Mon, where $I_{\rm pol}/I$ decreases more rapidly with wavelength than $Q_{\phi}/I$, although the overall trend remains qualitatively similar.

To quantitatively characterise the wavelength dependence of the polarised brightness, we computed the logarithmic gradient $\eta_{\lambda_2\lambda_1}$ of the polarised intensity between two bands (either both optical or optical and near-IR). This gradient provides a measure of the disc’s polarimetric colour:

\begin{equation}
    \eta_{\lambda_2\lambda_1}=\frac{\textrm{log}(Q_{\phi}/I_{tot})_{\lambda_1}-\textrm{log}(Q_\phi/I_{tot})_{\lambda_2}}{\textrm{log}(\lambda_2/\lambda_1)},    
\end{equation}

where $\lambda_1 < \lambda_2$, and $-0.5 < \eta < 0.5$ is classified as grey colour, $\eta < -0.5$ as blue and $\eta > 0.5$ as red  \citep{Tazaki2019MNRAS.485.4951T, Ma2023A&A...676A...6M}.

The resulting optical ($\eta_{VI}$) and optical-IR ($\eta_{VH}$) polarimetric colour for each target binary system are presented in Table~\ref{tab:colours}. Notably, three out of the five post-AGB systems (HR\,4226, V709\,Car, and AR\,Pup) exhibit results consistent with a grey disc colour, while U\,Mon and HR\,4049 display borderline blue colour.

A multi-wavelength polarimetric imaging study of the post-AGB system IRAS\,08544-4431 revealed variations in disc size and morphology across different wavelengths \citep{Andrych2024IRAS08}. To investigate whether similar effects are present in our targets, we first compared the morphology of the smallest resolved section of the circumbinary disc close to the central binary (the bright `ring'). Unlike the results for IRAS\,08544-4431, we did not find any significant variation (beyond the uncertainty) in the apparent inclination of the discs with wavelength (see Figure\ref{fig:paper3_ellipse_fit} and Table~\ref{tab:paper3_disc_orient}).

\begin{figure*} 
    \includegraphics[width=0.3\linewidth]{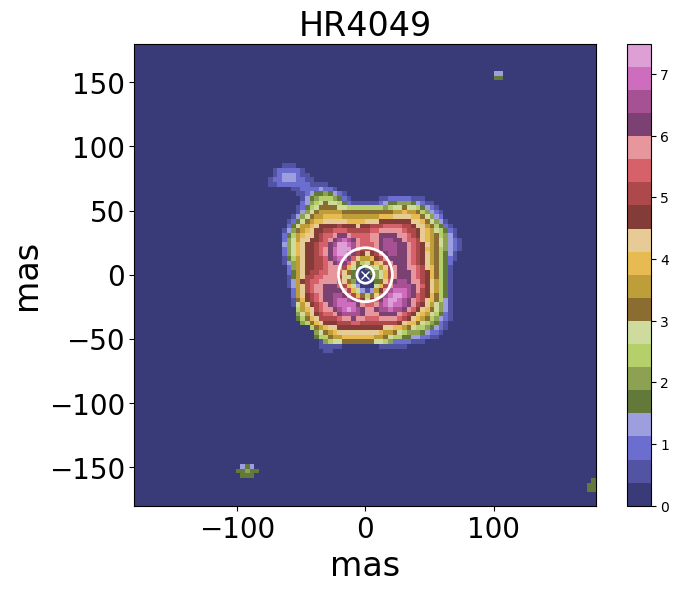}
    \includegraphics[width=0.3\linewidth]{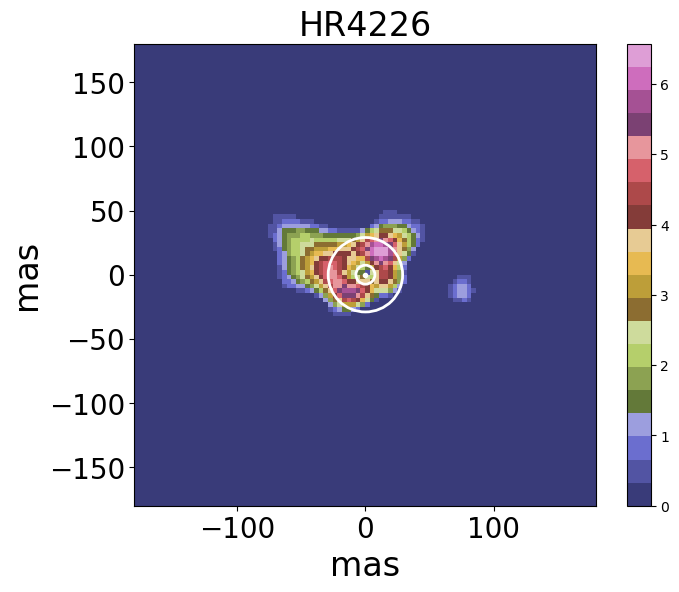}
    \includegraphics[width=0.3\linewidth]{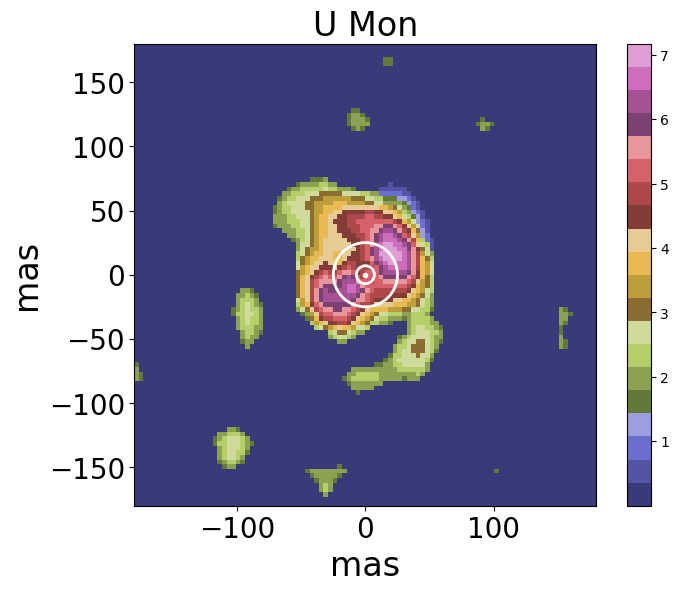}
    \includegraphics[width=0.3\linewidth]{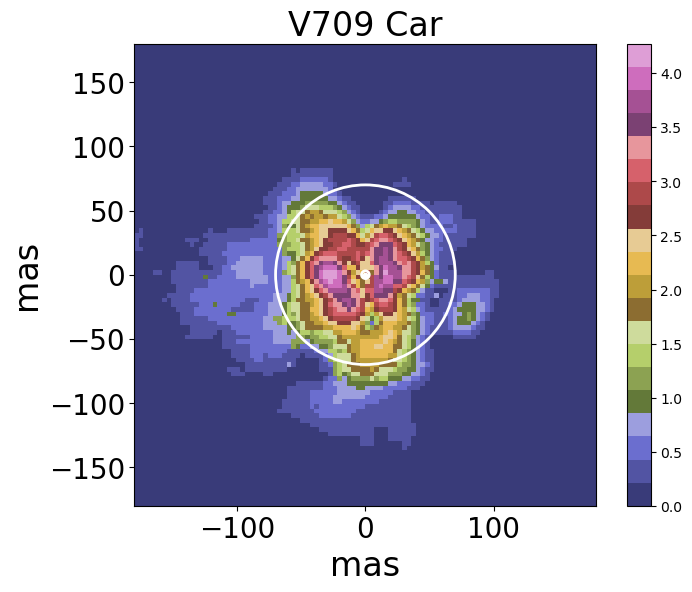}
    \includegraphics[width=0.3\linewidth]{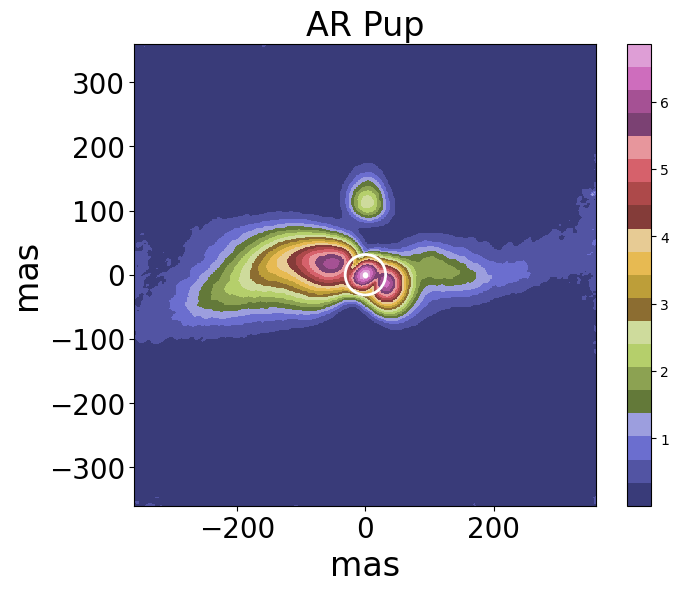}
    \caption{Combined $Q_\phi$ polarised disc morphology for all targets using $V$ and $I-$band data from this study, along with the SPHERE/IRDIS $H-$band results for HR\,4049 and U\,Mon \citep[adapted from][]{Andrych2023MNRAS.524.4168A}. The white dot represents the position of the binary star. Before combining the $Q_\phi$ images in each band, they were normalised to the total intensity. The white contour represents the disc model obtained by geometrical modelling of mid-IR interferometric data with VLT/MIDI \citep{Hillen2017}. Images are presented on an inverse hyperbolic scale and normalised to highlight the intensity change along the discs. See Section~\ref{sec:paper3_wavelength} for more details.}
    \label{fig:paper3_combined}
\end{figure*}

We also combined polarimetric images of all our targets in the $V$ and $I'-$band data (from this study), along with $H-$band data for HR\,4049 and U\,Mon \citep[adapted from][]{Andrych2023MNRAS.524.4168A}. For consistency, the $I_{\rm pol}$ images (or $Q_\phi$ in the case of HR\,4049; see Section~\ref{sec:paper3_data_reduction}) were normalised to the corresponding total intensity of each system. In Figure~\ref{fig:paper3_combined} we present the combined polarimetric images for all targets across the $V$, $I'$, and $H$ bands, where available. To highlight spatial variations in brightness along the discs, we used an inverse hyperbolic scaling and a discrete colour map. We find that substructures are generally more pronounced in the $I'-$band compared to the $V-$band for HR\,4049, AR\,Pup, and HR\,4226, but a noticeable difference in disc size is observed only for V709\,Car, where the disc appears more extended in the $I'-$band. Additionally, while comparing SPHERE/ZIMPOL optical results with SPHERE/IRDIS near-IR images, we found that U\,Mon shows a more extended disc in the near-IR, whereas HR\,4049 does not exhibit any change in size. We further discuss the results for each binary system individually in Section~\ref{sec:paper3_indiv_cases}.

\subsection{Individual case studies}
\label{sec:paper3_indiv_cases}

In this section, we provide a comprehensive analysis of the SPHERE/ZIMPOL polarimetric imaging results for each evolved binary system in our sample. For each target, we present detailed observational findings, including the resolved disc structures and their scattering and polarisation properties, which are also summarised in Tables~\ref{tab:paper3_disc_orient},~\ref{tab:colours}. We also compare our results, where possible, with previous studies based on high-resolution imaging data, IR and radio interferometry. While this section focuses on presenting the results, their broader implications are discussed in Section~\ref{sec:paper3_discussion}.

\subsubsection{AR Pup}
\label{sec:paper3_ar_pup}

AR\,Pup was the first post-AGB binary for which the extended disc structure was resolved with SPHERE \citep{Ertel2019AJ....157..110E}. However, that study lacked polarimetric data, as the SPHERE polarimetric mode was not available at the time. Here, we present the first multi-wavelength polarimetric results for this target.

Similar to other targets in the sample, the polarimetric images of AR\,Pup include contributions from unresolved central polarisation, which arises from both the unresolved portion of the disc and interstellar polarisation (see Section~\ref{sec:paper3_data_reduction}). Although we estimated the degree and orientation of the unresolved component (see Table~\ref{tab:paper3_unresolved}), the orientation of the disc leads to a significant overestimation of the unresolved polarised signal. This complicates the subtraction of unresolved polarisation without affecting the resolved disc structure. Therefore, we did not apply the correction for unresolved polarisation for this data. We note that both the radial and linear brightness profiles of AR\,Pup polarimetric images ($I_{\rm pol}$) exhibit clear peaks at the position of the central binary, corresponding to the unresolved polarisation, as well as additional peaks associated with the extended disc polarimetric emission (see Figure~\ref{fig:paper3_profiles_ar_pup}, \ref{sec:paper3_ap_profiles}).

AR\,Pup displays the highest polarimetric disc brightness in our sample (4.4$\pm$0.3\% in the $V$-band and 5.1$\pm$0.2\% in the $I'-$band, see Table~\ref{tab:paper3_disc_orient}), likely due to its disc orientation, which partially obscures the host binary and leads to an underestimation of the stellar intensity. This aligns with AR\,Pup's known RVb phenomenon, a long-term periodic variation in mean magnitude \citep[period of 1194 days,][]{Kiss2017}, typically explained by an inclined circumbinary disc that periodically obscures the pulsating primary star \citep{Vega2017ApJ...839...48V, Manick2017}. AR\,Pup is the only target in the sample for which the polarimetric disc brightness slightly increases with wavelength. However, this increase may be attributed to the greater number of scattered photons reaching the observer due to the decrease in dust absorption at longer wavelengths \citep{Kirchschlager2014A&A...568A.103K, Tazaki2018ApJ...860...79T}, rather than an increase in polarimetric efficiency. Investigating the deconvolved DoLP map of the resolved scattered emission of AR\,Pup (see bottom panel of Figure~\ref{fig:paper3_AR_Pup} and Section~\ref{sec:paper3_scatter} for the details) we found a maximum degree of resolved polarisation to be $\sim0.7$ in $V-$band and $\sim0.55$ in $I'-$band, in both cases the peak was located to the east from the position of the binary. For the northwestern disc surface, the polarisation reaches $\sim0.2$ in $V-$band and $\sim0.3$ in $I'-$band. 

\citet{Ertel2019AJ....157..110E} presented the first SPHERE high-resolution imaging of AR\,Pup, revealing a distinct `double-bowl' structure separated by a dark band. They interpreted this resolved morphology as an edge-on flared disc with an optically thick mid-plane, where the two `bowls' correspond to the disc surfaces scattering stellar light. Their study also found that the southeastern disc surface appears brighter than the northwestern one, suggesting that the southeastern side is oriented toward us, allowing starlight to scatter directly into our line of sight, while the northwestern surface is partially obscured by the disc itself. Our results support these findings and interpretations, as AR\,Pup exhibits the same `double-bowl' structure in both $V-$ and $I'-$band polarimetric images as well as in total intensity frames (see Figure~\ref{fig:paper3_AR_Pup}). From the location of the disc mid-plane in our polarimetric images, we estimate its size to be approximately 80 mas, with an inclination of $75\pm10^\circ$ and a PA of $50\pm5^\circ$, consistent with the values reported by \citet[$75\pm15^\circ$ and $45\pm10^\circ$,][]{Ertel2019AJ....157..110E}. In addition to confirming the disc geometry, our observations provide further insight into the polarisation properties of the system. We argue that the disc orientation proposed by \citet{Ertel2019AJ....157..110E} explains why the northwestern 'bowl' appears significantly dimmer in azimuthally polarised $Q_\phi$ images compared to total polarised intensity frames, whereas the southeastern `bowl' remains consistently bright. If the northwestern surface is oriented away from us, the stellar light scattered there likely undergoes multiple scattering through the disc material, disrupting the polarisation orientation and reducing the expected azimuthal $Q_\phi$ signal typically seen after single scattering. The suggested disc orientation is further supported by the fact that the northwestern `bowl' is brighter in the $I'-$band than in the $V-$band. This pattern is consistent with the behaviour of light as it passes through the disc, where dust absorption decreases with increasing wavelength, allowing more light to penetrate.

We also note that the brightness asymmetry within the southeastern disc surface differs between our 2018 observations and those from 2016 presented in \citet{Ertel2019AJ....157..110E} (see Figure~\ref{fig:ar_pup_ertel} in \ref{sec:paper3_appendix_arpup}). In the earlier observations, the brighter arc extended along the southern side of the `bowl', whereas in our data, it appears more prominent on the eastern side. This change is consistent with the binary's orbital motion, which, over the two-year interval between observations, would have moved the post-AGB star by approximately 60\% of its orbit (orbital period of $\sim1200$ days; see Table~\ref{tab:paper3_sample}). This result highlights the potential for high-resolution imaging to trace the dynamic response of the circumbinary disc to stellar motion.

\subsubsection{HR 4049}
\label{sec:paper3_hr4049}

HR\,4049 was observed twice, on January 7 and 8, 2019, due to not ideal weather conditions during the first observation. To increase the signal-to-noise ratio, we initially reduced the polarimetric cycles from each observation separately, selecting the highest-quality images before combining them. We then mean-combined all frames after reduction and performed the analysis on the combined set.

We found that HR\,4049 shows the smallest unresolved polarisation component in the sample (see Table~\ref{tab:paper3_unresolved}, Section~\ref{sec:paper3_data_reduction}). This can be attributed to HR\,4049's position above the Galactic midplane, which reduces polarisation from the diffuse interstellar medium along the line of sight. Additionally, geometrical modelling of near-IR interferometric data for HR\,4049 revealed a binary and Gaussian ring with inner rim diameter of $\sim16$ mas \citep[][]{Kluska2019A&A...631A.108K}, suggesting that the unresolved portion of the disc is relatively small. Moreover, the unresolved central polarisation we measured for HR\,4049 falls within the uncertainty range of the total polarisation detected from the reference star (HD96314), both in value and orientation (see Figure~\ref{fig:paper3_wave_dep_unres}). Therefore, we conclude that the unresolved central polarisation observed for HR\,4049 is predominantly caused by interstellar polarisation rather than the unresolved portion of the circumbinary disc.

We measured a borderline blueish polarimetric colour index of resolved disc brightness for the system ($\eta_{VI}=-0.4\pm1$ and $\eta_{VH}=0.6\pm0.5$). The unusually large uncertainty in $\eta_{VI}$ is primarily due to a ‘cross-shaped’ observational artefact in polarimetric intensity observed in the $V-$ and (to less extent) $I'-$bands (see Section~\ref{sec:artefact}). This intensity drop also disrupts the azimuthal brightness profiles in both $V-$ and $I'-$bands (see Figure~\ref{fig:paper3_profiles_hr4049}, \ref{sec:paper3_ap_profiles}).  

HR\,4049 shows a bright resolved disc surface in both $V-$ and $I'-$bands. We obtained a disc inclination of $\sim29^\circ$  and $23^\circ$, with PA of $109^\circ$ and $138^\circ$ in $V-$ and $I'-$bands, respectively (see Section~\ref{sec:paper3_disc_orient}). These values align with previously identified disc orientations in near-IR $H$-band polarimetric imaging \citep[inclination of $17^\circ$ and PA of $174^\circ$,][]{Andrych2023MNRAS.524.4168A}, suggesting a nearly face-on disc orientation. However, this low inclination contradicts the observed RVb phenomenon of HR\,4049 \citep{Waelkens1991A&A...242..433W} and also differs from the results of IR interferometric studies \citep[inclination of 49$^\circ$, PA of 63$^\circ$,][]{Kluska2019A&A...631A.108K}. While we cannot fully explain this discrepancy, we suggest that the RVb phenomenon may be caused by a high disc scale-height, a potentially misaligned inner part of the disc, or other unresolved dynamical processes closer to the central binary. We did not detect any significant change in the size of the resolved disc of HR\,4049 across the $V-$, $I'-$, and $H$-band polarimetric data, with extended resolved emission observed up to $\sim80$ mas from the central star (see Figure~\ref{fig:paper3_combined}). 
Although the $V-$ and $H-$band polarimetric images do not reveal any significant substructures, the $I'-$band images display an arc-like feature to the northeast of the binary position. To determine whether this substructure might be an observational artefact, we compared separately reduced observations from January 7 and 8, 2019, and found that this structure is consistently visible in both datasets. While additional observations of the disc midplane, such as dust continuum observations with ALMA, are necessary to draw definitive conclusions about the physics of this feature, we speculate that it may be part of a spiral structure.

\subsubsection{U Mon}
\label{sec:paper3_umon}

For U\,Mon, we measured an unresolved polarisation degree of $0.52\pm0.05$\% in the $V$-band and $0.77\pm0.04$\% in the $I'$-band, both of which exceed the polarised intensity measured for the reference star (see Figure~\ref{fig:paper3_wave_dep_unres}). Since U\,Mon is located in the Galactic plane, we expect a quite significant effect of interstellar polarisation for this system (see Section~\ref{sec:paper3_data_reduction}). However, the reference star HD71253 lies above the Galactic midplane and, therefore, cannot serve as a reliable estimate of interstellar polarisation for U\,Mon. Thus, we cannot determine the relative contributions of the unresolved circumbinary disc polarisation and interstellar polarisation for U\,Mon.

We measured the polarised disc brightness of U\,Mon to be 1.8$\pm$0.2\% of the total intensity in the $V$-band and 1.4$\pm$0.07\% in the $I'-$band, resulting in a blueish polarimetric colour index of $\eta_{VI}=-0.6\pm0.4$ and $\eta_{VH}=-0.9\pm0.3$. 

U\,Mon shows a clear bright `ring' in  $V-$band and two arcs separated by a dark strip in $I'-$band. However, we note that this dark strip is an artefact from the subtraction of unresolved central polarisation, including some polarised emission from the disc. We estimated a disc inclination of $\sim 45^\circ$  and PA of $\sim130^\circ$ in both $V-$ and $I'-$bands (see Section~\ref{sec:paper3_disc_orient}). While the PA of the disc aligns with the previously identified value from near-IR $H$-band polarimetric imaging ($\sim140^\circ$), our inclination estimate is significantly higher \citep[previously estimated at $25^\circ$ in the $H$-band,][]{Andrych2023MNRAS.524.4168A}. However, the higher inclination value based on $V-$ and $I'-$band data is consistent with values obtained from geometric modelling using near-IR interferometric observations \citep[$\sim 45^\circ$][]{Kluska2019A&A...631A.108K} and with the observed RVb phenomenon \citep{Kiss2017}. We note that the north-western side of the disc appears brighter than other areas in both the $V$- and $I'$-bands (see Figure~\ref{fig:paper3_profiles_umon}). We suggest that this brightness peak is likely caused by the forward scattering of stellar light on dust grains that are comparable in size to the observational wavelength \citep[e.g.,][]{Ginski2023ApJ...953...92G}. This effect also indicates that the northern part of the disc is inclined toward the observer. While an asymmetric dust distribution could, in principle, produce such a bright region, generating a feature this prominent would likely require significant disc disruption (such as from a companion or localised instability) and should leave detectable signatures in other observations, such as infrared interferometry. To our knowledge, no such features have been reported for U\,Mon.

While we do not observe any significant change in the resolved disc size or morphology across the $V$- and $I'$-band polarimetric data, we note that the near-IR $H$-band polarimetric image reveals a more extended disc (see Figure~\ref{fig:paper3_combined}). A similar effect has been reported for the post-AGB system IRAS\,08544-4431 \citep{Andrych2024IRAS08}. As suggested for IRAS\,08544-4431, we propose that the larger apparent disc size and lower estimated inclination in $H$-band polarimetric images may result from SPHERE probing slightly deeper disc layers at longer wavelengths, as dust opacity decreases from optical to near-IR \citep{Kirchschlager2014A&A...568A.103K, Tazaki2018ApJ...860...79T}.

\subsubsection{HR 4226}
\label{sec:paper3_hr4226}

HR\,4226 is located in the Galactic plane, however, we measured a relatively low degree of unresolved polarisation for this target, with DoLP of $0.15\pm0.06$\% in $V-$band and $0.16\pm0.04$\% in $I'-$band (see Section~\ref{sec:paper3_data_reduction}). Although the degree of unresolved polarisation is similar to that detected from the reference star (HD98025), the orientation differs (see Figure~\ref{fig:paper3_wave_dep_unres}). Thus, we cannot determine the relative contribution of the unresolved circumbinary disc polarisation and interstellar polarisation for HR\,4226.

For the resolved disc brightness in polarised light of HR\,4226, we measured  1.9$\pm$0.1\% of the total intensity in the $V$-band and 1.75$\pm$0.06\% in the $I'-$band, resulting in a grey polarimetric colour index of $\eta_{VI}=-0.3\pm0.2$. We also note that similarly to HR\,4049, the $V$-band polarimetric observations of HR\,4226 suffer from a reduction artefact. Although the effect is less pronounced, it causes a ‘cross-shaped’ intensity decrease that disrupts the azimuthal brightness profile of the disc (see Figure~\ref{fig:paper3_profiles_hr4049} and Section~\ref{sec:paper3_data_reduction}).  

HR\,4226 shows a clear bright `ring' in both $V-$ and $I'-$bands. We estimated a disc inclination of $51^\circ$  and $40^\circ$, with PA of $104^\circ$ and $106^\circ$ in $V-$ and $I'-$bands, respectively (see Section~\ref{sec:paper3_disc_orient}). We did not detect any significant change in the size of the resolved disc of HR\,4226 across the $V$-, $I'-$band polarimetric data with the extended emission of up to $\sim90$ mas from the central star (see Figure~\ref{fig:paper3_combined}). We also note that azimuthal brightness profiles of the disc show two distinct peaks along the major axis (see middle panel of Figure~\ref{fig:paper3_profiles_hr4226}, \ref{sec:paper3_ap_profiles}), commonly attributed to the projection of an inclined circular disc onto the sky. These findings complement geometric modelling results based on IR interferometric data, which trace emission from warm dust closer to the central binary. A near-IR interferometric study with VLTI/PIONIER found that the circumbinary material was over-resolved, resulting in a binary model without a detectable ring \citep{Kluska2019A&A...631A.108K}. Geometric modelling of mid-IR interferometric data from VLT/MIDI revealed a ring-like structure, with a half-light radius of $18\pm1$ mas and an outer diameter of $58\pm3$ mas \citep{Hillen2017}. To develop a more complete understanding of this system, we conducted a thorough literature review, including spectroscopic studies and radial velocity (RV) monitoring results. While no elemental abundance studies are currently available for HR\,4226, RV monitoring indicates an orbital period of $527\pm6$ days with small velocity variations ($\Delta V \simeq 1$ km/s), leading to an exceptionally low mass function of $1.4\times10^{-6}$ \citep{Maas2003PhDT.......265M}. These results, combined with the results of our SPHERE study, raise doubts on the nature of HR\,4226, which are further explored in Section~\ref{sec:paper3_discussion}.

\subsubsection{V709 Car}
\label{sec:paper3_v709_car}

For V709\,Car, we measured a high degree of unresolved polarisation with DoLP of $1.7\pm0.16$\% in $V-$band and $2.6\pm0.01$\% in $I'-$band (see Section~\ref{sec:paper3_data_reduction}). These values are much higher than polarisation detected from the reference star HD94680, located closely in the Galactic plane. Therefore, we suggest that the unresolved central polarisation measured for V709\,Car is predominantly caused by the unresolved part of the circumbinary disc rather than polarisation in the diffuse interstellar medium along the line of sight.

We measured the resolved disc brightness in polarised light of V709\,Car to be the smallest in the sample (0.6$\pm$0.09\% in the $V$-band and 0.49$\pm$0.07\% in the $I'-$band, see Table~\ref{tab:paper3_disc_orient}), resulting in a borderline blueish polarimetric colour index of $\eta_{VI}=-0.5\pm0.8$. We also note that the polarimetric observations of V709\,Car suffer from instrumental artefacts that disrupt the azimuthal brightness profile of the disc (see Figure~\ref{fig:paper3_profiles_v709car} and Section~\ref{sec:artefact}). However, this effect is visible only in $I'$-band data. 

Similar to other targets in the sample, V709\,Car also shows a `ring' structure in both $V-$ and $I'-$bands, which we interpret as a circumbinary disc surface. We estimated a disc inclination of $\sim30^\circ$ and PA of $\sim120^\circ$ in both $V-$ and $I'-$bands (see Section~\ref{sec:paper3_disc_orient}).  We also note that $V-$band azimuthal brightness profile of the disc shows two distinct peaks along the major axis (see middle panel of Figure~\ref{fig:paper3_profiles_v709car}, \ref{sec:paper3_ap_profiles}), commonly attributed to the projection of an inclined circular disc onto the sky. V709\,Car is the only target in our sample that shows a significant difference in the size of the resolved extended emission in reaching $\sim150$ mas from the central star in $I'$-band polarimetric image, while appearing much smaller in $V-$band (see Figure~\ref{fig:paper3_substr}). However, we note that due to the low level of resolved signal and alignment of the radial brightness profiles with the expected r$^{-2}$ illumination drop-off (see bottom panel of Figure~\ref{fig:paper3_profiles_hr4226}, \ref{sec:paper3_ap_profiles}), these values have to be treated with caution. The resolved disc size is likely constrained by the telescope's sensitivity, as the scattered emission farther from the central star may be too faint to detect, rather than being limited by the disc's true extent or shadowing effects. 

Geometric modelling of near-IR interferometric data reveals a structure dominated by the primary star and a circumbinary ring with remarkably high temperature of $7700\pm1900 K$ and outer radius of $\sim2$ mas \citep{Kluska2018A&A...616A.153K}. Geometric modelling of mid-IR interferometric data with VLT/MIDI indicates a ring with a half-light radius of $75\pm4$ mas, $\sim115$ times larger than its near-IR size \citep{Hillen2017}.  The Spitzer survey of \citet{Gielen2011A&A...533A..99G} indicates that the spectrum of this source is dominated by amorphous silicates with no crystalline dust features. These literature results, combined with outcomes of this study, raise questions about the evolutionary stage of the primary star. We have explored this further in Section~\ref{sec:paper3_discussion}. 

\section{Discussion}
\label{sec:paper3_discussion}

In this section, we interpret the obtained results to better understand the morphology and dust characteristics in post-AGB circumbinary discs. We accomplish this by: i) characterising the dust properties on the disc surfaces within our sample, ii) comparing the circumstellar environments of AGB and post-AGB systems, iii) examining the similarities between circumbinary discs around post-AGB binaries and PPDs around YSOs and iv) investigating two targets whose evolutionary nature is unclear.

\subsection{Dust properties and disc morphology of post-AGB binaries}
\label{sec:dust_properties}

The appearance of circumstellar discs in scattered light depends on their geometry and the optical properties of the dust particles within them. These characteristics can be inferred by analysing the intensity and polarisation degree of reflected stellar light, as well as its brightness distribution and wavelength dependencies \citep[e.g.,][]{Tazaki2023ApJ...944L..43T, Benisty2022arXiv220309991B}.

Four of five evolved systems in our sample (AR\,Pup, HR\,4049, HR\,4226, and U\,Mon) exhibit high polarimetric disc brightness, with more than 1\% of the total intensity of the system in optical wavelengths ($V$ and $I'$-bands). For AR\,Pup, we also directly measured the maximum degree of resolved polarisation of $\sim0.7$ in the $V-$band and $\sim0.55$ in the $I'-$band (see Section~\ref{sec:paper3_scatter}). We interpret the brighter region in U\,Mon resolved disc surface located to the northwest of the binary as evidence of forward scattering, which would be similar to findings for the post-AGB system IRAS\,08544-4431 \citep{Andrych2024IRAS08} and is indicative of anisotropic scattering. However, forward scattering signatures are less evident for HR\,4049 and HR\,4226, likely due to their orientation, instrumental artefacts, and limited spatial resolution. The optical polarimetric colour across the sample tends toward grey or slightly blue ($\eta_{VI} \sim -0.6 to -0.3$), except for AR\,Pup, which exhibits a borderline grey-red colour ($\eta_{VI} \sim0.4$). In AR\,Pup, total polarimetric brightness slightly increases with wavelength, likely due to dominant dust absorption effects caused by the high inclination of the system and significant disc flaring (see Fig~\ref{fig:paper3_AR_Pup}). 

Numerical models by \citet{Tazaki2019MNRAS.485.4951T} show that the degree and colour of scattered polarised light are sensitive to dust grain structure. Single monomers, which are small compared to the scattering wavelength, produce highly blue optical to IR polarimetric colours ($\sim-2.4$), while highly porous aggregates (composed of such monomers but large compared to the wavelength) produce more moderate colours ($\sim-0.5$ to $-0.2$), and compact (less porous) aggregates tend toward grey or reddish colours ($\sim0$ to $0.9$). These results suggest that the observed grey or slightly blue colours in our sample are consistent with porous aggregates. Building on this, \citet{Tazaki2022A&A...663A..57T} conducted a detailed parameter study using the T-matrix method to compute the DoLP of light scattered by such aggregates. Their results show that the DoLP strongly depends on parameters such as monomer size, aggregate size, porosity, and dust composition, with porous aggregates producing maximum polarisation degrees in a broad range from $\sim$10\% to $\sim$90\% at optical wavelengths. In this context, our measured maximum DoLP values for AR\,Pup fall within this theoretical range. Our measurements are also consistent with expectations for aggregates composed of amorphous carbon with moderate porosity ($\sim$0.5–0.7), which typically exhibit a decrease in maximum DoLP with increasing wavelength across the optical regime \citep[$\sim 5-20\%$ from the $V$- to $I'$-band,][]{Tazaki2022A&A...663A..57T}. Radiative transfer (RT) models by \citet{Min2012A&A...537A..75M} illustrate that polarimetric intensity distributions in inclined discs with porous dust grains align well with the polarimetric morphology observed in AR\,Pup's disc. While these polarimetric observations are consistent with porous dust aggregates, we acknowledge that they do not uniquely indicate this grain structure. Similar observables could also arise from a size distribution of compact grains or monomers with sizes comparable to the observed wavelengths. To place our results in a broader context, we refer to mid- and far-infrared studies of post-AGB discs. High-resolution Spitzer spectra of 21 Galactic post-AGB systems analysed by \citet{Gielen2008A&A...490..725G} reveal strong dust processing and grain growth, with irregularly shaped grains and typical sizes exceeding 2$\mu$m. These findings were reinforced by \citet{Gielen2011A&A...533A..99G}, who analysed a larger sample of 57 systems in both the Galaxy and the Large Magellanic Cloud, detecting high crystallinity fractions (20–60\%) and large grain sizes through spectral fitting of Spitzer infrared spectra (5–37$\mu$m). Additionally, \citet{Scicluna2020MNRAS.494.2925S} analysed far-infrared and sub-mm fluxes of 46 post-AGB discs, finding shallow spectral indices ($-2.5 \lesssim \alpha \lesssim -2.0$) indicative of grain size distributions extending to several hundred microns. 
Radiative transfer modelling of near- and mid-IR VLTI interferometric data combined with SED fitting points to grains up to millimetre sizes \citep{Hillen2015A&A...578A..40H} and a higher fraction of large grains (>1$\mu$m size) in the inner disc, consistent with ongoing grain growth \citep{Corporaal2023IRAS08}. These independent studies provide direct evidence for the presence of large (micron- to millimetre-sized), irregular grains and extensive grain processing in post-AGB discs. Taken together with our polarimetric data, this multi-wavelength evidence makes the presence of porous dust aggregates in the disc surface layers more likely than a population dominated solely by monomers. Nevertheless, our conclusions from polarimetric data are based on qualitative comparisons with theoretical models rather than direct constraints on dust grain properties. A definitive determination of grain sizes and structures would require detailed radiative transfer modelling across a broad parameter space, which is beyond the scope of this study. 

Polarimetric imaging of all observed post-AGB systems reveals a bright central disc region and more extended emission with resolved substructures in both $V-$ and $I'-$bands. We note that substructures appear brighter in the $I'-$band. A comparison with SPHERE/IRDIS near-IR imaging results shows that U\,Mon shows a larger disc size in near-IR compared to optical, similar to IRAS\,08544-4431 \citep{Andrych2024IRAS08}. In contrast, HR\,4049's disc size remains consistent across studied wavelengths. These variations and a range of resolved substructures in the post-AGB discs indicate a diversity in disc geometries, orientation and potentially dust composition within our sample. Additionally, we note that all targets studied to date with SPHERE in multi-wavelength polarimetric imaging belong to the `full disc candidate' category (with dust inner rims likely located at the theoretical dust sublimation radius based on the IR excess of SED, see Section~\ref{sec:paper3_intro}). To thoroughly investigate the dust properties and geometry of post-AGB circumbinary discs, future multi-wavelength polarimetric imaging observations should also include `transition' discs, which contain large inner cavities \citep{Corporaal2023A&A...674A.151C}.

\subsection{Post-AGB discs in the broader context of circumstellar evolution} 

Advances in high-contrast imaging over the last decade have enabled direct observations of dusty circumstellar environments across various evolutionary stages. These include eruptive stars \citep[e.g.,][]{Zurlo2024A&A...686A.309Z}, planet-forming discs around YSOs \citep[e.g.,][]{Garufi2024A&A...685A..53G}, debris discs \citep[e.g.,][]{Esposito2020AJ....160...24E, Crotts2024ApJ...961..245C}, as well as environments around AGB stars \citep[e.g.,][]{Khouri2020A&A...635A.200K}, and post-AGB binary systems \citep[][]{Ertel2019AJ....157..110E}. These studies provide critical insights into the complex processes of dust formation, evolution, and the dynamical interactions between the star and its circumstellar material throughout stellar evolution. While systematic observational studies have extensively explored the circumstellar environments of YSOs \citep[e.g.,][]{Avenhaus2018ApJ...863...44A, Garufi2020A&A...633A..82G, Ginski2021ApJ...908L..25G}, fewer high-resolution polarimetric studies have focused on the later evolutionary stages. To bridge this gap, we compare polarimetric observations of post-AGB circumbinary discs with those of AGB stars and YSOs, providing new insights into the evolution of dust and scattering properties in these environments.

\subsubsection{Parallels with circumstellar environments around YSOs}
\label{sec:disc_yso}

\begin{figure*}     
    \includegraphics[width=0.5\linewidth]{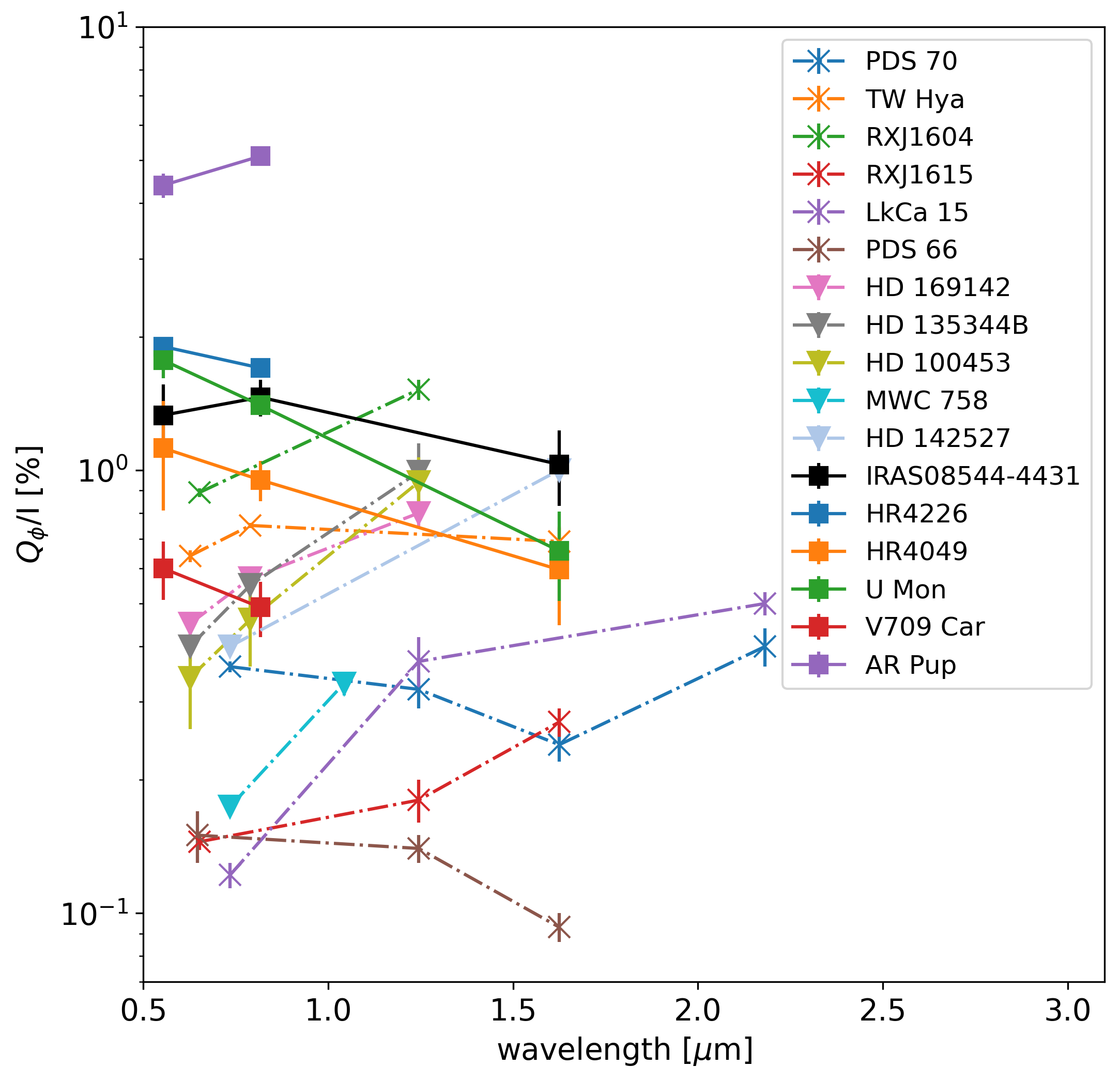}
    \includegraphics[width=0.43\linewidth]{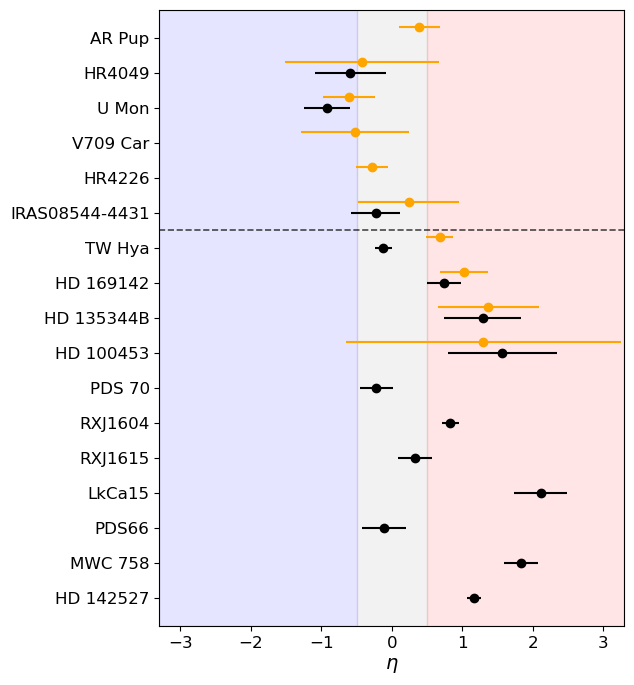}

    \caption{Comparison of polarised disc brightness (left panel) and colour ($\eta$, right panel) measurements for the post-AGB sample \citep[from this study and][]{Andrych2024IRAS08} and a sample of young stellar objects \citep[T Tauri and Herbig stars,][]{Ma2024A&A...683A..18M}. In the left panel, post-AGB targets are represented by squares with solid lines, while T Tauri star discs are marked with x and Herbig star discs with $\triangledown$. For a given disc, the results for different wavelength bands are connected with lines. In the right panel, post-AGB targets are positioned above the black horizontal dashed line, while young stellar objects are located below it. Orange points indicate colours in the visible wavelength, while black points represent colours between visible and near-IR bands. The shading represents the used definition of blue, grey, and red disc polarimetric colour. See Section~\ref{sec:disc_yso} for more details.}
    \label{fig:disc_comp}
\end{figure*}

To facilitate a direct comparison between post-AGB circumbinary discs and well-characterised PPDs, we refer to the results of \citet{Ma2024A&A...683A..18M}, who conducted a precise quantitative polarimetric analysis of T Tauri and Herbig stars. Their study accounted for the same instrumental effects as ours, ensuring consistency in the assessment of disc-polarized brightness (see Section~\ref{sec:paper3_data_reduction}). In Figure~\ref{fig:disc_comp}, we present the polarised brightness and polarimetric colour of post-AGB systems from this study, alongside post-AGB binary IRAS\,08544-4431 \citep{Andrych2024IRAS08} and a sample of T Tauri and Herbig stars. Despite their smaller angular sizes in scattered light ($\sim$0.2–0.6$''$ for post-AGB discs compared to $\sim$1–10$''$ for YSOs), post-AGB discs exhibit significant polarimetric brightness at optical and near-IR wavelengths, often exceeding 1\% of the system’s total intensity. This places post-AGB discs among the brightest PPDs in terms of relative polarised flux.

While PPDs around YSOs typically display grey to red polarimetric colours, post-AGB discs show slightly blue to grey. However, our observations reveal comparable wavelength-dependent polarimetric behaviour between post-AGB circumbinary discs and certain PPDs, such as those surrounding T Tauri stars like TW Hya, PDS 70, and PDS 66 \citep{Ma2024A&A...683A..18M}.  This resemblance suggests that similar dust-scattering processes play in both environments, even though the dust in post-AGB systems has never undergone the icy ($\sim$10 K) conditions of a molecular cloud,  potentially leading to fundamentally different dust properties. 
Red polarimetric colours in some PPDs may also result from additional reddening caused by dust absorption close to the central star — a scenario proposed for GG Tau based on HST observations \citep{krist2005AJ....130.2778K}. Such reddening may be a relatively common feature in PPDs. In contrast, this mechanism is likely suppressed in post-AGB systems due to disc clearing around the central binary and efficient dust-gas separation, as evidenced by the spectroscopic depletion of refractory elements in the primary star’s photosphere (see Section~\ref{sec:paper3_intro}). Overall, variations in polarimetric colour among PPDs are likely driven by a complex interplay of dust composition, grain structure, and local environmental conditions. However, current studies have not yet identified specific system parameters that fully account for these differences \citep{Ma2024A&A...683A..18M}. 

The complex morphologies observed in some post-AGB circumbinary discs, including arcs, cavities, and flared disc surfaces \citep[this study; see also][]{Ertel2019AJ....157..110E, Andrych2023MNRAS.524.4168A, Andrych2024IRAS08}, resemble those commonly found in protoplanetary discs \citep[][]{Benisty2022arXiv220309991B}. These structural similarities, along with their shared polarimetric properties, suggest post-AGB circumbinary discs as valuable analogues for studying disc evolution across stellar evolutionary stages. However, the greater distances towards post-AGB binaries limit the achievable spatial resolution.

\subsubsection{Parallels with circumstellar environments around AGBs}

Polarimetric studies of AGB and post-AGB systems provide crucial insights into the evolution of circumstellar dust and the role of binarity in shaping these environments. A recent SPHERE/ZIMPOL study by \citet{Montarges2023A&A...671A..96M}, which examined 14 single AGB stars, revealed that their dusty circumstellar environments are clumpy, with isolated dust features. The observed maximum degree of polarisation does not exceed $\sim$0.4 and decreases with wavelength, suggesting the presence of small (0.01–0.1 $\mu$m) or possibly intermediate-sized (0.01–1 $\mu$m) grains.

Our study finds that post-AGB circumbinary discs exhibit higher maximum degree of resolved polarisation ($\sim$0.7 in the $V$-band and $\sim$0.55 in the $I'$-band) However, we note that our values are based on PSF-deconvolved images, whereas \citet{Montarges2023A&A...671A..96M} did not apply PSF correction in their analysis. The maximum DoLP for AR\,Pup post-AGB disc in non-deconvolved images reaches $\sim$0.3 in the $V$-band, placing it among the highest in the \citet{Montarges2023A&A...671A..96M} sample. Only one AGB star in their study (W\,Aql) shows a higher DoLP (0.39 in the $V$-band). Authors note that W\,Aql also exhibits the strongest mid-infrared excess and most spatially extended polarised emission, suggesting especially efficient dust production due to grain size, composition, or distribution. It is also important to note that AR\,Pup is viewed at a high inclination, and the full scattering surface of the disc is not resolved in scattered light. As a result, the measured DoLP likely represents a lower limit. 

The presence of a companion during the AGB phase can also lead to disc formation, as observed in the highly inclined AGB binary system L2 Pup \citep{Kervella2015EAS....71..211K}. L2 Pup hosts a dusty disc with a maximum degree of polarisation of 0.46 in the $V$-band and 0.61 in the $R$-band.  We note that the disc configuration of L2 Pup is remarkably similar to that of post-AGB binary AR\,Pup in this study, with both targets displaying extremely flared disc surfaces. However, AR\,Pup shows a maximum degree of polarisation that decreases with wavelength, whereas L2 Pup exhibits the opposite trend, increasing with wavelength. This contrast may reflect differences between AGB and post-AGB discs in dust properties (such as size and porosity of dust grains) or disc characteristics (such as optical depth).

Similarities between binary AGB and post-AGB systems emphasise a key open question: how quickly can a stable circumbinary disc be formed in evolved binary systems? Understanding the transition from an AGB wind-driven environment to a structured post-AGB disc is critical for constraining disc formation timescales and the role of binary interactions. A broader, systematic study of circumstellar environments across AGB, post-AGB, and planetary nebulae evolutionary stages is essential to address these questions. By probing dust formation, processing, and the influence of characteristics of the host star (such as stellar parameters and binarity), such research will provide a more complete picture of the life cycle of circumstellar discs and their role in late stellar evolution.

\subsubsection{Investigating targets whose post-AGB nature is unclear}

As noted in Sections~\ref{sec:paper3_hr4226} and \ref{sec:paper3_v709_car}, the evolutionary status of HR\,4226 and V709\,Car remains ambiguous. While both systems show evidence of circumbinary discs and were initially classified as post-AGB stars based on their near-IR excesses, several of their properties deviate from those typically seen in confirmed post-AGB binaries. In this section, we examine their characteristics in more detail and explore the alternative possibility that these stars could be in transition between AGB and post-AGB stages.

HR\,4226 was initially classified as a post-AGB star based on its SED and a fitted effective temperature of 4275$^{+600}{-550}$ K \citep{Hillen2017}, which is at the lower end of the post-AGB range \citep[typically $\sim$4000–9000 K;][]{Kamath2014MNRAS.439.2211K, Kamath2015MNRAS.454.1468K}. Although near-IR interferometric observations showed only over-resolved circumbinary material \citep{Kluska2019A&A...631A.108K}, geometric modelling of mid-IR interferometric data revealed a resolved ring-like structure \citep{Hillen2017}. Our polarimetric imaging reveals a well-defined disc structure with a high degree of azimuthally polarised brightness (>1.5\%; see Table~\ref{tab:paper3_disc_orient} and Section~\ref{sec:paper3_hr4226}). This level of azimuthally polarised signal is consistent with single scattering off circumstellar dust and suggests an optically thick disc, similar to those found around confirmed post-AGB binaries. Despite these indicators, several aspects of HR\,4226 are atypical for a post-AGB binary. The radial velocity amplitude is exceptionally low ($\sim$0.1 km/s), leading to a minimum mass function consistent with a substellar companion \citep[$\sim 8$ Jupiter masses, assuming a 0.6 M$\odot$ primary;][]{Maas2003PhDT.......265M}. This is significantly lower than the typical companion masses observed in post-AGB binary systems, which generally peak around 1 M$\odot$ with a standard deviation of 0.62 M$\odot$ \citep{Oomen2018}. While the 572-day period reported for HR\,4226 \citep{Maas2003PhDT.......265M} could, in theory, be linked to Mira-like pulsations, the observed radial velocity amplitude is much smaller than that typically seen in classical Mira variables \citep[$\geq$10 km/s;][]{Nowotny2010A&A...514A..35N}. This strongly suggests that the variability is more likely caused by orbital motion than pulsations. Additionally, \citep{Kiss2007MNRAS.375.1338K} conducted a photometric analysis, revealing that the $V$-band light curves exhibit considerable scatter, with two distinct minima occurring approximately 260 days apart. HR\,4226 characteristics straddle both post-AGB binary and AGB binary scenarios, possibly hinting at a system in transition between the AGB and post-AGB phases. However, we need a comprehensive spectroscopic analysis to accurately determine the stellar atmospheric parameters and elemental abundances, providing a definitive conclusion on the target's evolutionary nature.

Similar to HR\,4226, V709\,Car was initially classified as a post-AGB star based on the near-IR excess in its SED. However, its effective temperature \citep[$3500\pm175$ K,][]{Hillen2017, Kluska2022}, derived from SED fitting, is lower than the typical value for post-AGB systems and more consistent with AGB stars. Moreover, spectroscopic analysis using Spitzer survey indicates a lack of crystalline dust features in the circumstellar environment of V709\,Car \citep{Gielen2011A&A...533A..99G}, which is an unusual composition for post-AGB discs. Although a $\sim$320-day orbital period has been proposed for V709,Car based on radial velocity monitoring \citep{Maas2003PhDT.......265M}, the data exhibit substantial scatter, preventing a reliable orbital solution and making it difficult to confidently confirm the system’s binary nature. Near-IR interferometric study of the target reveals a high circumstellar temperature ($>7000$ K), indicating that the observed emission is not thermal radiation from dust \citep{Kluska2019A&A...631A.108K}. Geometrical modelling of mid-IR interferometric observations results in a ring with a half-light radius of $150\pm8$ mas \citep{Hillen2017}, $\sim 115$ times larger than the near-IR size, revealing a striking discrepancy in spatial scales. Our polarimetric imaging shows that V709\,Car exhibits notably lower resolved polarised brightness and more diffuse polarised emission compared to other targets in our sample (see Section~\ref{sec:paper3_indiv_cases}). However, polarimetric observations alone do not definitively distinguish this target from the rest of the sample. Results of IR interferometric studies, combined with the relatively low effective temperature of the primary star and a lack of crystalline dust features in the circumstellar environment, suggest that V709\,Car may still be in the AGB phase. If this is the case, the observed polarimetric brightness and disc morphology of V709\,Car may be shaped by ongoing stellar wind-driven mass loss rather than a well-defined, dense disc. However, a definitive conclusion on the evolutionary stage of V709\,Car requires a comprehensive spectroscopic analysis to accurately determine stellar atmospheric parameters and elemental abundances, as well as optical photometry and radial velocity measurements to better constrain its orbital properties.

\section{Conclusions}
\label{sec:paper3_conclusion}

We present a multi-wavelength polarimetric imaging study of five evolved binary systems (AR\,Pup, HR\,4049, HR\,4226, U\,Mon, and V709\,Car) using the VLT/SPHERE instrument. Our study aimed to i) assess whether the polarimetric properties and surface morphologies of circumbinary discs are consistent across the post-AGB sample, ii) explore the dust properties inferred from both the intensity and degree of polarisation of the scattered light and iii) compare our findings with known observations for AGB and YSO systems. This multiwavelength study of post-AGB binary systems leads to several key conclusions:
\begin{itemize}
    \item All studied post-AGB systems exhibit high polarimetric disc brightness of more than 1\% of the total intensity in optical $V-$ and $I'-$bands. For the post-AGB binary AR\,Pup, we also resolved the circumbinary disc in non-polarised scattered light and directly measured the maximum fractional polarisation of the scattered light of $\sim0.7$ in the $V-$band and $\sim0.55$ in the $I'-$band.
    
    \item The polarimetric colour across the sample tends toward grey or slightly blue.
    
    \item  The observed high polarimetric brightness, colour, and intensity distribution in post-AGB circumbinary discs are consistent with theoretical models of porous aggregates composed of small monomers as the dominant surface dust composition. While similar polarimetric signatures could also be produced by a distribution of compact grains with sizes comparable to the observed wavelengths (submicron size), the porous aggregate scenario is further supported by multi-wavelength evidence of irregular, micron- to millimetre-sized dust grains and extensive dust processing in post-AGB discs, as revealed by infrared and sub-mm spectroscopic studies \citep{Gielen2008A&A...490..725G, Gielen2011A&A...533A..99G, Scicluna2020MNRAS.494.2925S}. However, a definitive determination of dust grain sizes requires detailed radiative transfer modelling, which is beyond the scope of this study.
    
    \item Comparing our results with SPHERE/IRDIS near-IR imaging, we find that U\,Mon exhibits a larger disc size in the near-IR compared to the optical, similar to IRAS,08544-4431 \citep{Andrych2024IRAS08}. In contrast, HR\,4049 maintains a consistent disc size across wavelengths. These variations, along with the diverse resolved substructures in post-AGB discs, indicate a range of disc geometries, orientations, and potentially different dust compositions within our sample.
    
    \item The resolved size of the polarimetric emission for one system in our sample (V709\,Car) appears significantly larger in the $I'$-band compared to the $V$-band, potentially resulting from self-shadowing effects due to higher dust absorption at shorter wavelengths.

    \item Post-AGB discs show a higher degree of polarisation compared to typical values for single AGB stars. However, binary AGB stars, such as L2 Pup, also show dusty discs remarkably similar to those observed around post-AGB binaries.
    
    \item Our findings further support the similarities between post-AGB circumbinary discs and PPDs around YSOs, including complex disc morphologies, high polarimetric brightness, comparable wavelength dependence of polarised emission, and similar dust sizes and compositions.
    
    \item Combining our findings with existing literature studies, we question the classification of two systems in our sample, HR 4226 and V709 Car, as post-AGB binaries. Our polarimetric observations alone are insufficient to confirm their evolutionary stage, highlighting the need for detailed spectroscopic analysis to accurately determine their stellar atmospheric parameters and elemental abundances.
    
\end{itemize}

A comprehensive approach that integrates multi-wavelength and multi-technique observations with advanced radiative-transfer modelling will provide deeper insights into disc dynamics and evolution. This will further clarify the influence of stellar parameters and binarity in shaping the dusty circumstellar environment, potentially facilitating planet formation in evolved systems. Ultimately, this approach will enhance our understanding of disc evolution across the Hertzsprung-Russell diagram.

\begin{acknowledgement}
DK and KA acknowledge the support of the Australian Research Council (ARC) Discovery Early Career Research Award (DECRA) grant (DE190100813). KA and DK, HVW acknowledge the support from the Australian Research Council Discovery Project DP240101150. TDP acknowledges support of the Research Foundation - Flanders (FWO) under grant 11P6I24N. This study is based on observations collected at the European Southern Observatory under ESO programmes 101.D-0752(A), 0102.D-0696(A), and 0101.D-0807(B). This work has made use of the High Contrast Data Centre, jointly operated by OSUG/IPAG (Grenoble), PYTHEAS/LAM/CeSAM (Marseille), OCA/Lagrange (Nice), Observatoire de Paris/LESIA (Paris), and Observatoire de Lyon/CRAL, and supported by a grant from Labex OSUG@2020 (Investissements d’avenir – ANR10 LABX56).
\end{acknowledgement}

\paragraph{Data Availability Statement}
The data underlying this article are stored online in the ESO Science Archive Facility at http://archive.eso.org, and can be accessed by program IDs.
%\endnote in some journals will behave like \footnote; \printendnotes will not output anything. 
\printendnotes
\bibliography{mnemonic,references}
%\printbibliography

\appendix

\section{Data reduction results for all targets}
\label{sec:paper3_ap_reduct}

\begin{figure*}
     \includegraphics[width=1\columnwidth]{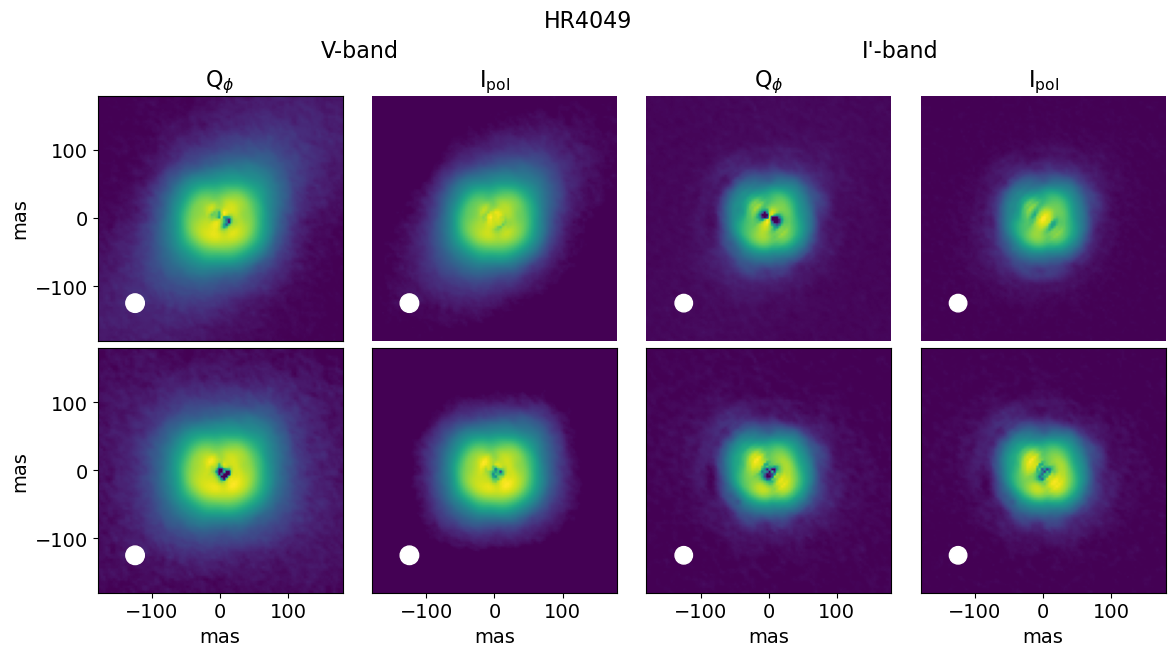}
    \caption[The polarised signal of HR\,4049 before and after subtraction of the unresolved central polarisation.]{The polarised signal of HR\,4049 before (top row) and after (bottom row) subtraction of the unresolved central polarisation in $V$ (first and second column) and $I'$ (third and fourth column) bands. $Q_\phi$ represents azimuthal polarised intensity, while $I_{\rm pol}$ represents the total polarised signal. White circles in the left corner of each image represent the size of the resolution element. All images are presented on an inverse hyperbolic scale and oriented North up and East to the left. See \ref{sec:paper3_ap_reduct} for more details.}
    \label{fig:paper3_ap_reduct_HR4049}
\end{figure*}

\begin{figure*} 
    \includegraphics[width=1\columnwidth]{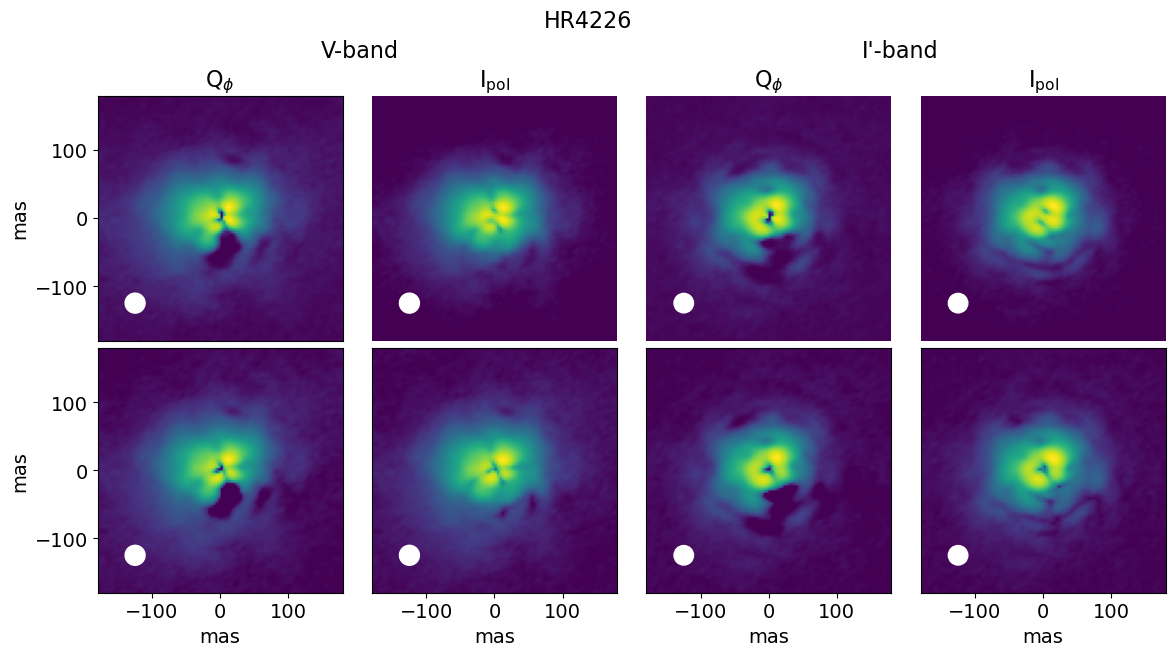}
    \caption{Same as Figure~\ref{fig:paper3_ap_reduct_HR4049} but for HR\,4226.}
    \label{fig:paper3_ap_reduct_HR4226}
\end{figure*}

\begin{figure*} 
     \includegraphics[width=1\columnwidth]{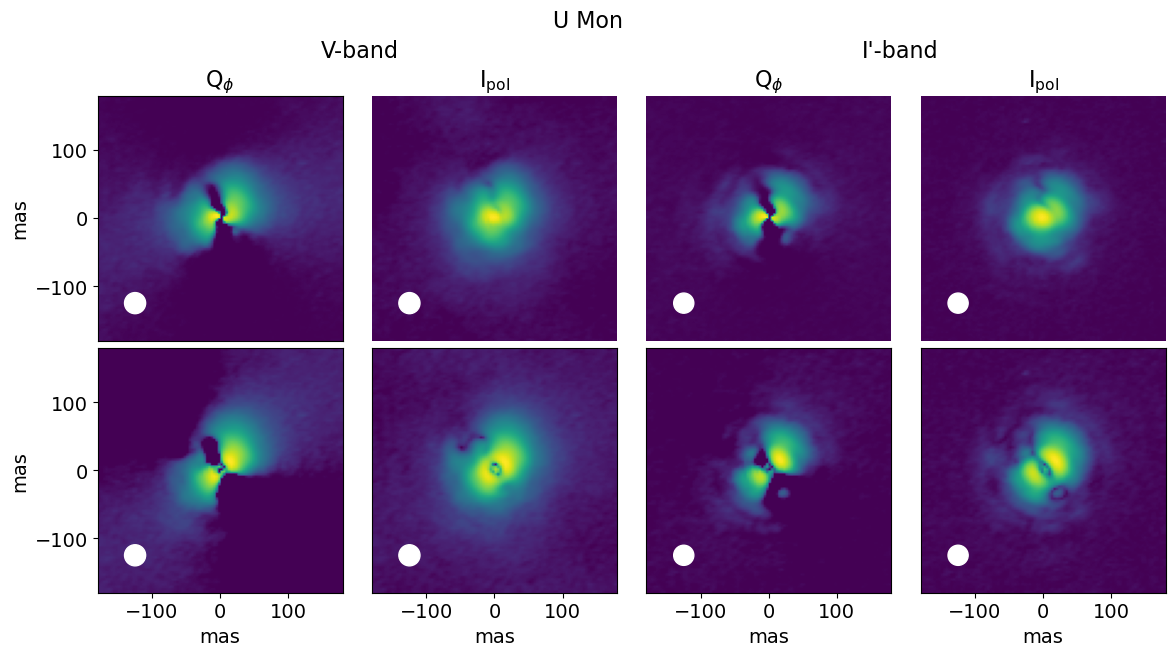}
    \caption{Same as Figure~\ref{fig:paper3_ap_reduct_HR4049} but for U\,Mon.}
    \label{fig:paper3_ap_reduct_umon}
\end{figure*}

\begin{figure*} 
    \includegraphics[width=1\columnwidth]{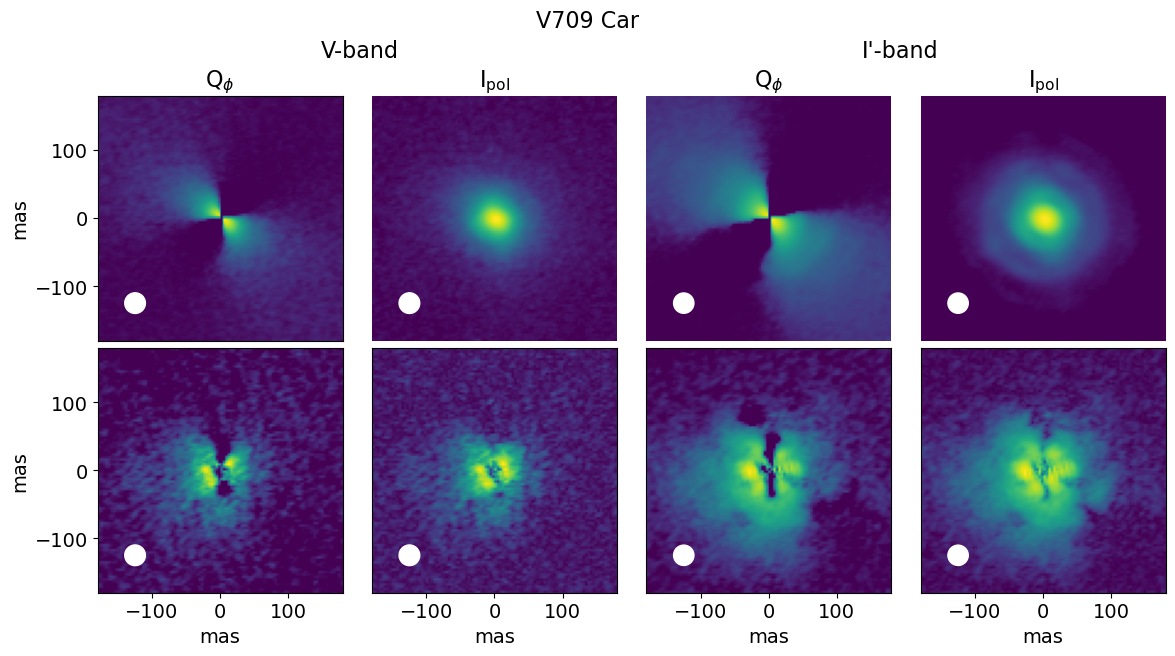}     
    \caption{Same as Figure~\ref{fig:paper3_ap_reduct_HR4049} but for V709\,Car.}
    \label{fig:paper3_ap_reduct_v709}
\end{figure*}

\begin{figure*} 
     \includegraphics[width=1\columnwidth]{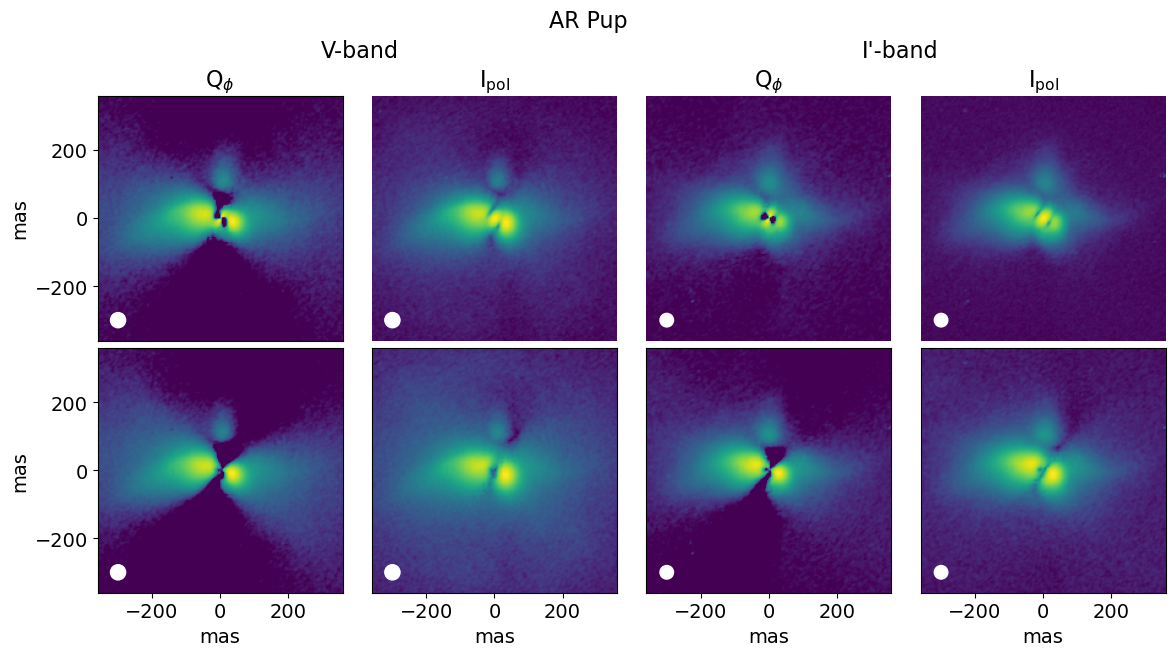}
    \caption{Same as Figure~\ref{fig:paper3_ap_reduct_HR4049} but for AR\,Pup.}
    \label{fig:paper3_ap_reduct_AR_pup}
\end{figure*}

In Figure~\ref{fig:paper3_ap_reduct_HR4049}, \ref{fig:paper3_ap_reduct_HR4226}, \ref{fig:paper3_ap_reduct_umon}, \ref{fig:paper3_ap_reduct_v709}, \ref{fig:paper3_ap_reduct_AR_pup}, we present azimuthal ($Q_\phi$) and total($I_{\rm pol}$) polarised intensities for each target before (top row) and after (bottom row) subtraction of the unresolved central polarisation in $V$ and $I'$  bands. More details on the methodology are presented in Section~\ref{sec:paper3_data_reduction}.

\section{Telescope correction}
\label{sec:paper3_ap_telesc_corr}

\begin{figure*}
     \includegraphics[width=0.33\textwidth]{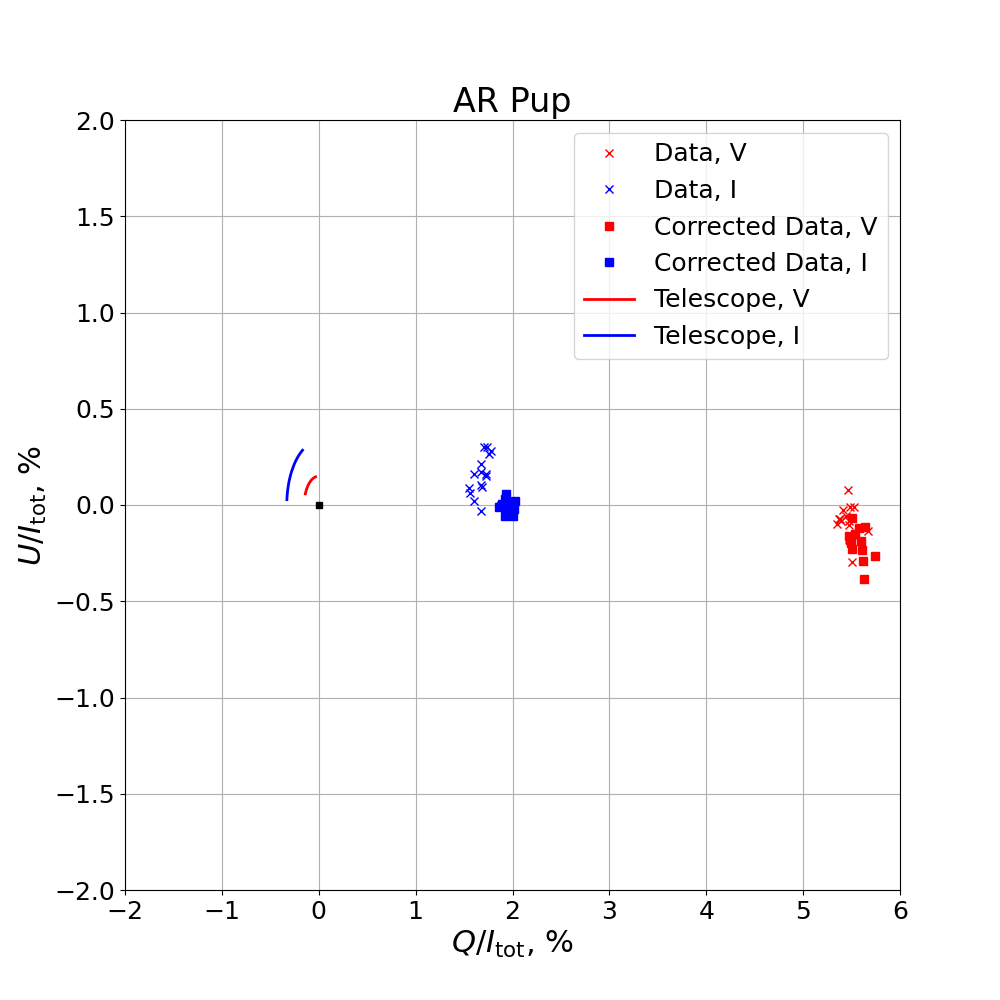}
     \includegraphics[width=0.33\textwidth]{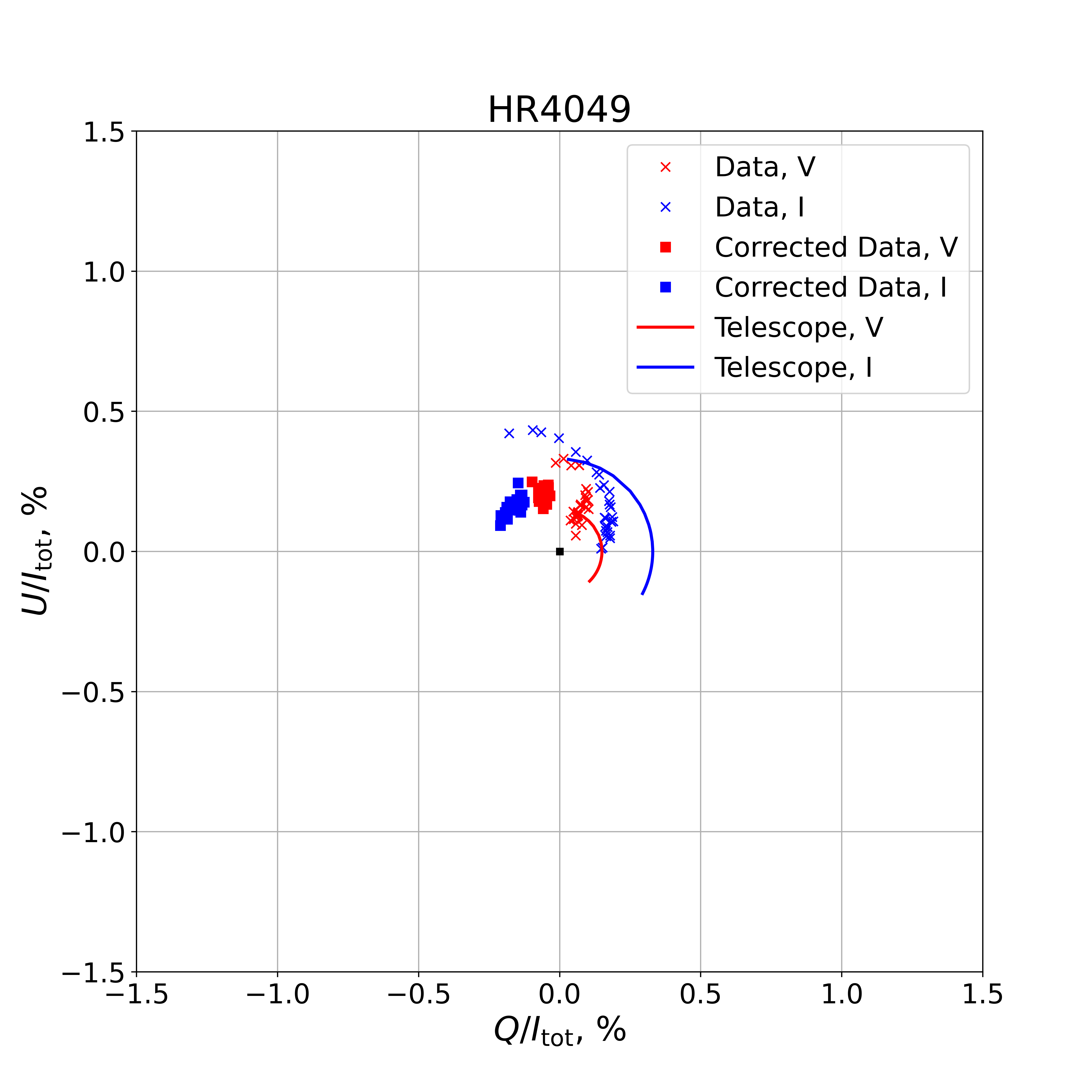}
     \includegraphics[width=0.33\textwidth]{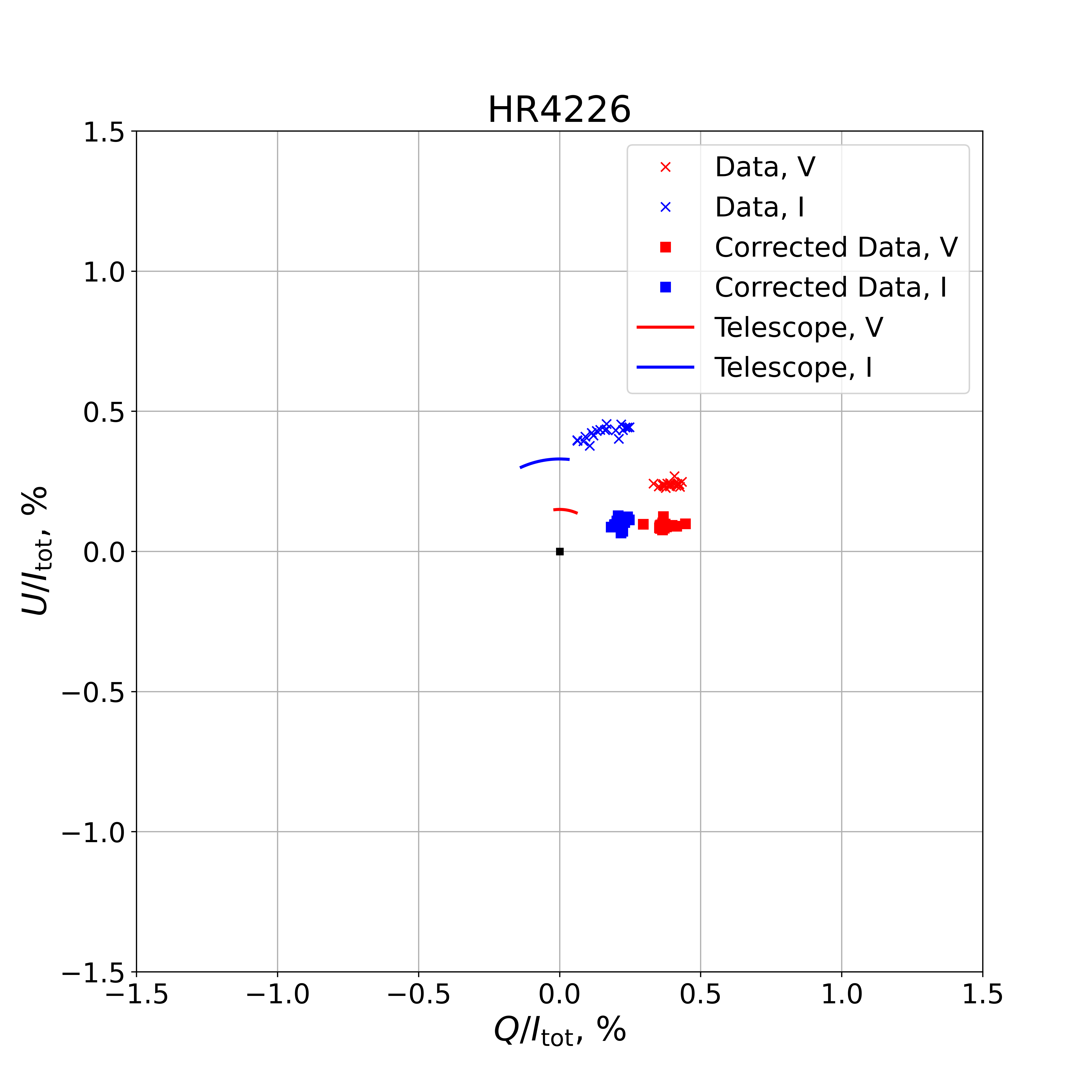}
     \includegraphics[width=0.33\linewidth]{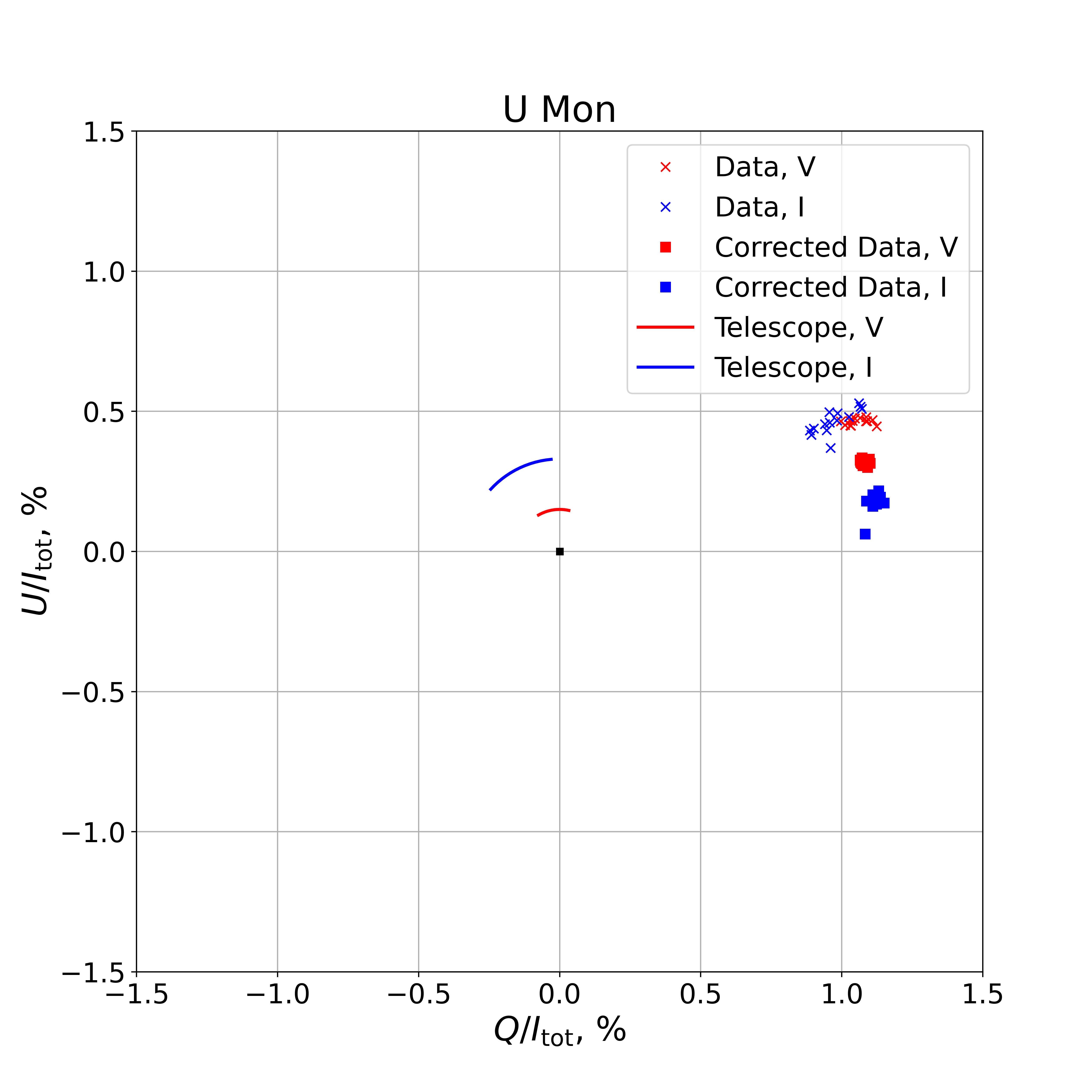} 
     \includegraphics[width=0.33\linewidth]{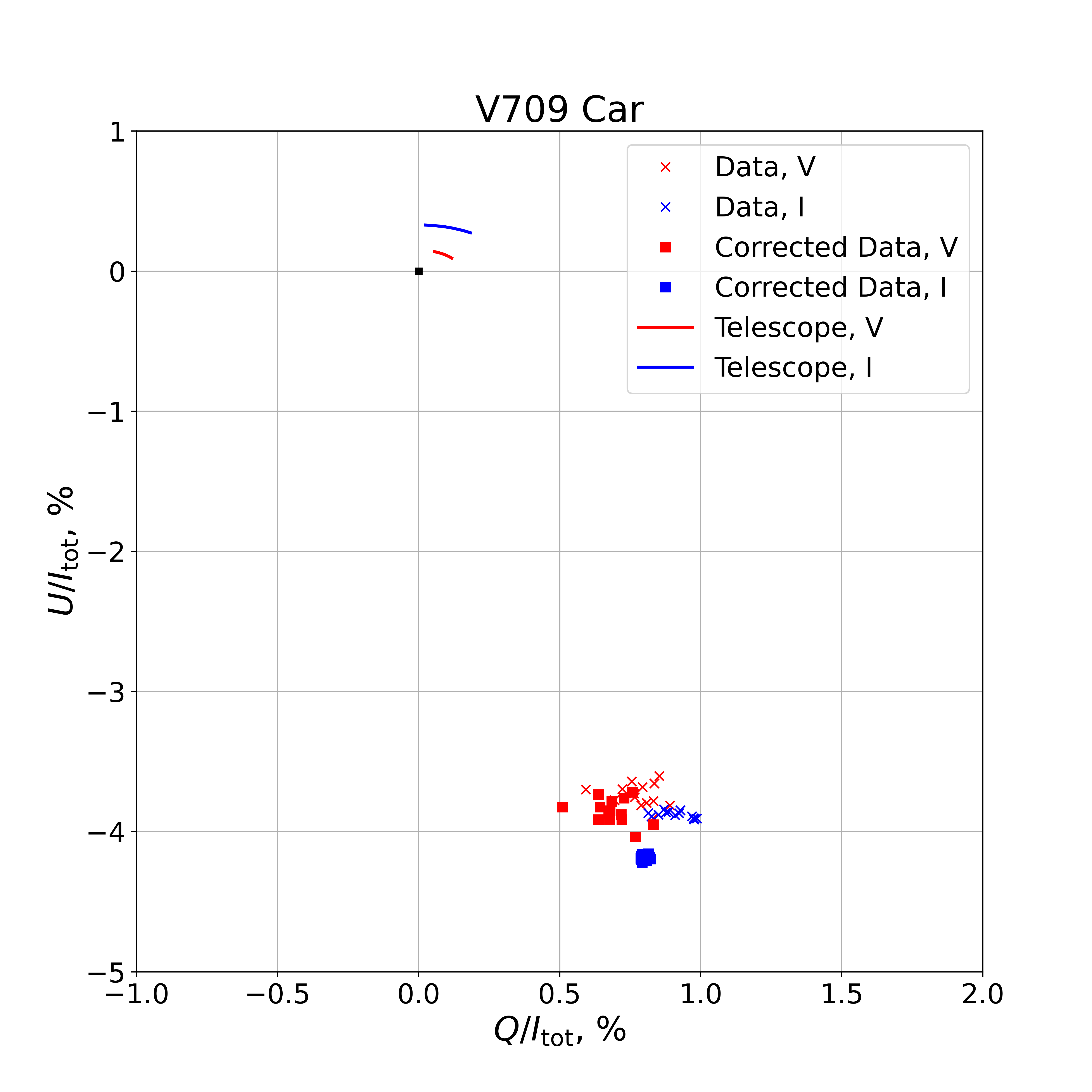}
    \caption[Measured fractional polarisation ($Q/I_{\rm tot}$ and $U/I_{\rm tot}$) for target post-AGB systems before and after the correction of the telescope polarisation.]{Measured fractional polarisation ($Q/I_{\rm tot}$ and $U/I_{\rm tot}$) for target post-AGB systems before (crosses) and after (squares) the correction of the telescope polarisation (lines) in $V$ (red) and $I'$ (blue) bands. See Section~\ref{sec:paper3_data_reduction} for more details.}
    \label{fig:telescope_corr}
\end{figure*}

In Figure~\ref{fig:telescope_corr}, we present the fractional polarisation ($Q/I_{\rm tot}$ and $U/I_{\rm tot}$) for each polarimetric cycle, both before and after applying the telescope polarisation correction, in the $V$ and $I'-$bands for all targets in our sample. For further details, refer to Section~\ref{sec:paper3_data_reduction}.

\section{Aperture polarimetry of fractional polarisation}
\label{sec:ap_aper_pol}

\begin{figure*} 

    \includegraphics[width=0.49\linewidth]{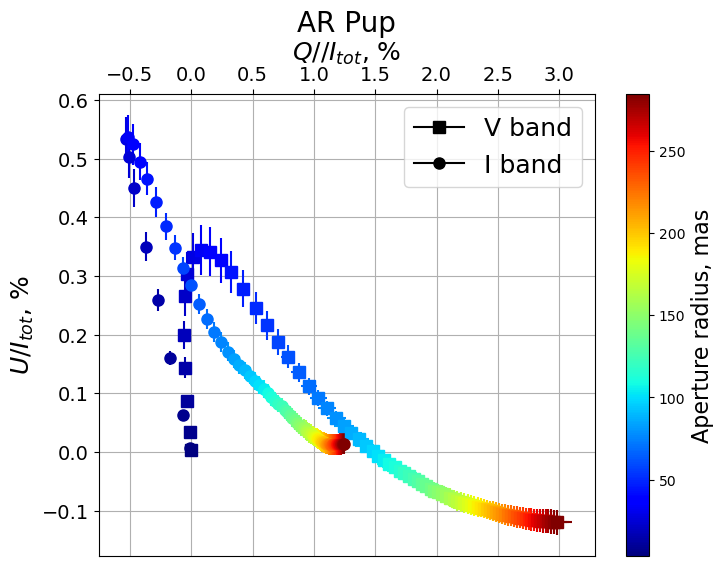}
    \includegraphics[width=0.49\linewidth]{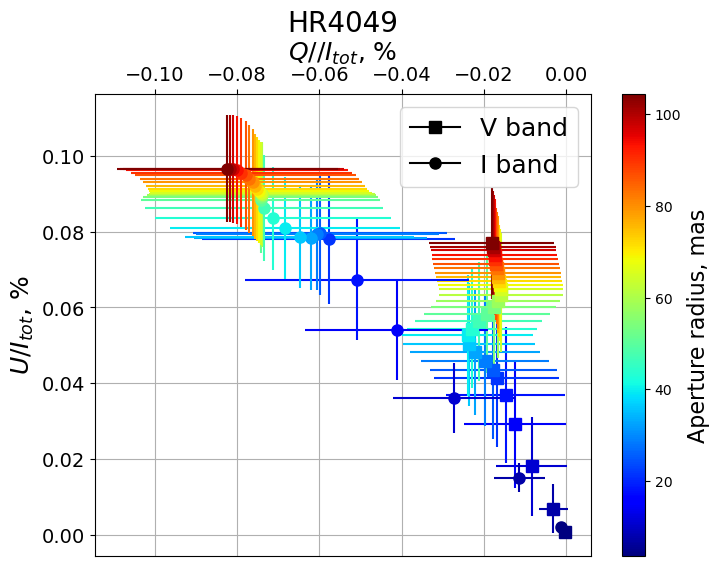}
    \includegraphics[width=0.49\linewidth]{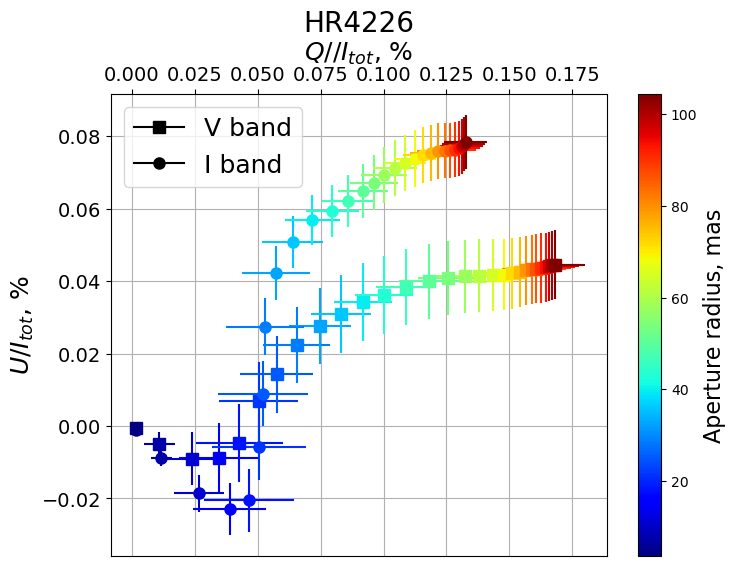}
    \includegraphics[width=0.49\linewidth]{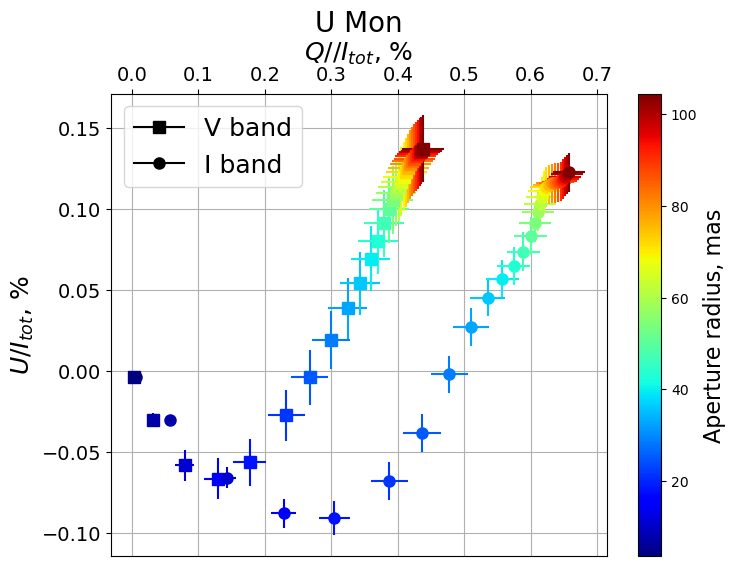}
    \includegraphics[width=0.49\linewidth]{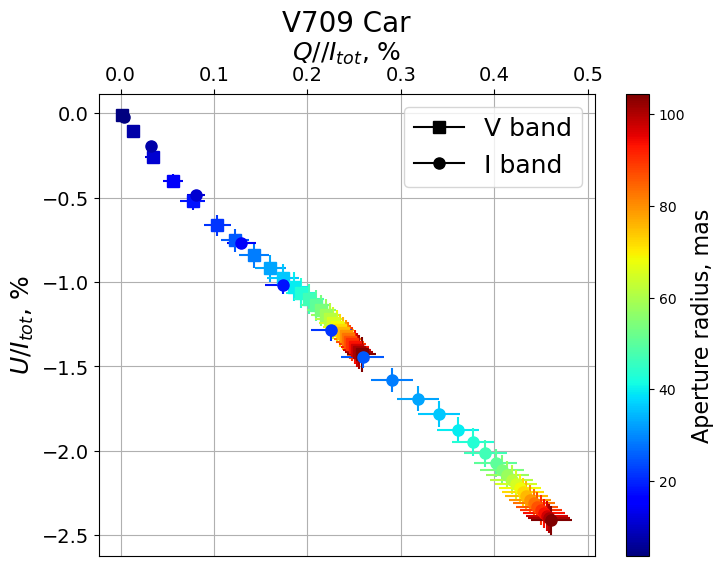}

    \caption[Fractional polarisation $Q/I_{\rm tot}$ and $U/I_{\rm tot}$ for a gradually enlarging aperture for both $V-$ and $I'-$bands for all targets.]{Fractional polarisation $Q/I_{\rm tot}$ and $U/I_{\rm tot}$ for a gradually enlarging aperture for both $V-$ and $I'-$bands for all targets. See Section~\ref{sec:paper3_aper_pol} for more details.}
    \label{fig:paper3_aper_pol}
\end{figure*}

In Figure~\ref{fig:paper3_aper_pol} we present total fractional polarisation ($Q/I_{\rm tot}, U/I_{\rm tot}$) as a function of distance from the central binary. The total fractional polarisation was measured using circular apertures with radii gradually increasing from 0.004'' to 0.11'' (0.25'' for AR\,Pup), with a step size of 0.0036'' (1 pixel). Further details on the methodology and results are provided in Section~\ref{sec:paper3_aper_pol}.

\section{Brightness profiles}
\label{sec:paper3_ap_profiles}

\begin{figure*} 
     \includegraphics[width=0.45\columnwidth]{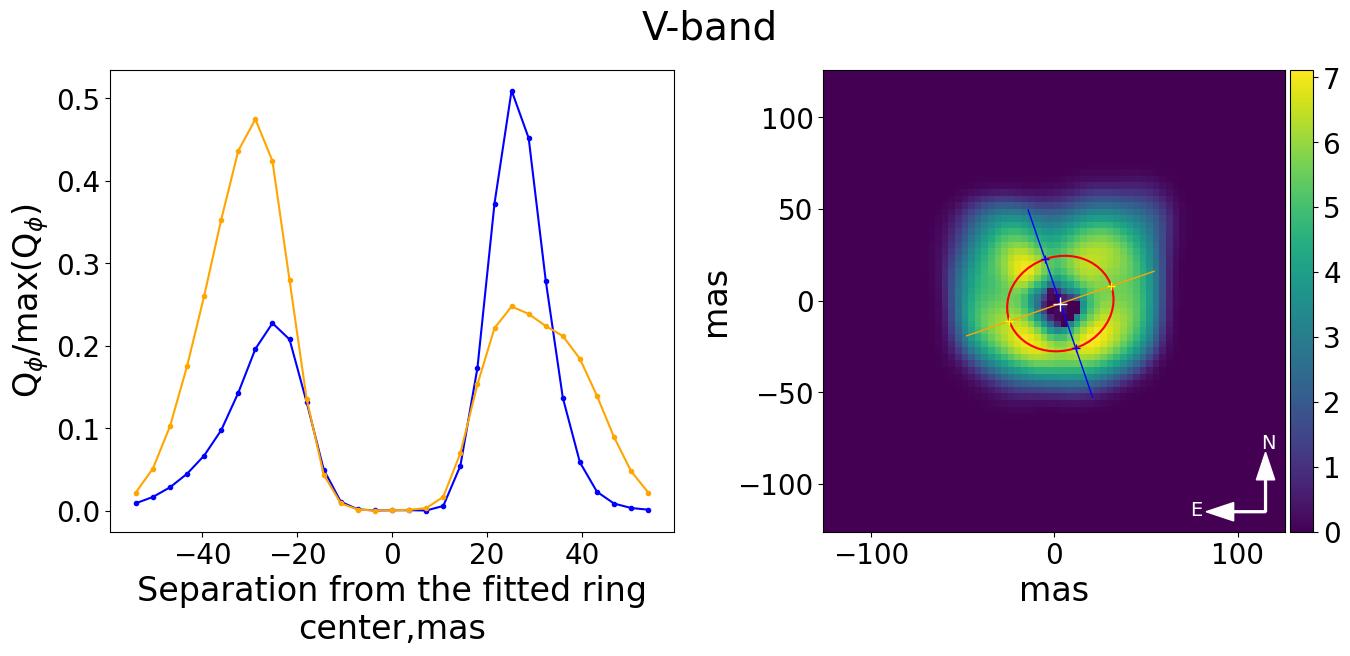}
     \includegraphics[width=0.45\columnwidth]{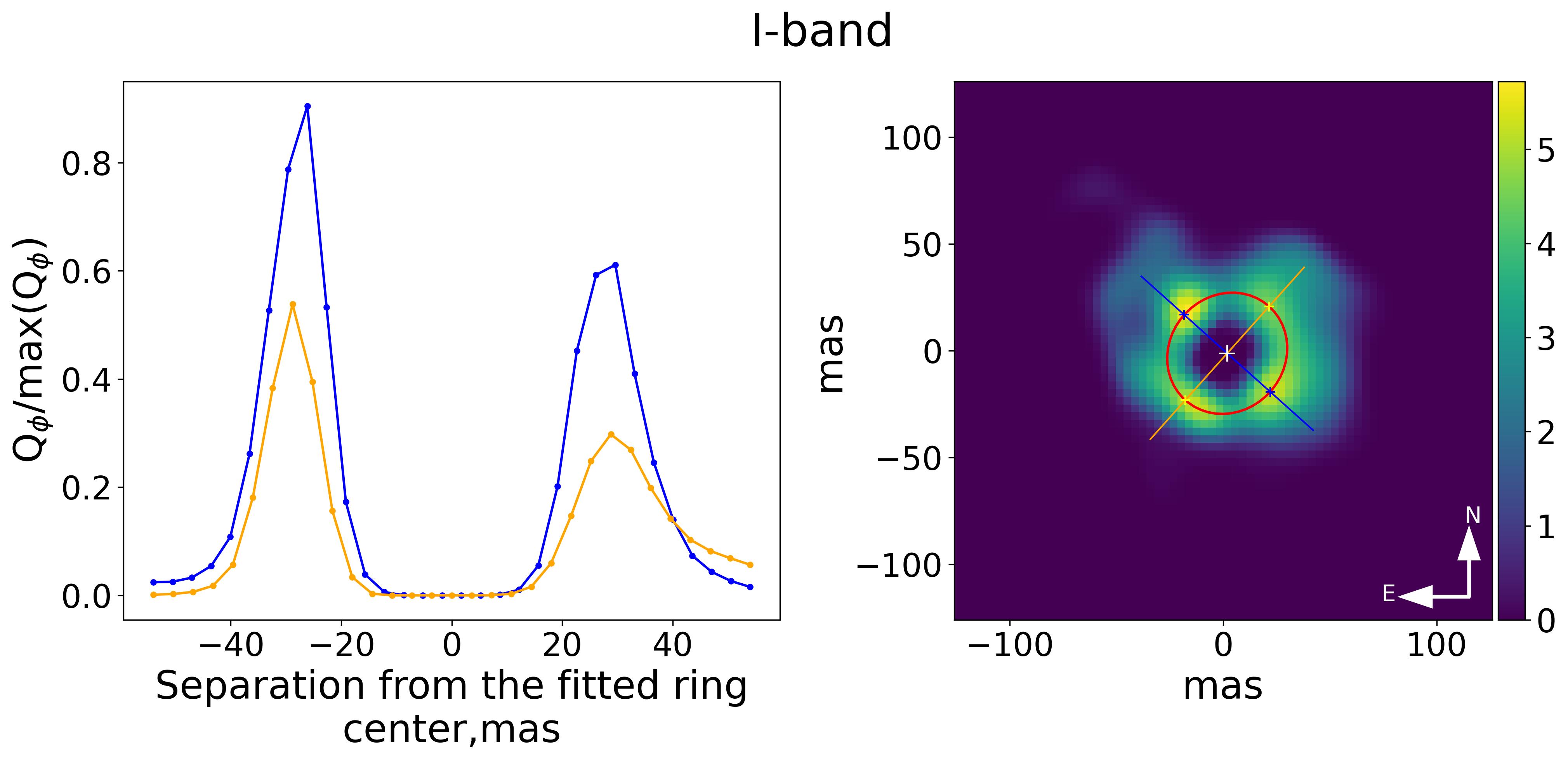}
    
    \includegraphics[width=0.455\columnwidth]{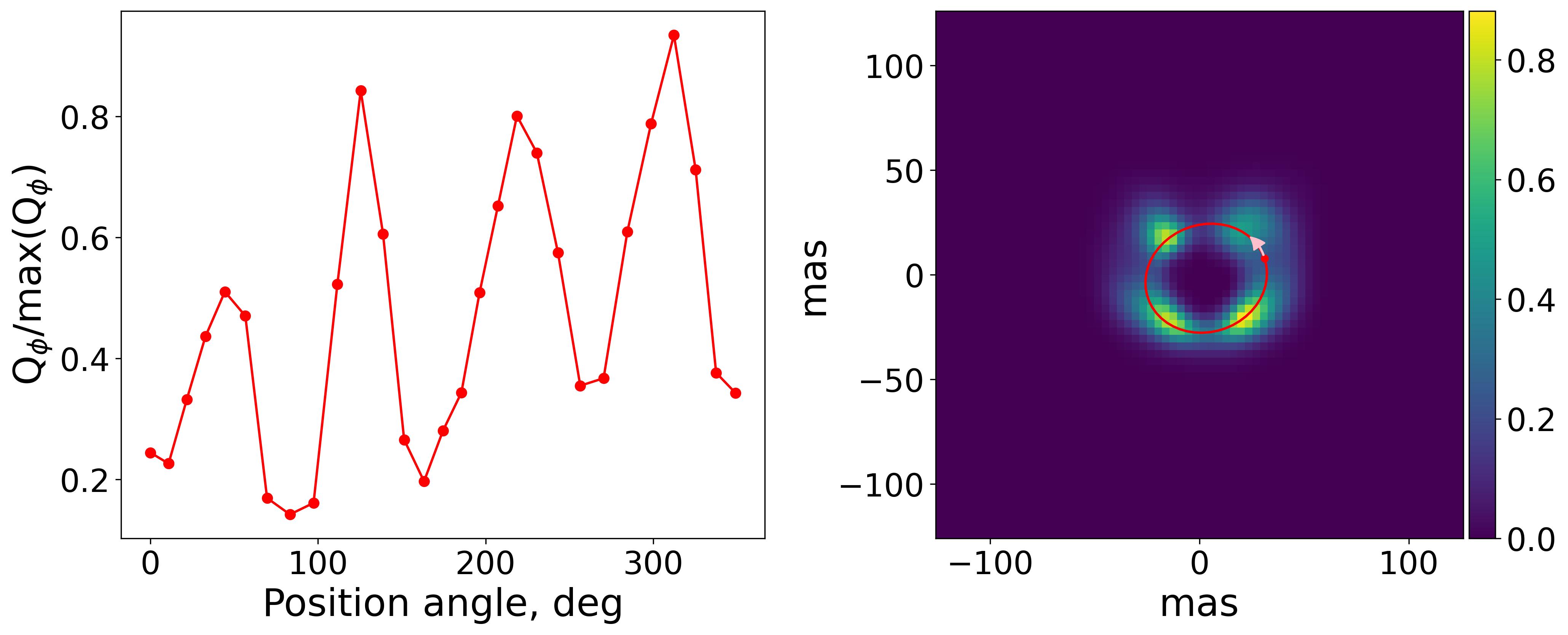}
     \includegraphics[width=0.455\columnwidth]{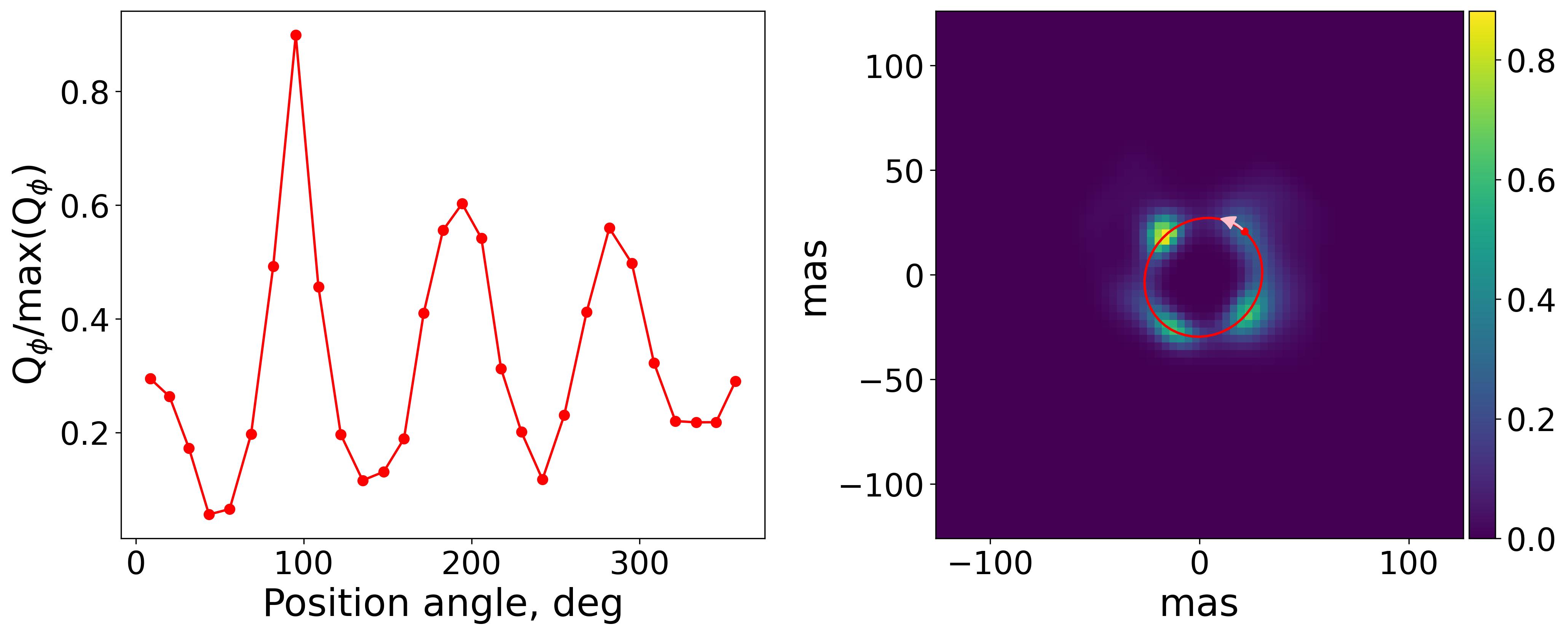}
     
     \includegraphics[width=0.45\columnwidth]{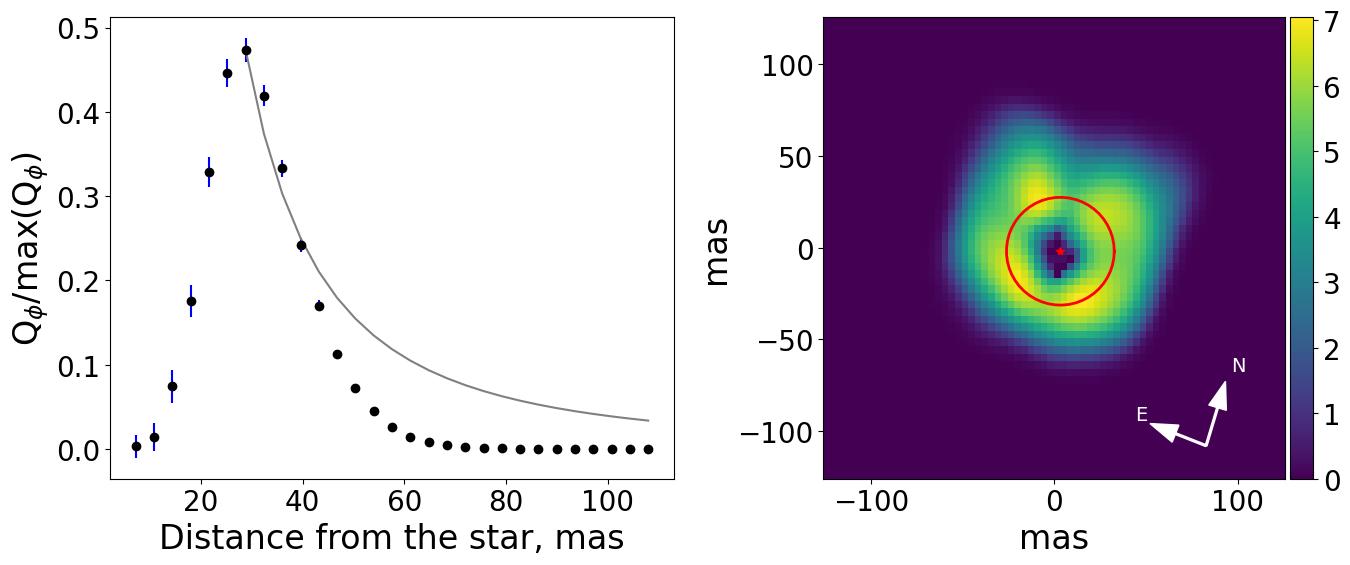} 
    \includegraphics[width=0.45\columnwidth]{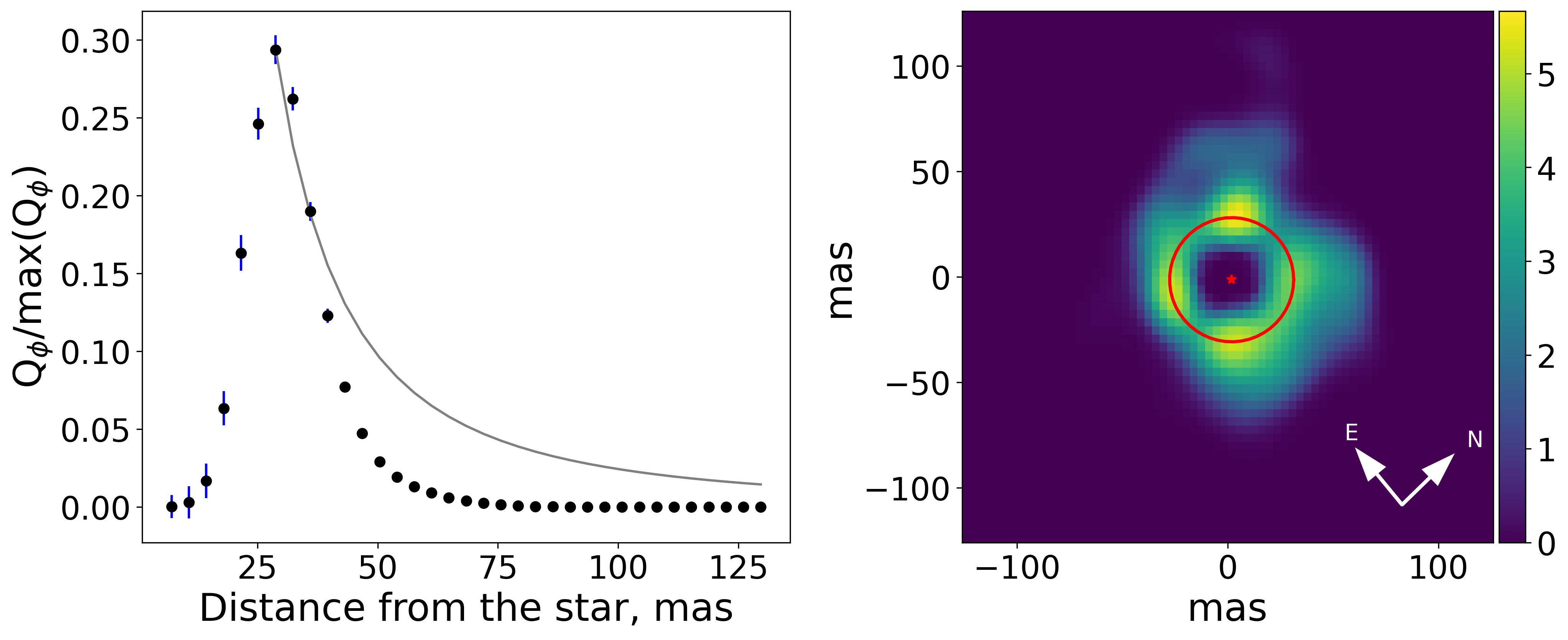}

    \caption[Brightness profiles for HR\,4049.]{Brightness profiles for HR\,4049 in $V$ (left panel) and $I'$ (right panel) bands (see \ref{sec:paper3_ap_profiles}) with corresponding  polarised images. In each panel, the left image displays the brightness profile, while the right image presents the corresponding polarised image. The top row shows linear brightness profiles of the polarised image along the major and minor axes of the fitted ellipse. The middle row represents the azimuthal brightness profiles of the polarised image. The red dot and arrow in the corresponding polarised image mark the starting point and direction of the azimuthal brightness profile calculation. The bottom row shows radial brightness profiles of the deprojected polarised image. In the radial brightness profile plots, grey solid lines are added to indicate a r$^{-2}$ drop-off, expected from a scattered light signal due to the dissipation of stellar illumination. Polarised images are presented on an inverse hyperbolic scale, with the middle-row images additionally normalised to the peak polarised intensity for clearer representation. The low intensity of the central 5x5 pixel region of each  polarised image is a reduction bias caused by correction of the unresolved central polarisation (see Section~\ref{sec:paper3_data_reduction}).}
    \label{fig:paper3_profiles_hr4049}
\end{figure*}

\begin{figure*} 
     \includegraphics[width=0.45\columnwidth]{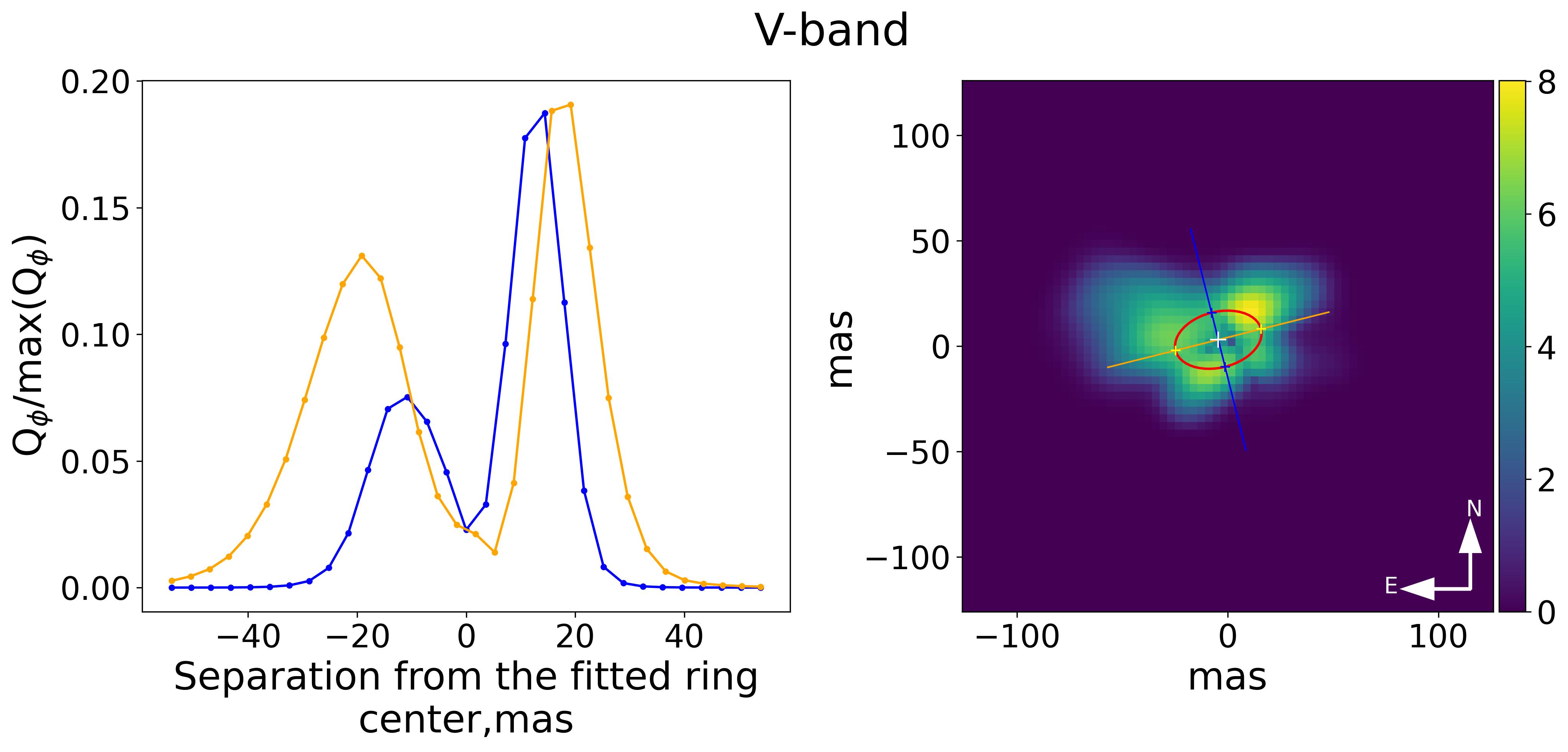}
     \includegraphics[width=0.455\columnwidth]{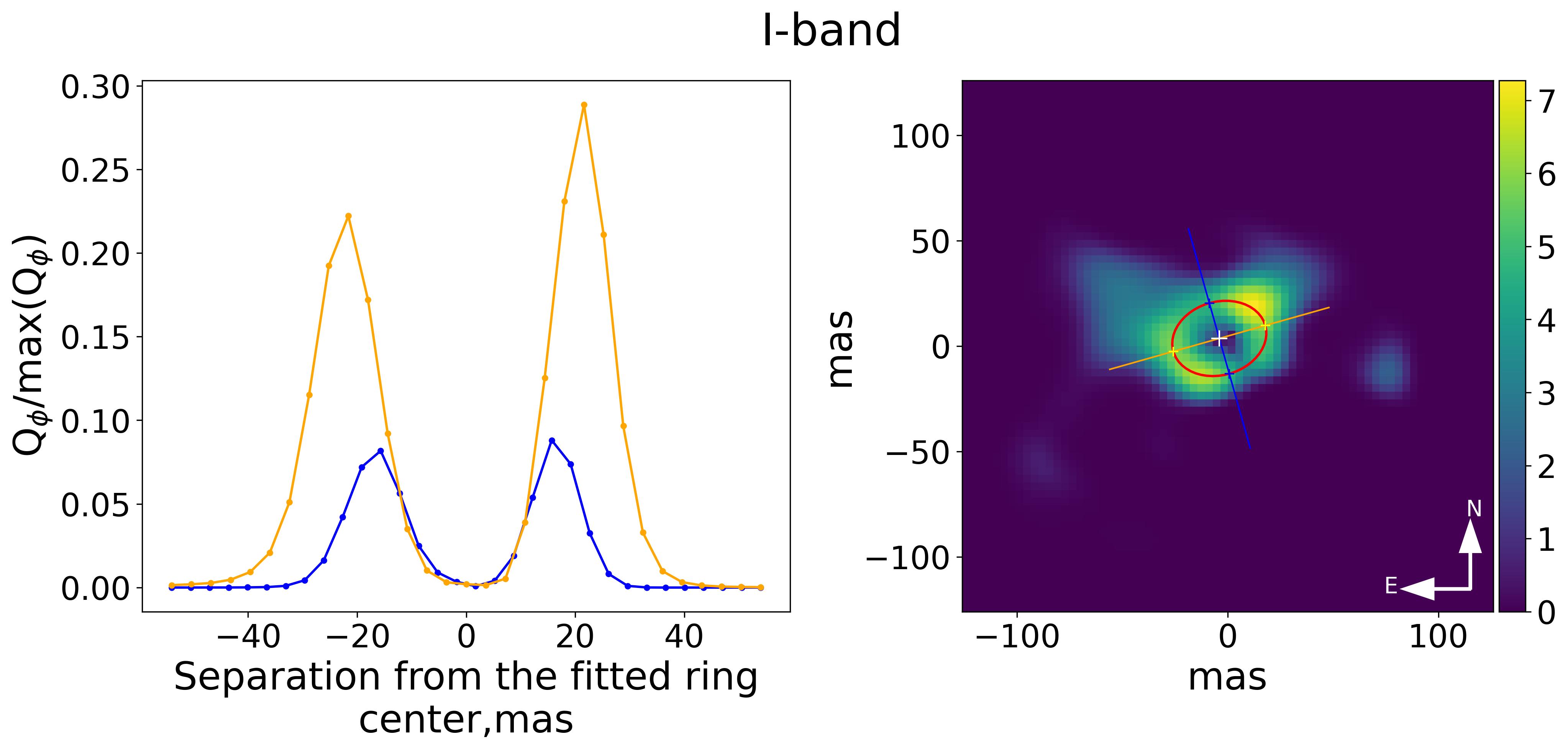}

    \includegraphics[width=0.46\columnwidth]{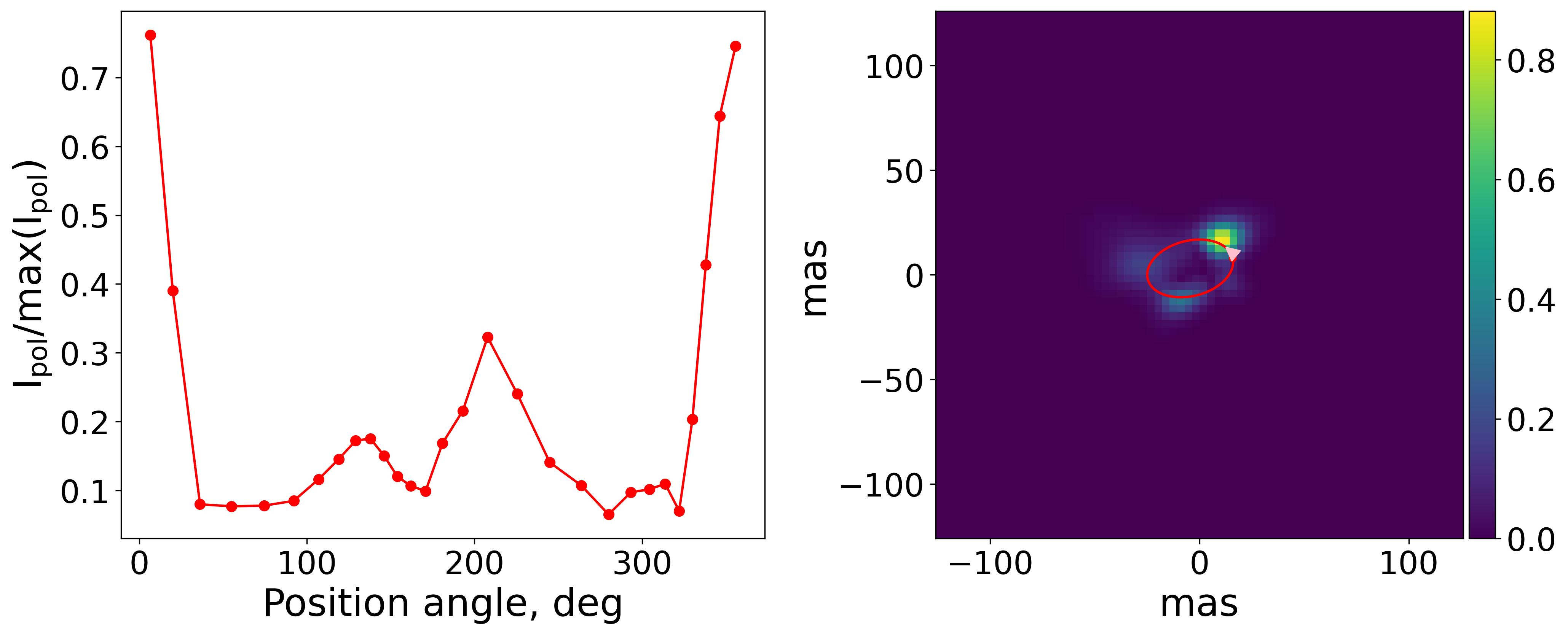}
    \includegraphics[width=0.46\columnwidth]{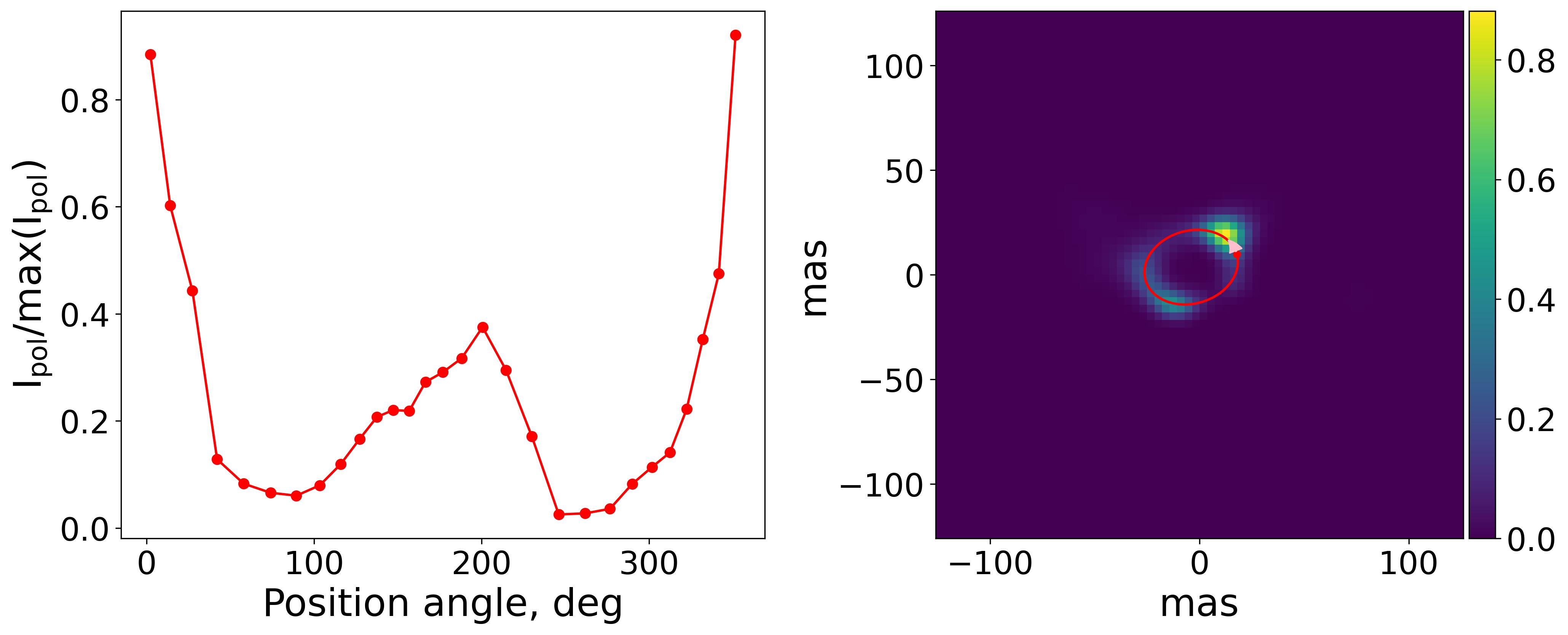}
     
     \includegraphics[width=0.455\columnwidth]{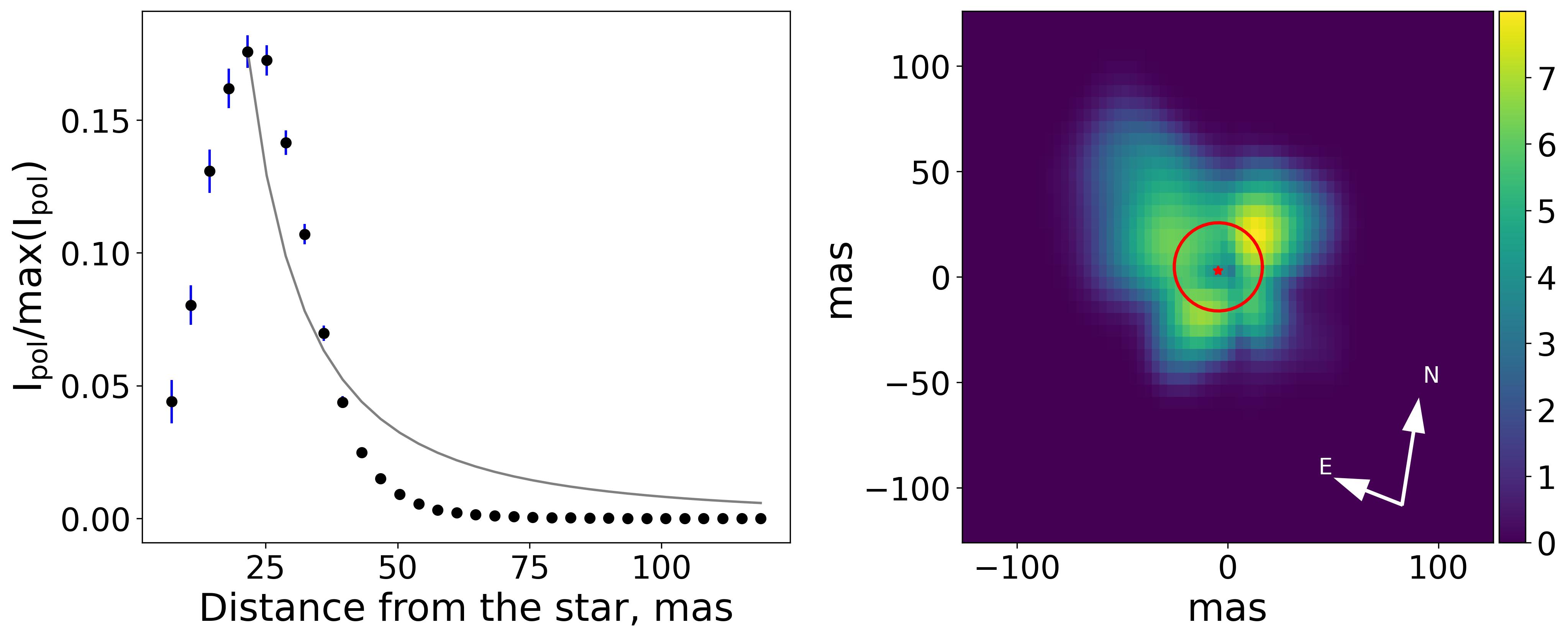} 
     \includegraphics[width=0.455\columnwidth]{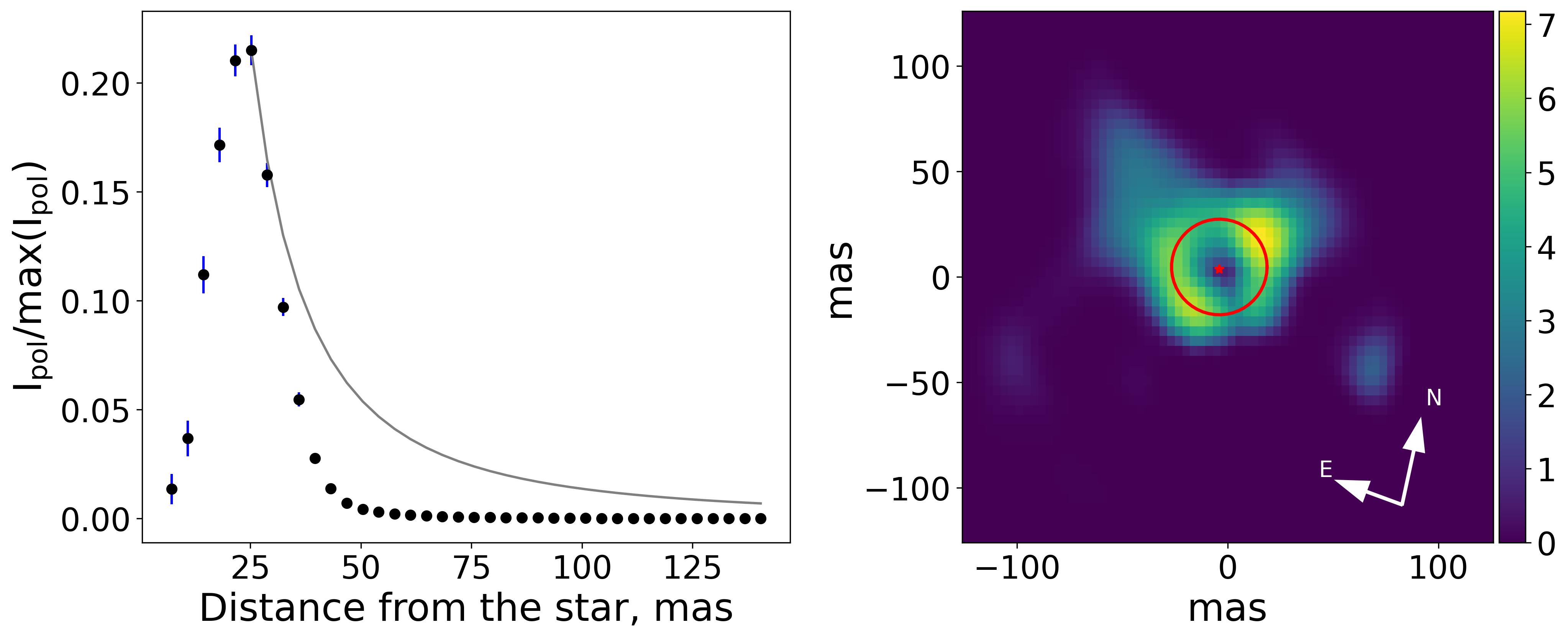}
    \caption{Same as Figure~\ref{fig:paper3_profiles_hr4049} but for HR\,4226.}
    \label{fig:paper3_profiles_hr4226}
\end{figure*}

\begin{figure*} 
     \includegraphics[width=0.45\columnwidth]{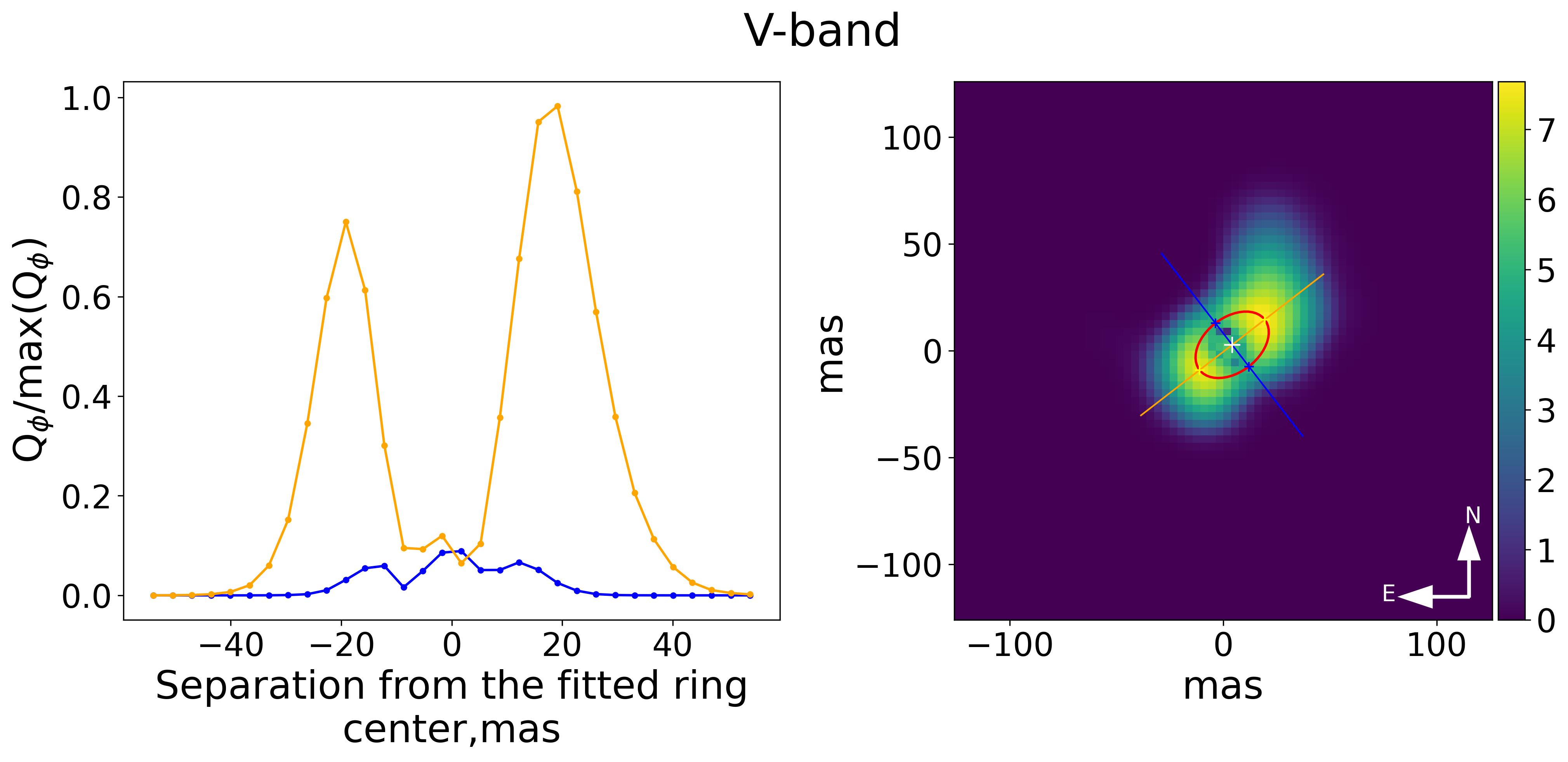}
     \includegraphics[width=0.455\columnwidth]{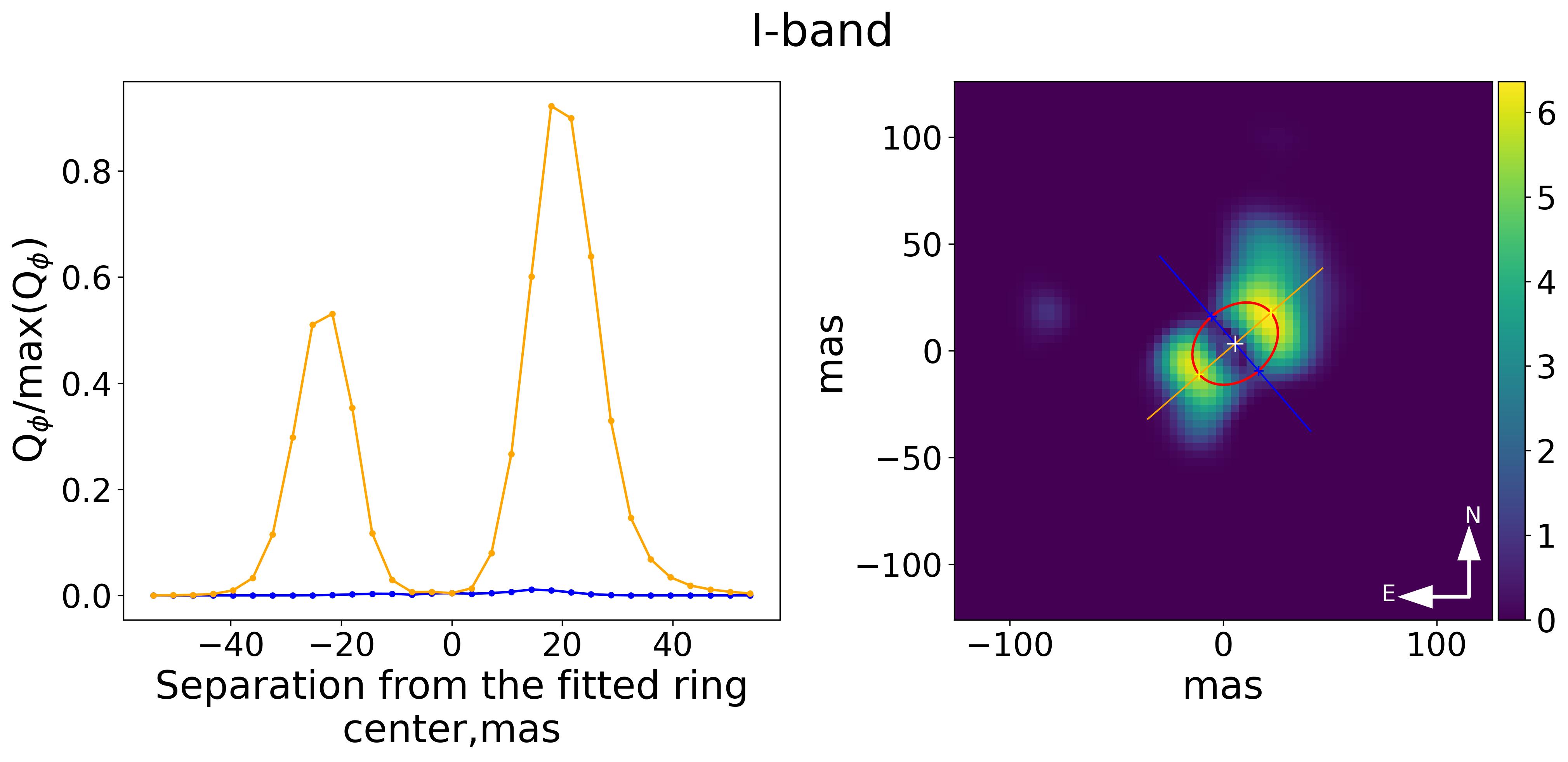}
     
    \includegraphics[width=0.46\columnwidth]{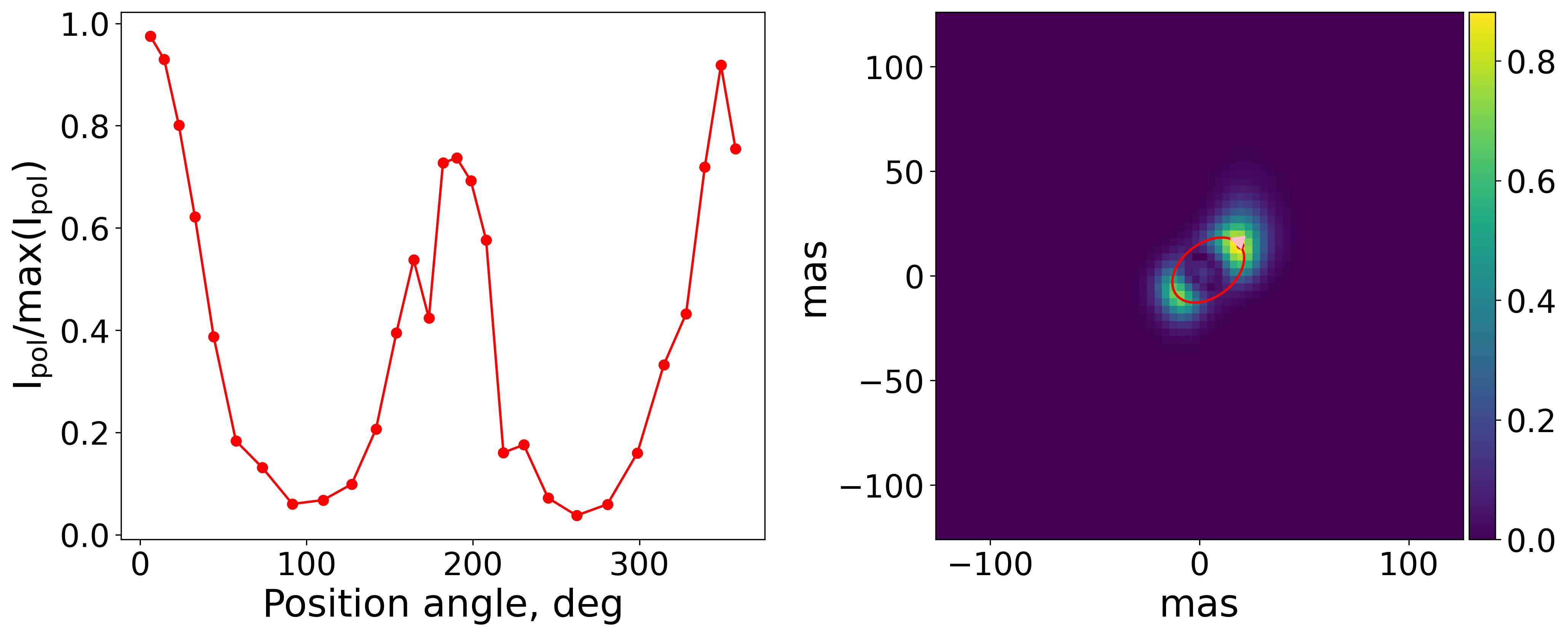}
     \includegraphics[width=0.46\columnwidth]{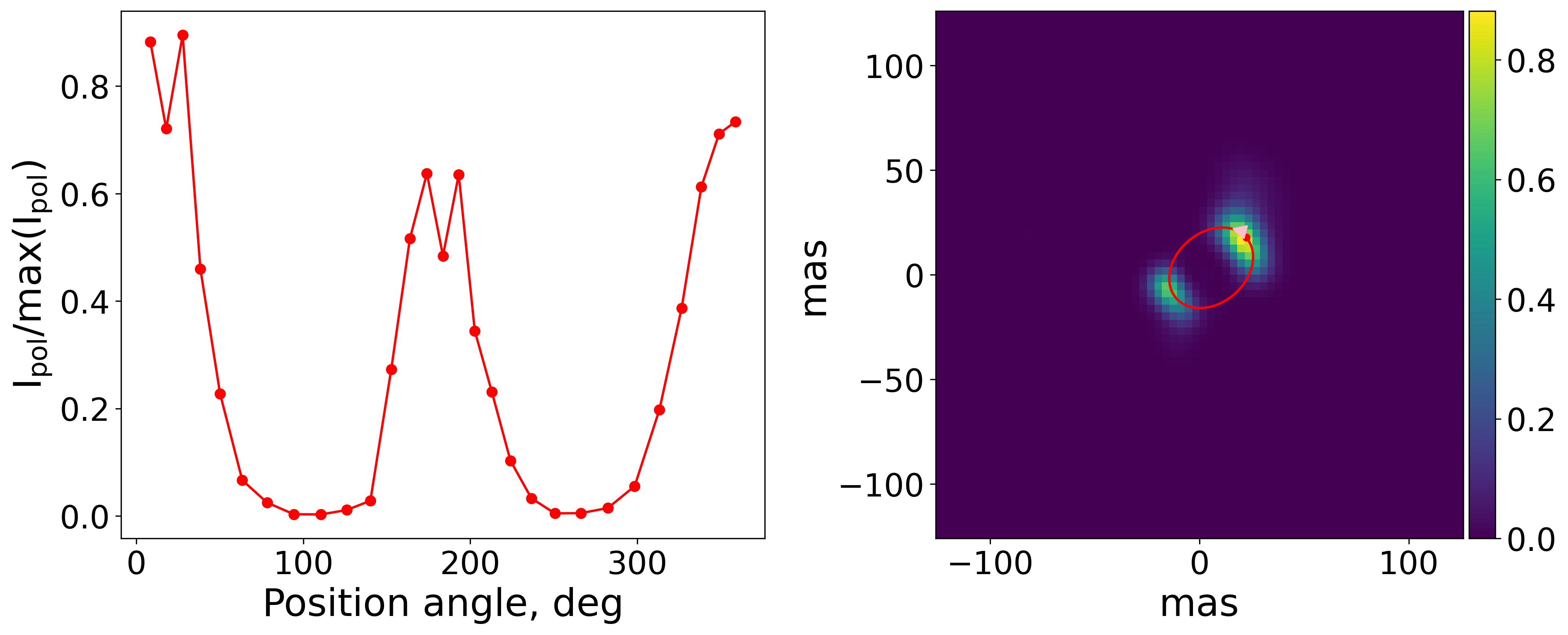}
     
     \includegraphics[width=0.455\columnwidth]{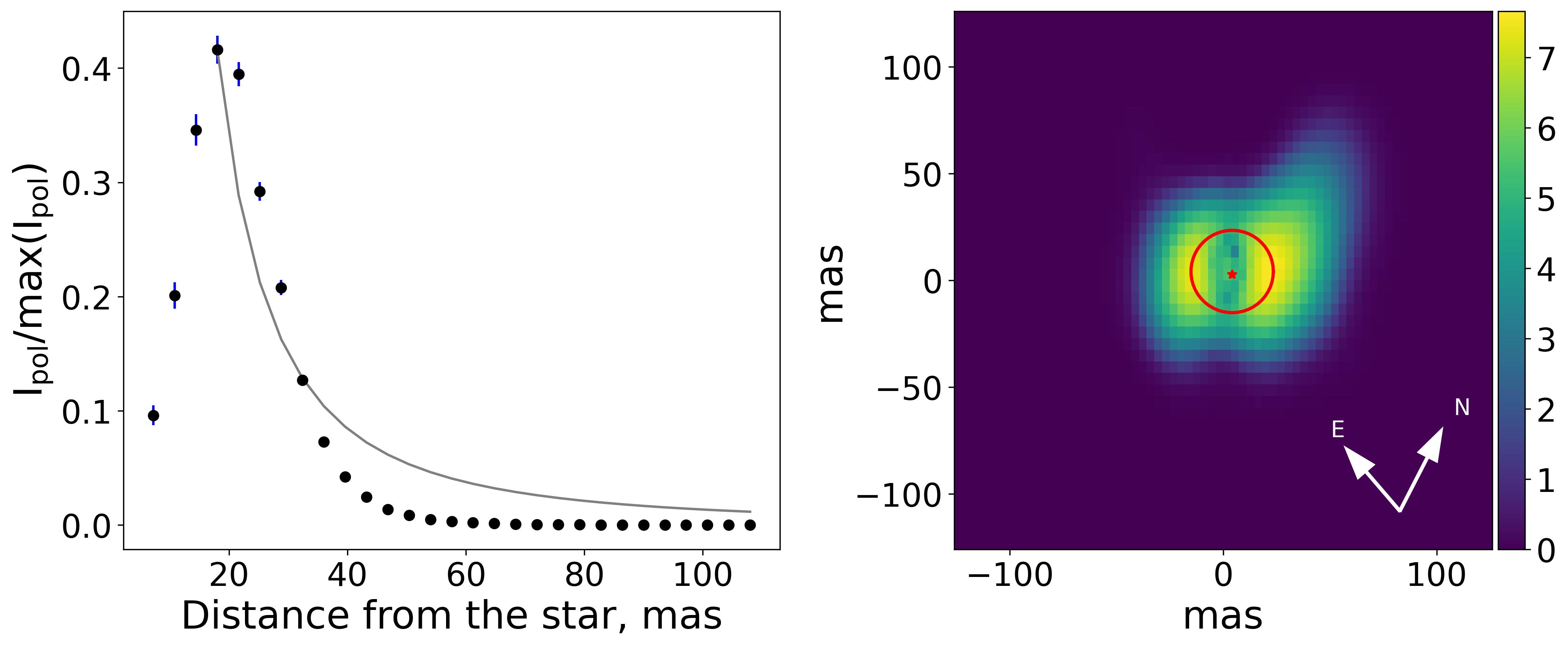} 
    \includegraphics[width=0.455\columnwidth]{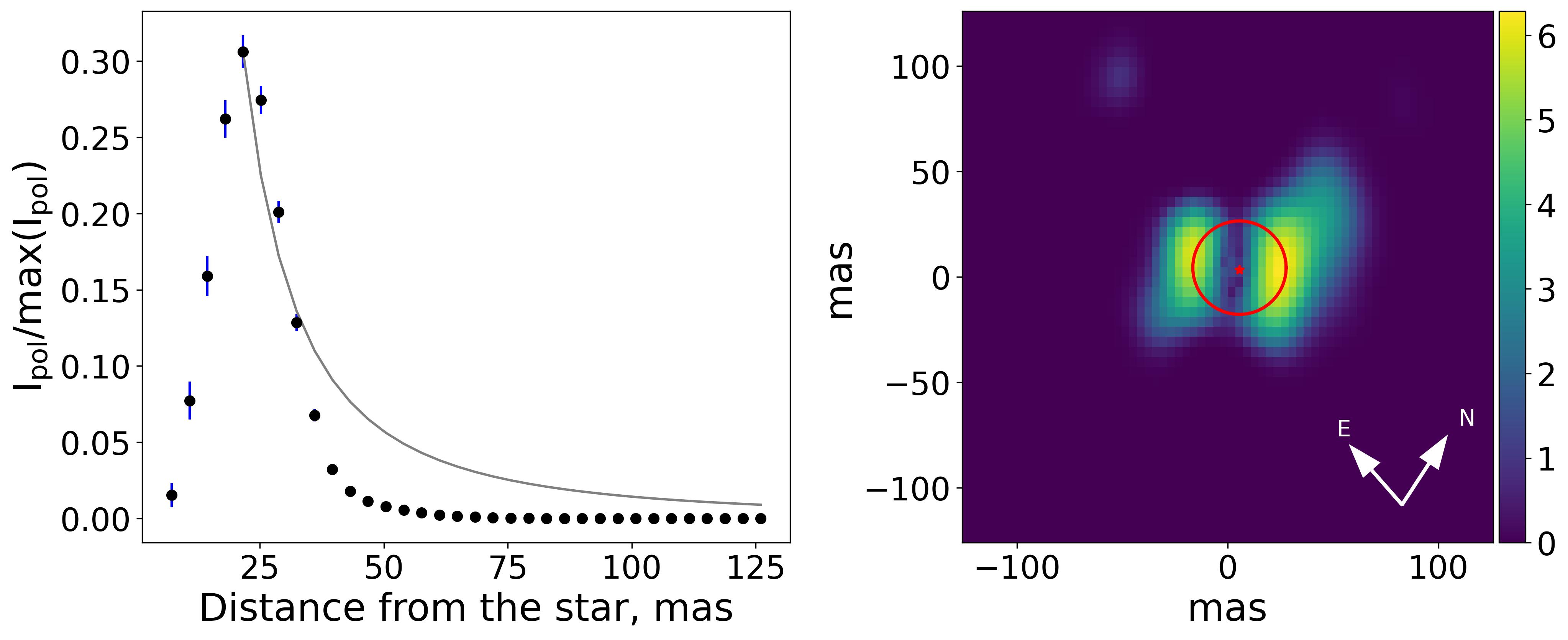}
    \caption{Same as Figure~\ref{fig:paper3_profiles_hr4049} but for U\,Mon}
    \label{fig:paper3_profiles_umon}
\end{figure*}

\begin{figure*} 
     \includegraphics[width=0.45\columnwidth]{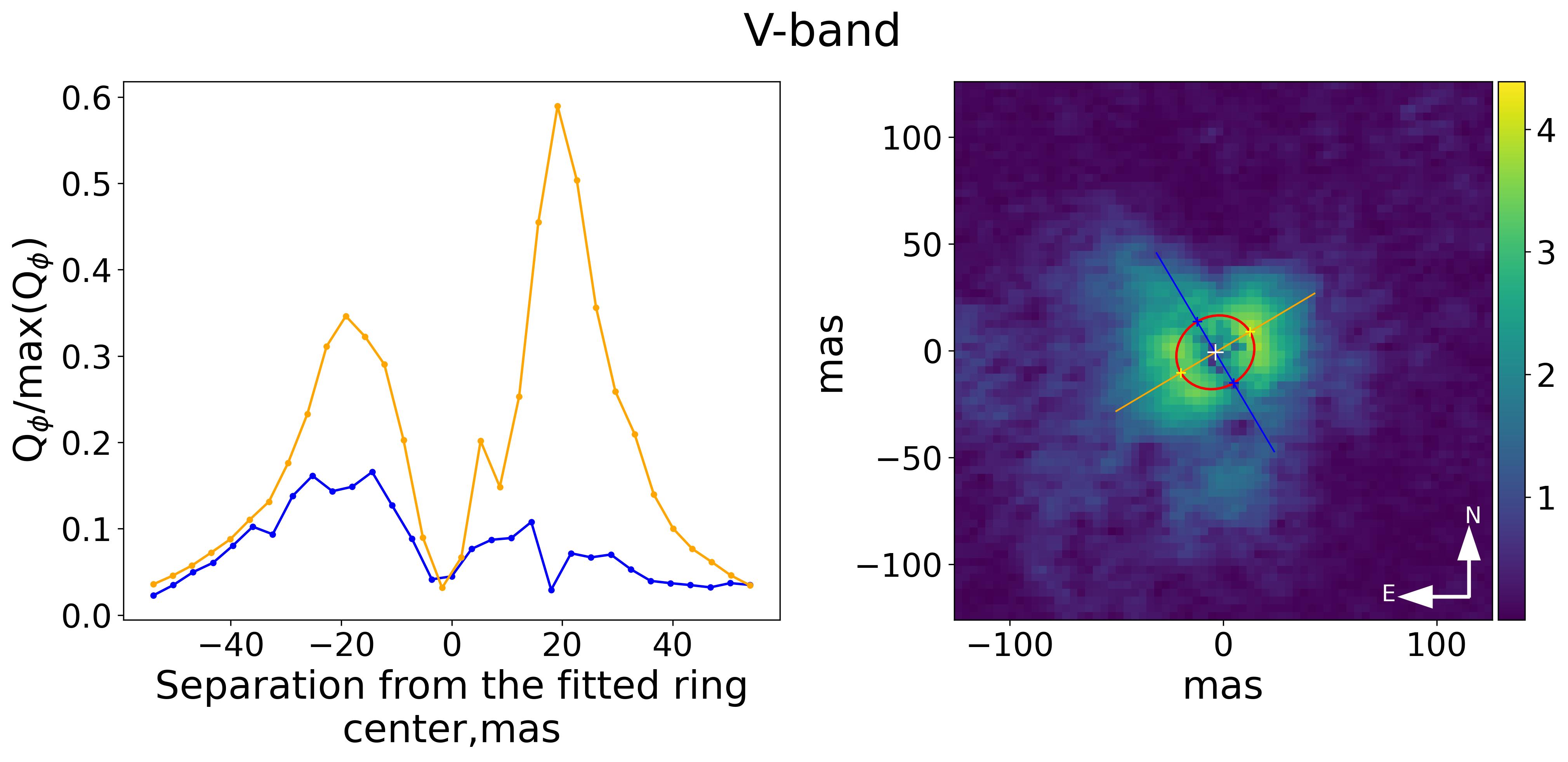}
     \includegraphics[width=0.45\columnwidth]{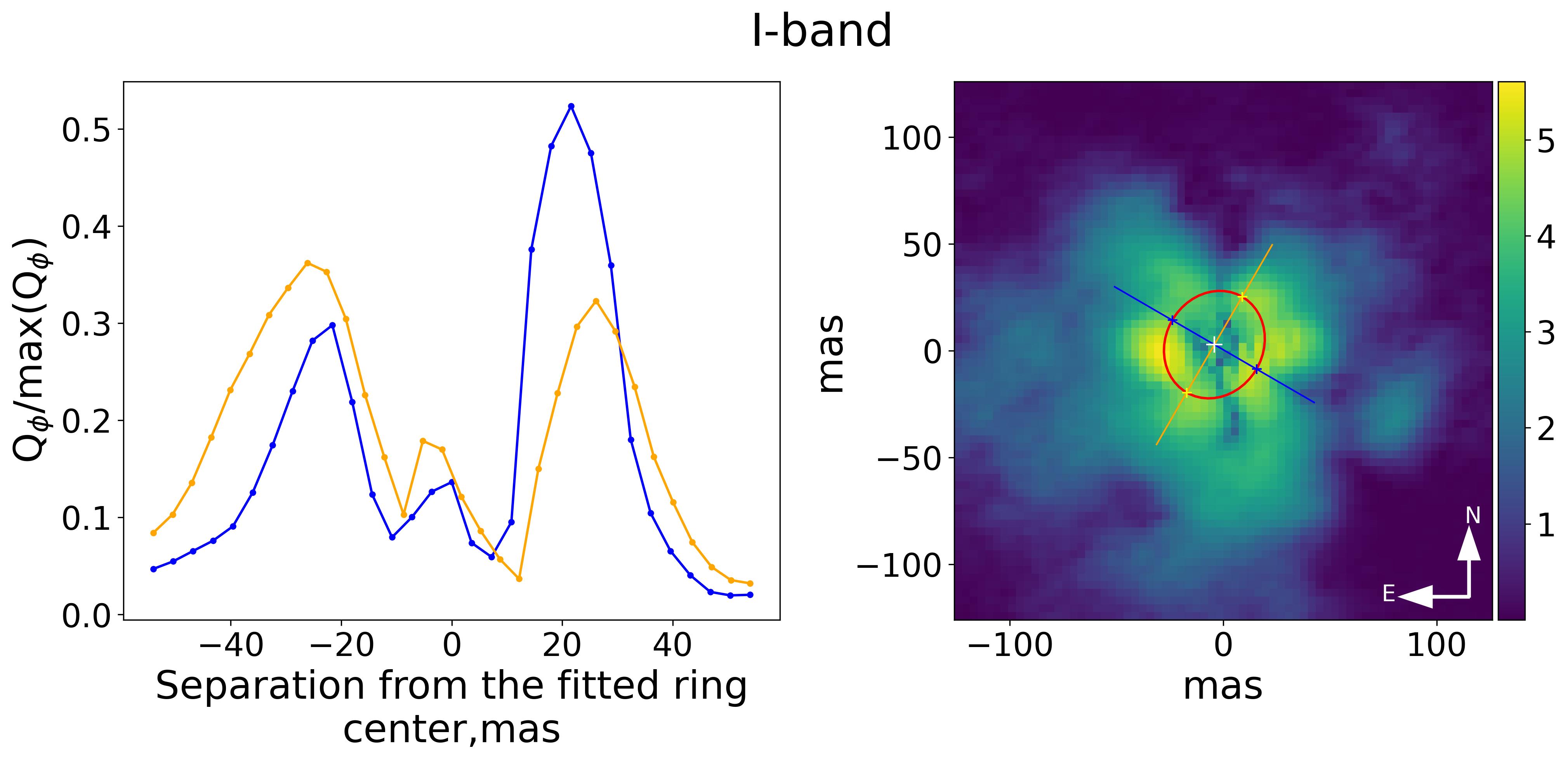}
    
    \includegraphics[width=0.46\columnwidth]{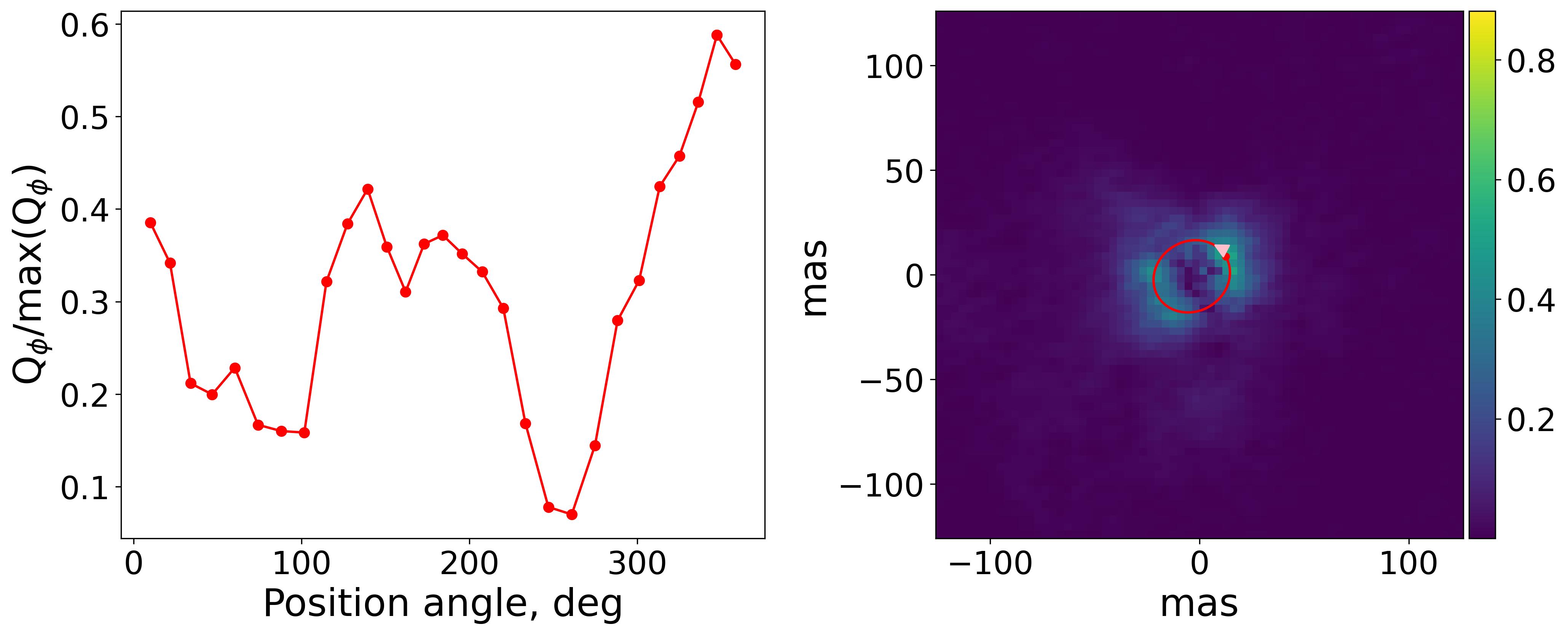}
     \includegraphics[width=0.46\columnwidth]{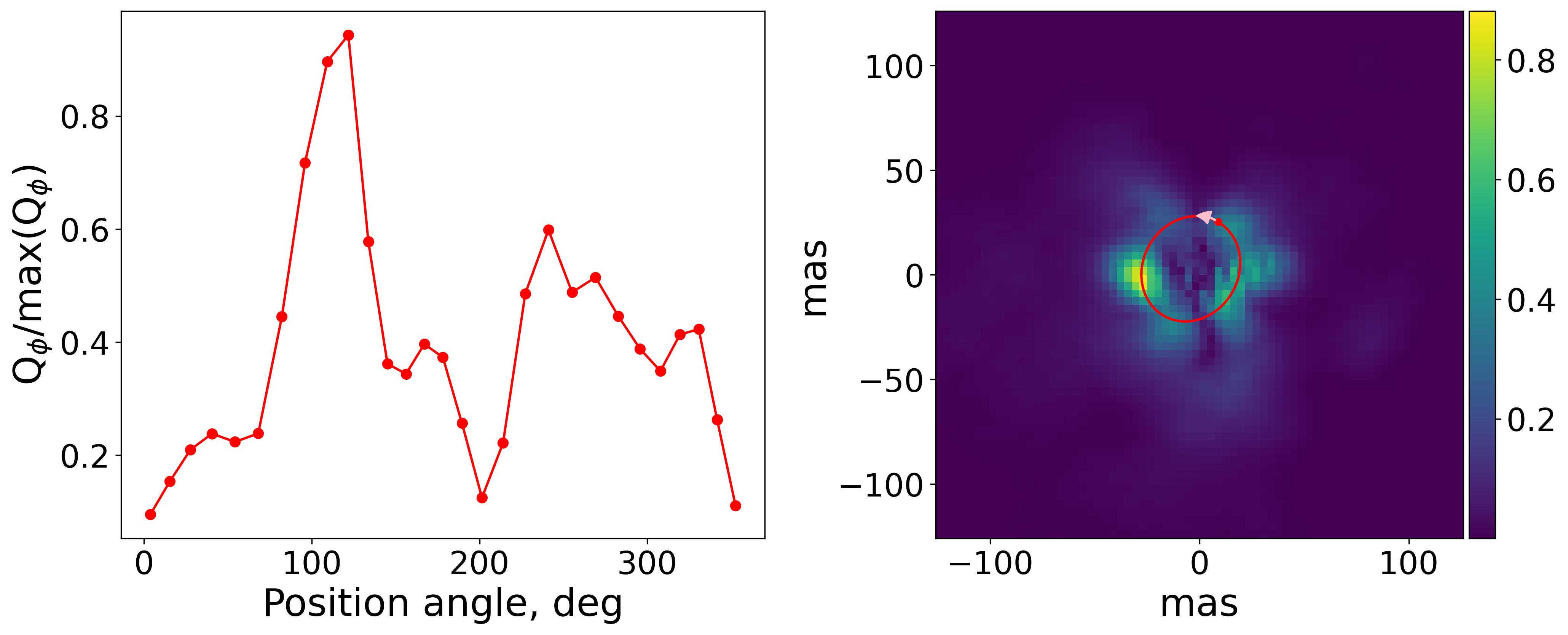}
     
     \includegraphics[width=0.45\columnwidth]{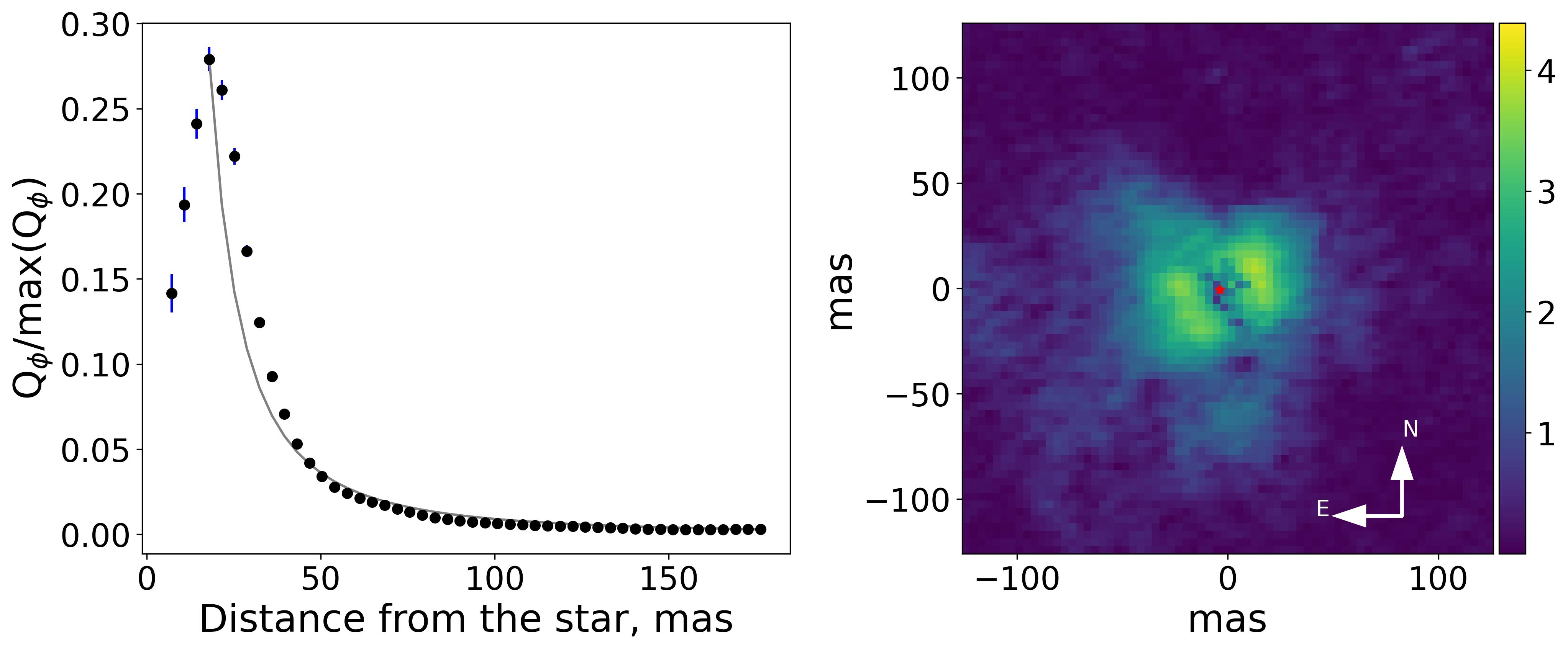}
     \includegraphics[width=0.45\columnwidth]{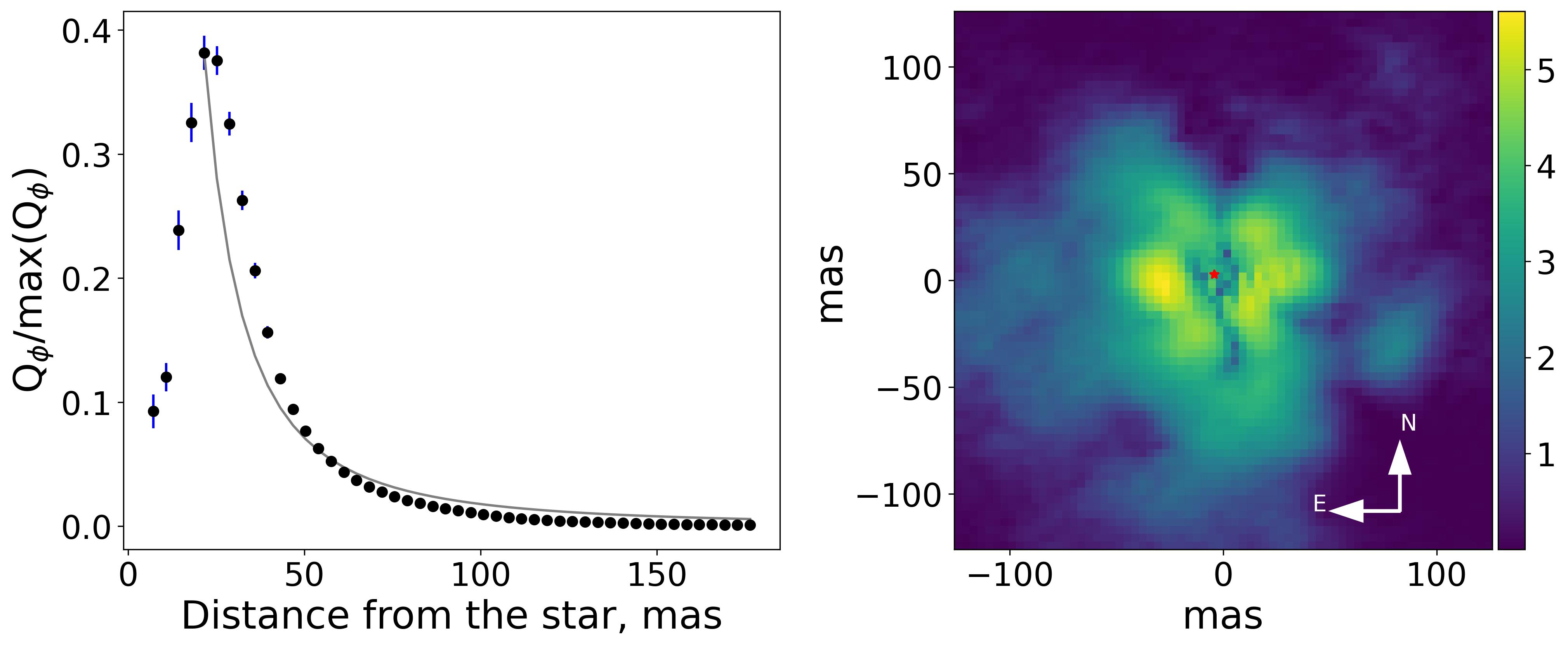}

    \caption{Same as Figure~\ref{fig:paper3_profiles_hr4049} but for V709\,Car.}
    \label{fig:paper3_profiles_v709car}
\end{figure*}

\begin{figure*} 
     \includegraphics[width=0.45\columnwidth]{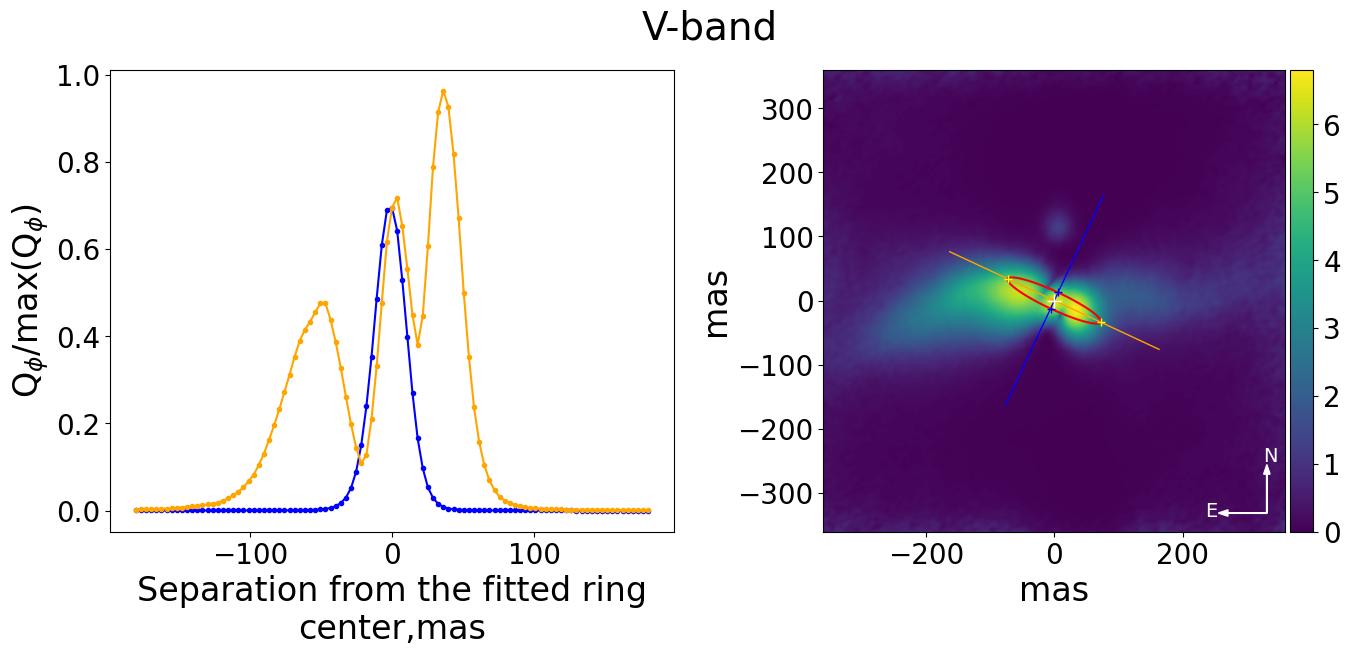}
     \includegraphics[width=0.45\columnwidth]{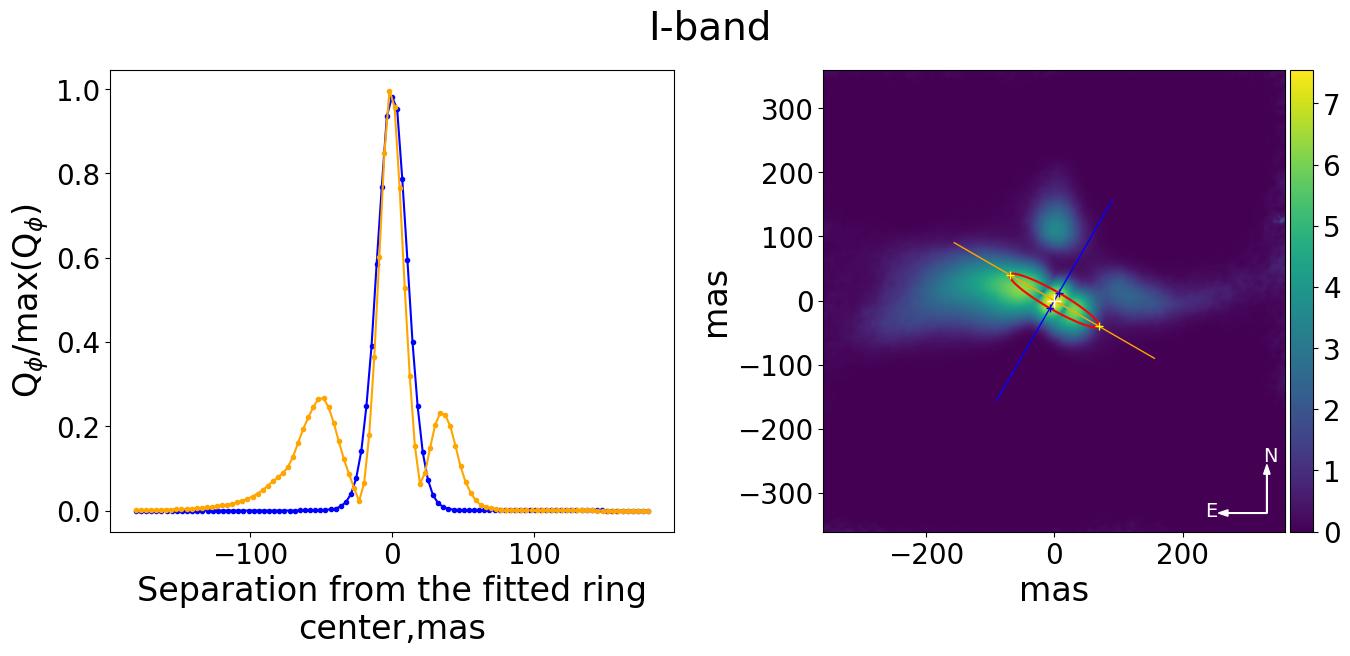}
     
     \includegraphics[width=0.455\columnwidth]{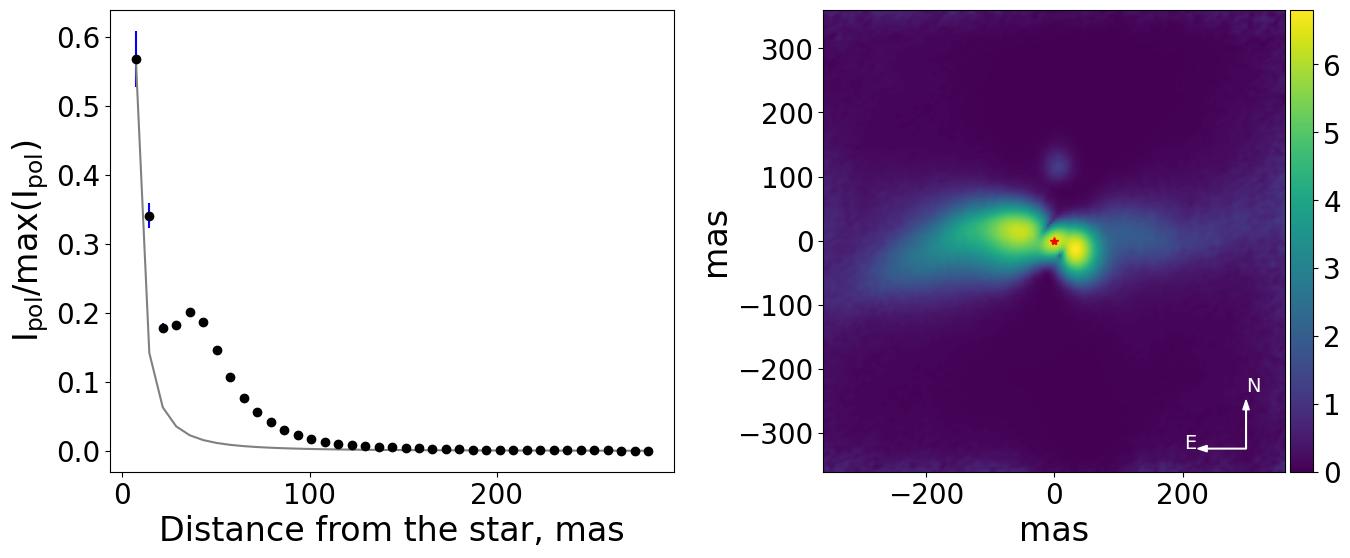}
     \includegraphics[width=0.455\columnwidth]{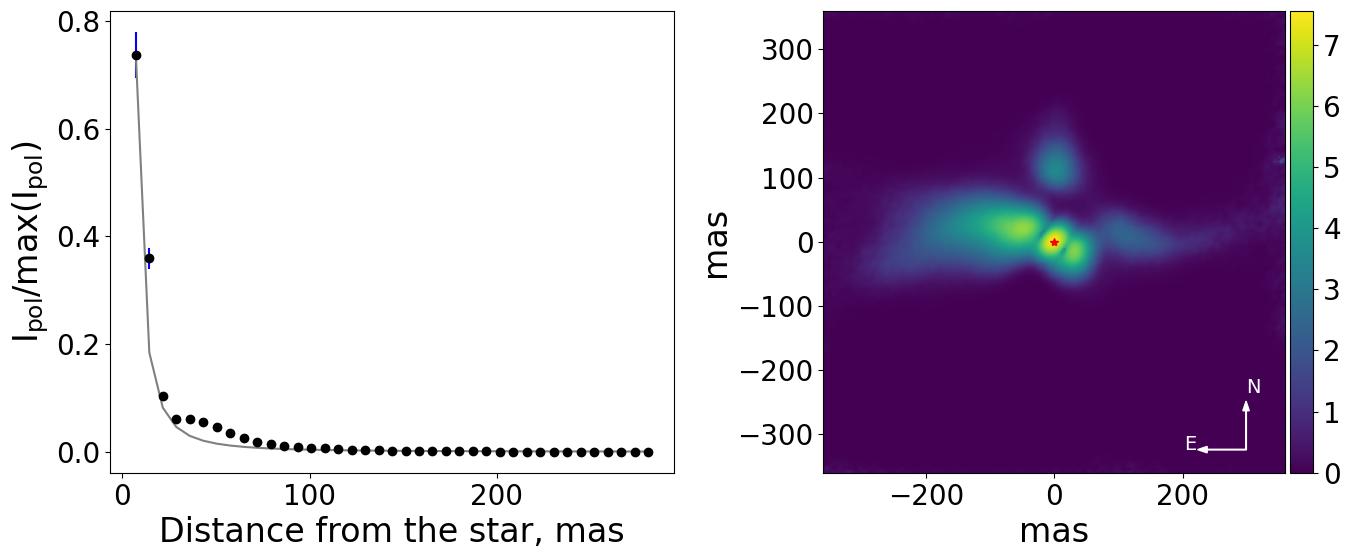}

    \caption[Brightness profiles for AR\,Pup.]{Brightness profiles for AR\,Pup in $V$ (left panel) and $I'$ (right panel) bands (see \ref{sec:paper3_ap_profiles}) with corresponding  polarised images. In each panel, the left image displays the brightness profile, while the right image presents the corresponding polarised image. The top row shows linear brightness profiles of the polarised image along the major and minor axes of the disc midplane. The bottom row presents radial brightness profiles of the deprojected polarised image. In the radial brightness profile plots, grey solid lines are added to indicate a r$^{-2}$ drop-off, expected from a scattered light signal due to the dissipation of stellar illumination. Polarised images are presented on an inverse hyperbolic scale. The low intensity of the central 5x5 pixel region of each  polarised image is a reduction bias caused by correction of the unresolved central polarisation (see Section~\ref{sec:paper3_data_reduction}).}
    \label{fig:paper3_profiles_ar_pup}
\end{figure*}

In Figure\ref{fig:paper3_profiles_hr4049}, \ref{fig:paper3_profiles_hr4226}, \ref{fig:paper3_profiles_umon}, \ref{fig:paper3_profiles_v709car}, \ref{fig:paper3_profiles_ar_pup}, we present three types of brightness profiles in the $V$ and $I'$ bands for each applicable target in our sample: top panels represent the linear profiles along the major and minor axes of the visible `ring' structure, middle panels represent the azimuthal brightness profiles along the `ring' surface, and bottom panels represent the radial profiles of the polarised disc intensity deprojected to a face-on orientation. 

The linear brightness profiles were calculated for all 5 sample post-AGB systems. However, we note that for AR\,Pup we use the estimation of the disc midplane orientation to define the position of major and minor axes (see Section~\ref{sec:paper3_ar_pup} for details).

The azimuthal brightness profiles were computed using bi-linear interpolation of the intensity from the four nearest pixels along the fitted ellipse in the linearly polarised image (see Section~\ref{sec:paper3_disc_orient}). These profiles trace variations in azimuthal brightness starting from the eastern end of the major axis and proceeding counterclockwise. We note that we do not calculate azimuthal brightness profiles for AR\,Pup because it does not show an elliptical `ring' due to the orientation of the system (see Section~\ref{sec:paper3_disc_orient} for more details).

The radial brightness profiles reflect variations in the observed radial intensity distribution and are therefore affected by disc orientation.
 To correct for this, we first reconstructed a 'face-on' view of the disc based on the estimated inclination of the resolved disc surface for both the $V-$ and $I'-$bands before calculating the radial brightness profile (see Section~\ref{sec:paper3_disc_orient} for details on the inclination). The deprojected polarised images were then subdivided into annuli, with widths increasing proportionally to $\sqrt{r}$, where $r$ is the corresponding radius. Finally, to plot the radial brightness profile, we calculated the mean brightness per pixel for these radially tabulated annuli, focusing only on statistically significant regions based on the SNR (see Section~\ref{sec:paper3_data_reduction}). To estimate the uncertainty, we used the variance of the pixel values in the same annuli of the $U_\phi$ image. Although this method overestimates the noise if any astrophysical signal is present in the $U_\phi$ image (which is the case for all targets in our sample, see Section\ref{sec:paper3_data_reduction}), it still allows us to establish an upper limit. The resulting deprojected polarised images are presented alongside corresponding radial brightness profiles in the bottom row of Figure~\ref{fig:paper3_profiles_hr4049}, \ref{fig:paper3_profiles_hr4226}, \ref{fig:paper3_profiles_umon} for HR\,4049, HR\,4226 and U\,Mon respectively. We also plot similar profiles for AR\,Pup and V709\,Car; however, we do not perform the deprojection for these targets (Figure~\ref{fig:paper3_profiles_v709car}, \ref{fig:paper3_profiles_ar_pup}). 
 
 Further details on the methodology are provided in \citet{Andrych2023MNRAS.524.4168A} and each target’s profile are discussed in Section~\ref{sec:paper3_indiv_cases}.

\section{Angle of Linear polarisation (AoLP)}
\label{sec:paper3_ap_aolp}

In Figure\ref{fig:paper3_aolp}, we present the local angles of linear polarisation (AoLP, in white) over the resolved structures in the deconvolved $I_{\rm pol}$ ($Q_\phi$ for HR\,4049, see Section~\ref{sec:paper3_data_reduction}) images for all targets in our sample. All images are presented on an inverse hyperbolic scale.

\section{Comparison of AR Pup total intensity images}
\label{sec:paper3_appendix_arpup}

\begin{figure*} 
     \includegraphics[width=0.34\columnwidth]{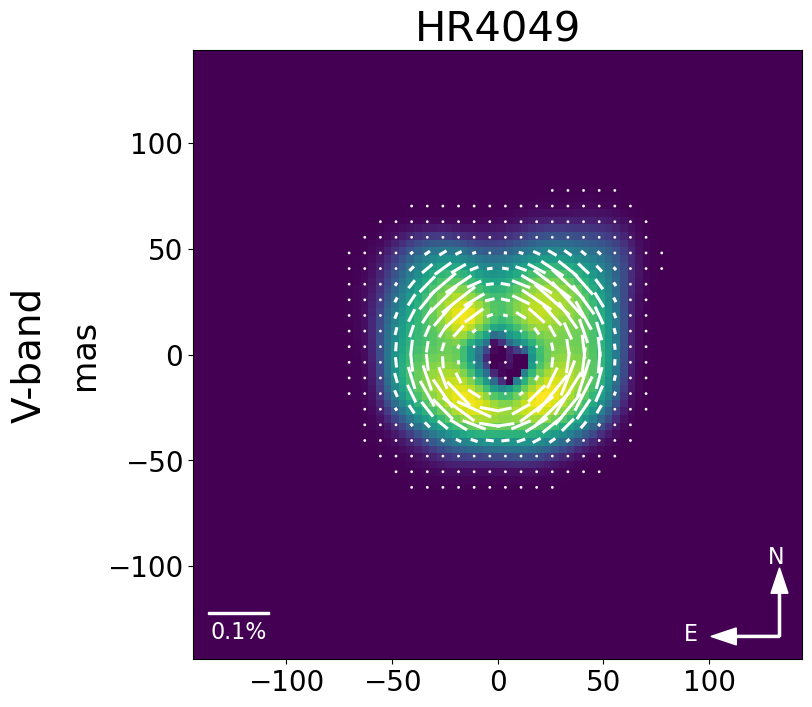}
    \includegraphics[width=0.3\columnwidth]{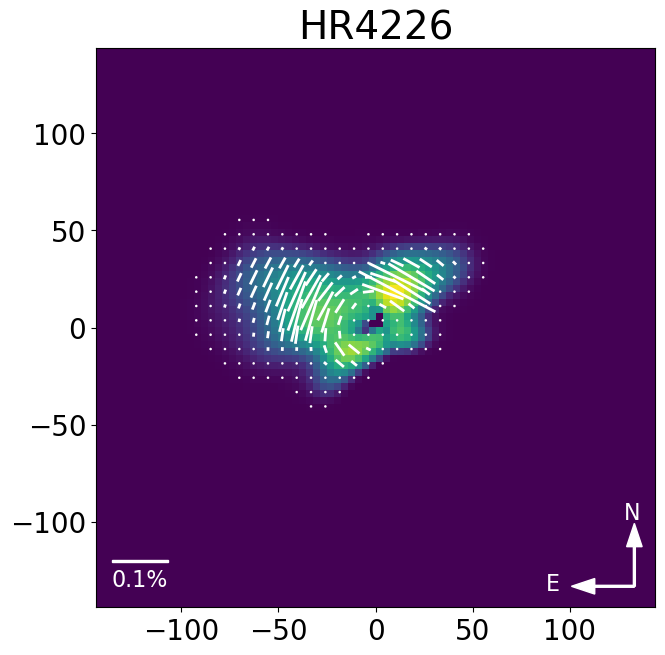}
    \includegraphics[width=0.3\columnwidth]{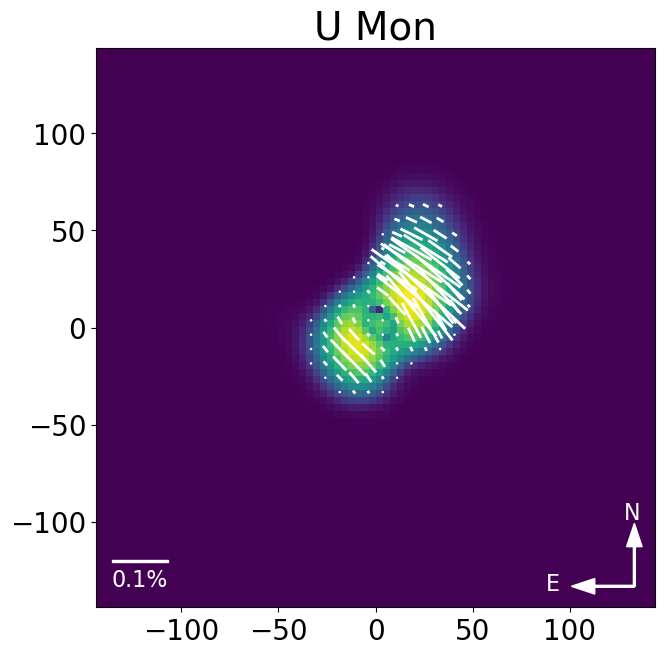}
    
     \includegraphics[width=0.34\columnwidth]{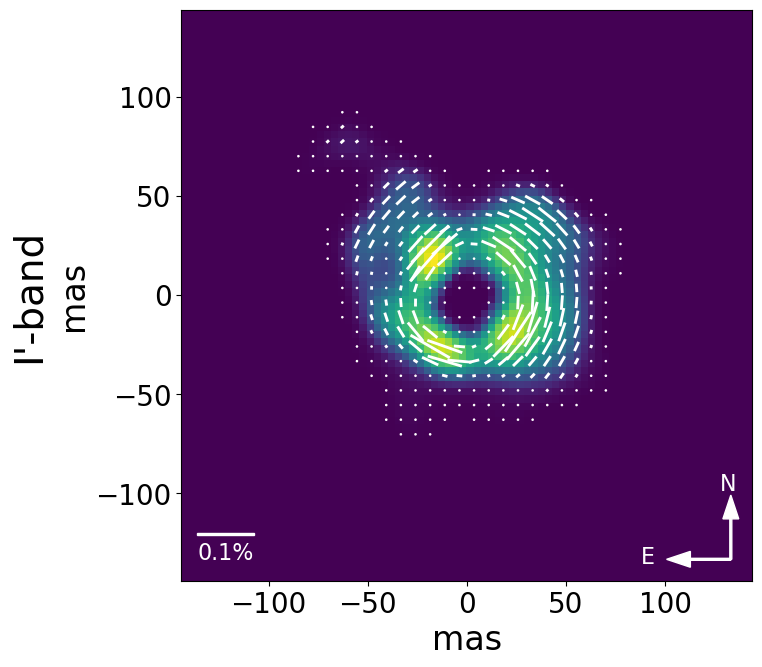}
    \includegraphics[width=0.3\columnwidth]{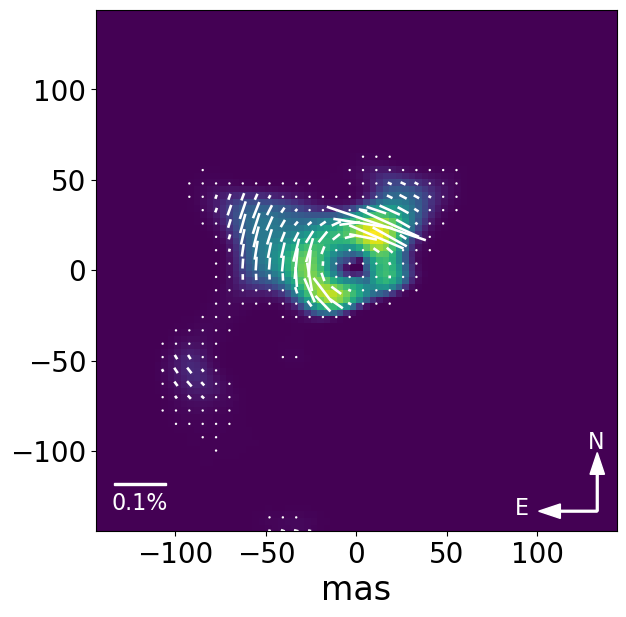}
    \includegraphics[width=0.3\columnwidth]{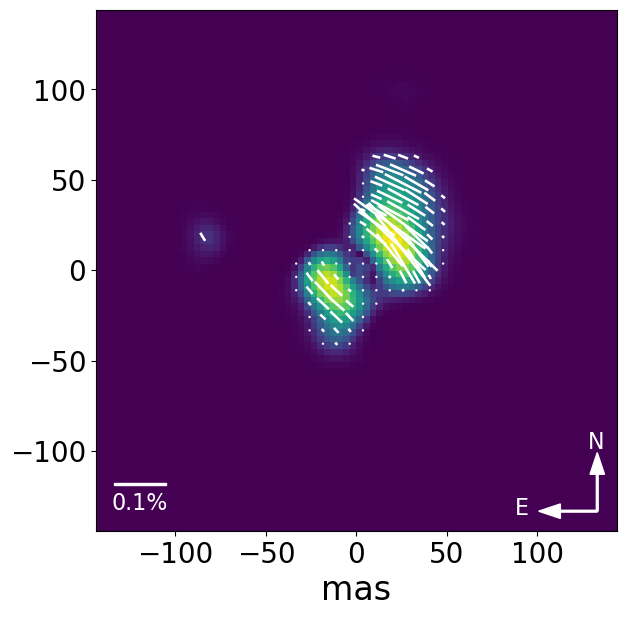}
     
     \includegraphics[width=0.34\columnwidth]{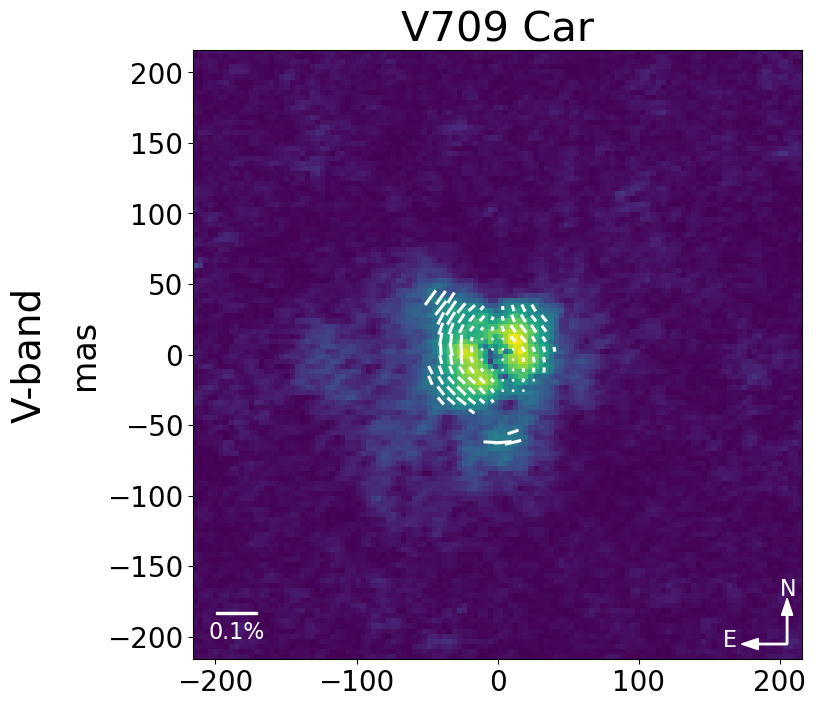}
    \includegraphics[width=0.3\columnwidth]{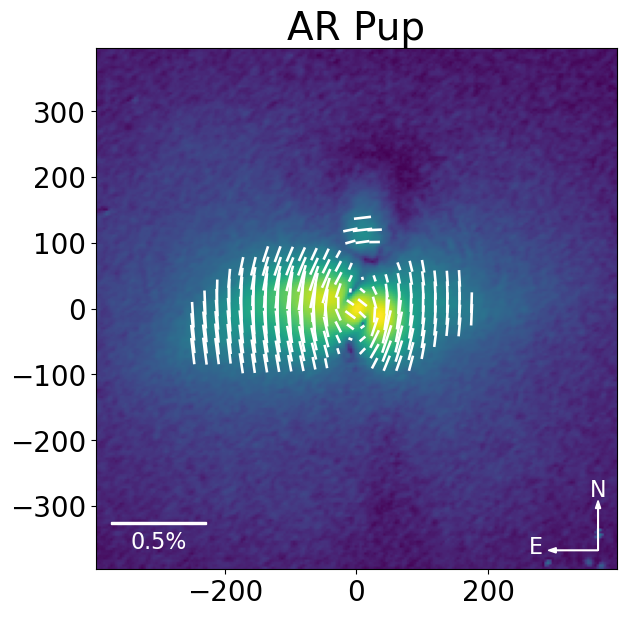}
    
     \includegraphics[width=0.34\columnwidth]{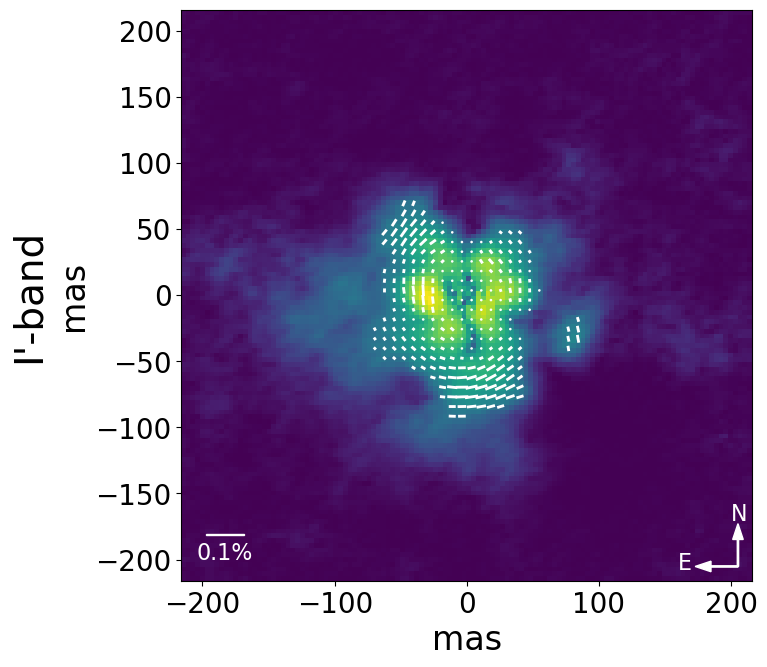}
    \includegraphics[width=0.3\columnwidth]{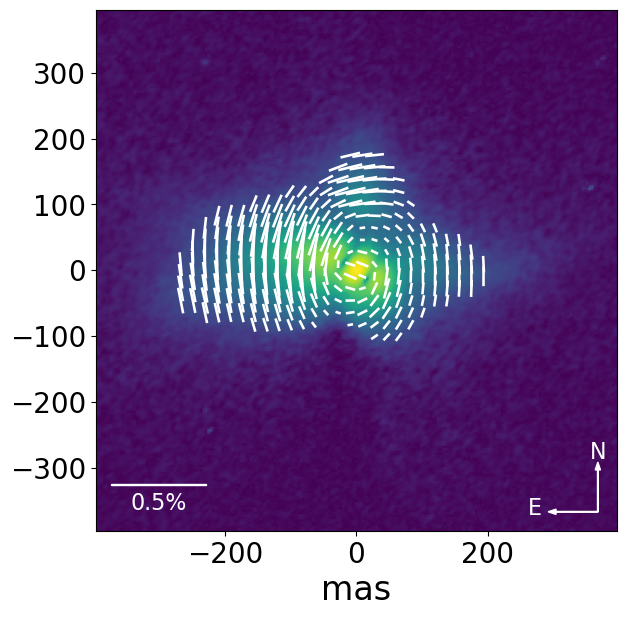}

    \caption[Local angles of linear polarisation for resolved polarised substructures for all targets in our sample.]{Local angles of linear polarisation (AoLP, in white) for resolved polarised substructures for all targets in our sample. All images are presented on an inverse hyperbolic scale. See \ref{sec:paper3_ap_aolp} for details.}
    \label{fig:paper3_aolp}
\end{figure*}

In Figure\ref{fig:paper3_aolp}, we present SPHERE/ZIMPOL total intensity images of AR Pup from this study and \citet{Ertel2019AJ....157..110E}, both taken in the $V$-band with a consistent orientation (North up, East to the left). These observations, separated by a two-year timespan, are further discussed in Section~\ref{sec:paper3_ar_pup}.

\begin{figure*}
    \includegraphics[width=0.9\linewidth]{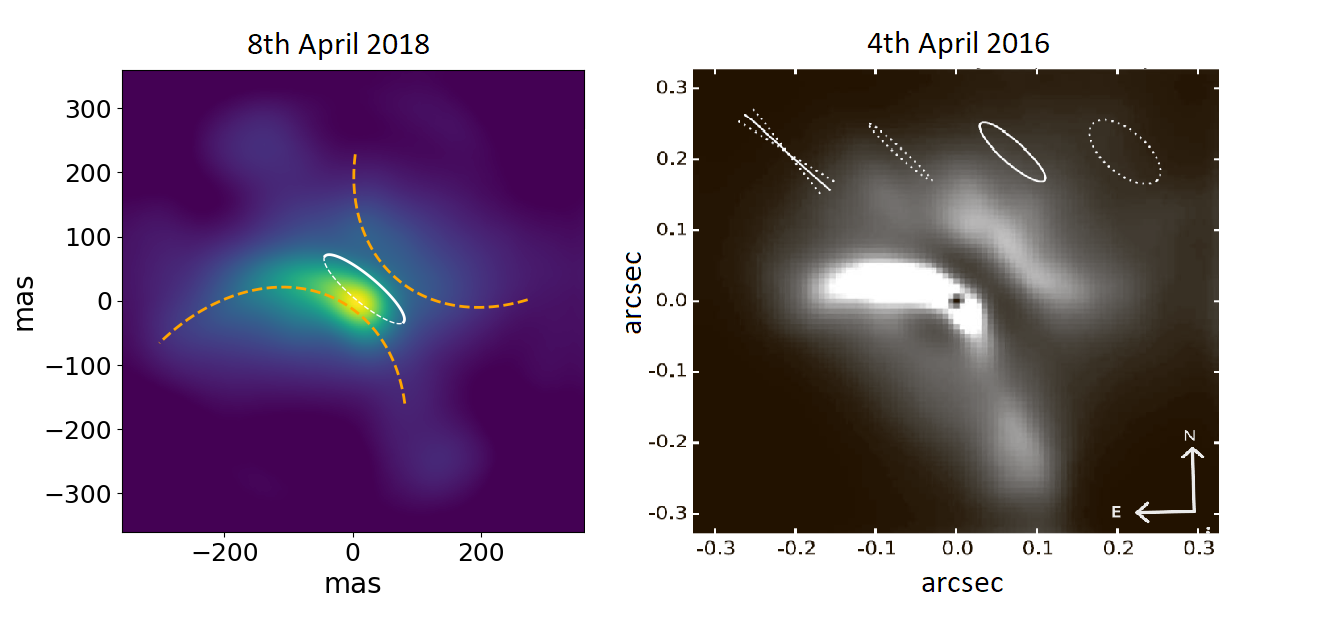}

    \caption{Comparison of SPHERE/ZIMPOL total intensity images of AR Pup from this study (left panel) and adopted from \citet{Ertel2019AJ....157..110E} (right panel). Both images are taken in $V-$band and oriented North up and East to the left. See Section~\ref{sec:paper3_ar_pup} for details.}
    \label{fig:ar_pup_ertel}
\end{figure*}

\end{document}